\documentclass[11pt,a4paper]{article}

\usepackage[margin=24mm]{geometry}
\usepackage{iftex}
\ifPDFTeX
  \usepackage[T1]{fontenc}
  \usepackage[utf8]{inputenc}
  \usepackage{lmodern}
\else
  \usepackage{fontspec}
\fi
\usepackage{amsmath,amssymb,booktabs,array,longtable,tabularx,graphicx}
\usepackage{microtype,xcolor,enumitem}
\usepackage[numbers,sort&compress]{natbib}
\usepackage{tikz}
\usetikzlibrary{arrows.meta,positioning,fit,calc}
\usepackage[unicode,colorlinks=true,linkcolor=teal,citecolor=teal,urlcolor=teal]{hyperref}
\usepackage{xurl}
\usepackage{soul}
\usepackage{soulutf8}
\renewcommand{\arraystretch}{1.18}
\newcolumntype{P}[1]{>{\raggedright\arraybackslash}p{#1}}

\newcolumntype{Y}{>{\raggedright\arraybackslash}X}
\title{Crystal Structure Prediction from Search to Learning:\\A 30-Year Expedition}
\usepackage{authblk}

\author[1,*]{Jianjun Hu}
\author[1]{Ying Feng}
\affil[1]{Department of Computer Science and Engineering, University of South Carolina, Columbia, SC 29201, USA}
\affil[*]{Corresponding author: \href{mailto:jianjunh@cse.sc.edu}{jianjunh@cse.sc.edu}}
\begin{document}
\maketitle
 
\begin{abstract}
Crystal structure prediction (CSP) seeks to determine periodic atomic or molecular arrangements from chemical composition and specified physical conditions. Over three decades, CSP algorithms have evolved from chemical construction and global energy optimization to hybrid approaches integrating symmetry, learned potentials, adaptive search, and deep generative models. This survey traces these developments through a curated corpus of method, software, review, and benchmark records published from 1996 to October 2026. We cover both inorganic and molecular-crystal prediction, including formula-conditioned diffusion, flow, and autoregressive models that preserve the prescribed composition. We organize the field by how methods construct or explore candidate structures. Our taxonomy comprises template and discrete construction, random search, evolutionary search, swarm and differential evolution, energy-landscape exploration, adaptive machine-learning-guided search, and learned generation. We also distinguish the roles of representation, relaxation, ranking, and physical qualification. Benchmark comparisons reveal no universally dominant approach. In a 46-target leaderboard snapshot, the best-performing template method achieves higher top-1 recovery than leading generative models. However, paired tests do not establish statistically significant differences between this method and several generative competitors. We further audit CSP predictions for 68 experimentally examined materials, of which 52 have structures confirmed by diffraction or microscopy. Predictions for 23 of the 68 materials preceded experiment, with most originating from first-principles global search. Within the reviewed evidence, we identify no independently confirmed discovery in which a deep generative model predicted the structure before experiment. We conclude by identifying the principal limitations of current CSP algorithms and outlining priorities for future research.

\end{abstract}
\tableofcontents
\clearpage
\section{Introduction}
Crystal structure prediction maps a chemical composition to its structure, which has become increasingly critical to new materials discovery \cite{CSP090} as almost all materials properties depend on the periodic atomic arrangement. Currently, CSP remains one of the hardest problems in materials science: the number of candidate structures grows exponentially with the number of atoms per cell, the energy landscape is rugged with many near-degenerate minima, and the thermodynamically stable structure is often separated from competing polymorphs by only a few meV/atom.
Besides handling many adjustable coordinates, a structure search must distinguish equivalent crystal descriptions, explore competing structural basins, select suitable cell sizes, and compare candidates with an energy or free-energy model of adequate accuracy. A chemically plausible atomic configuration can be far from a local minimum, and a locally minimized structure can still be irrelevant to the equilibrium phase or to an experimentally observed polymorph. These distinctions have shaped both inorganic and molecular CSP \citep{CSP036,CSP047,CSP061,CSP090,CSP103,CSP129}.
 
This survey reviews 30 years of technical development of CSP algorithms beginning in 1996, when landscape-based global optimization approaches were used to determine promising structure candidates \citep{CSP002}.
It includes the development of inorganic genetic search, periodic real-space evolutionary algorithms, ab initio random structure searching, particle-swarm optimization, variable-cell landscape exploration \citep{CSP007,CSP026,CSP027,CSP028,CSP042,CSP043}, and recently emerging deep generative CSP algorithms.
USPEX (Universal Structure Predictor: Evolutionary Xtallography)\citep{CSP027,CSP030}, the first widely used CSP algorithm, was reported in 2006 . Earlier foundations in energy-minimization approaches to molecular packing, exemplified by Williams's 1965 work \cite{Williams1965} and subsequent developments that introduced systematic searches across alternative packings
\cite{Dzyabchenko1984} and simulated-annealing approaches, including
early inorganic structure prediction by Pannetier and colleagues
in 1990 \cite{Pannetier1990} are acknowledged through the reviews, but are not counted in the survey corpus.
Recent CSP methods act on different stages of prediction: a machine-learning interatomic potential accelerates evaluation and relaxation, a Bayesian acquisition policy decides which structures receive expensive energy calculations, and a diffusion or flow generative model changes the candidate distribution itself. These contributions are complementary, and successful workflows often combine them \citep{CSP059,CSP075,CSP085,CSP102,CSP106,CSP128,CSP146}.
 
This review organizes CSP algorithms and models around candidate construction and search while treating representations, evaluators, conditioning information, and validation as separate dimensions. It covers both atomic and molecular crystals, including modern conditional generative models. The purpose is to explain how the algorithms work at the level supported by their reported formulations, where their assumptions differ, and what their evaluations establish.
 
This organization complements earlier reviews of energy minimization, machine-learning acceleration, and generative CSP. Graser, Kauwe, and Sparks~\citep{Graser2018Review} reviewed energy-minimization and machine-learning approaches and treated CSP as the classification of compositions into structure prototypes. Li et al.~\citep{Li2024MLCSPSimple} surveyed conventional search and its acceleration by graph neural networks and machine-learned force fields, together with early generative models. Chen et al.~\citep{Chen2025CSPMeetsAI} discussed generative models, including language models, for CSP. The present survey differs in three respects: it covers molecular as well as inorganic CSP, it organizes methods by how candidates are proposed and treats machine learning as one ingredient among several, and it examines how CSP methods are evaluated, including benchmark data, leakage, and whether generative models preserve a supplied formula.
 
\subsection{Literature scope and evidence}
The review draws on an initial curated, source-linked corpus of records spanning 1996 through 16 October 2026. The collection includes reusable algorithms, substantive workflow developments, selected software releases, reviews, benchmark studies, and enabling evaluators. Routine applications of an existing CSP algorithm to a new composition were not systematically collected. The initial corpus contains 136 atomic/inorganic records, 54 molecular/organic records, two cross-domain records, and four surface or cluster records included as methodological context. Fourteen further records were added after the initial corpus was assembled: seven papers on contact- and distance-constrained reconstruction, lattice and symmetry prediction, and quantitative evaluation, and seven on benchmarking, constrained molecular evolution, topology-based generation, graph-assisted search, and recent generative and agent-based CSP methods.
 
Records were collected from classical global search, learned and adaptive search, and molecular CSP and benchmarking, and extended by citation chaining and targeted searches on symmetry, discrete optimization, and recent generative methods. Bibliographic details were checked against Crossref where a DOI exists and otherwise against proceedings, author repositories, or preprint servers.
The supplementary catalogue preserves the initial 196-record corpus, with stable identifiers, method families, domains, and publication status. The expanded historical analysis in Figure~\ref{fig:literature-trends} includes 14 additional records, listed in the accompanying figure data, for a total of 210. The main text discusses representative and contrasting works, and cites software releases where a released interface differs from the original paper. Preprints and version-dependent software capabilities are identified where relevant.
 
Comparing these diverse CSP methods first requires a common definition of the prediction task: what information is supplied, which structural variables remain unknown, and what constitutes a successful prediction.
 
\section{Problem definitions and crystallographic variables}
\subsection{Fixed-composition prediction}
\begin{figure}[!htb]
\centering
\input{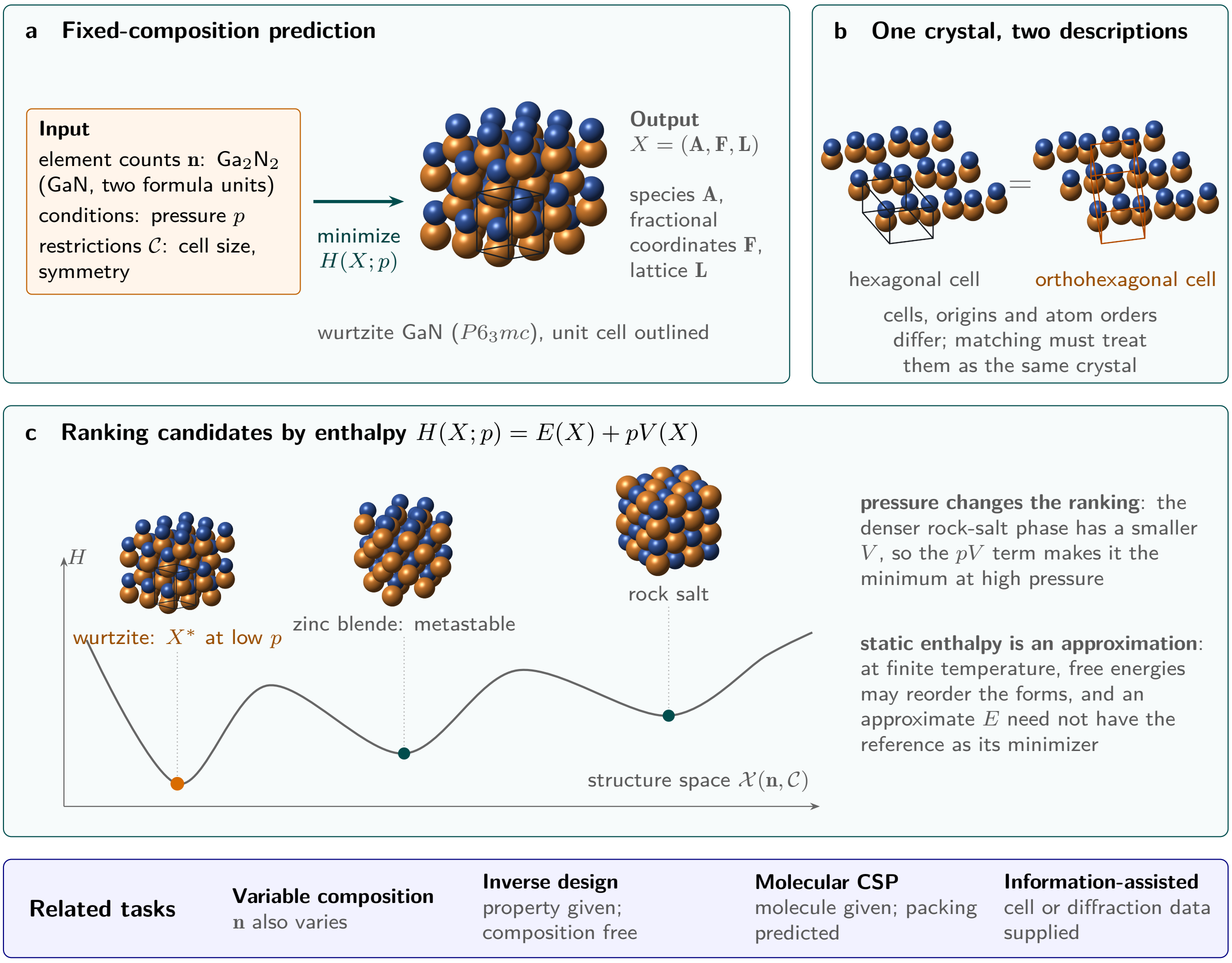}
\caption{Defining crystal structure prediction, illustrated with GaN. (a) Fixed-composition prediction: given the element counts of the cell (here Ga$_2$N$_2$, two formula units), the conditions, and any restrictions, find the structure $X=(\mathbf A,\mathbf F,\mathbf L)$ that minimizes the enthalpy; the output is wurtzite GaN. (b) The same crystal can be described by different cells, origins, and atom orders, so representations and structure matching must treat such descriptions as equivalent. (c) Candidate structures are ranked by $H(X;p)=E(X)+pV(X)$: wurtzite is the minimum at ambient pressure, zinc blende is a close metastable polymorph, and the denser rock-salt phase becomes the minimum at high pressure. Energies are schematic. Related tasks that need separate comparisons are listed at the bottom. Ga: orange; N: blue.}
\label{fig:problem-definition}
\end{figure}

An atomic periodic structure can be represented as
\begin{equation}
 X=(\mathbf A,\mathbf F,\mathbf L),\qquad
 \mathbf A=(a_1,\ldots,a_N),\quad
 \mathbf F\in[0,1)^{N\times3},\quad \mathbf L\in\mathbb R^{3\times3},
\end{equation}
where $\mathbf A$ contains atomic species, $\mathbf F$ contains fractional coordinates, and the rows of $\mathbf L$ are lattice vectors. Cartesian positions follow $\mathbf R=\mathbf F\mathbf L$. A supplied count vector $\mathbf n=(n_1,\ldots,n_K)$ determines the number of atoms of each element, with $N=\sum_k n_k$. A reduced formula determines ratios but not the number of formula units in the simulated cell. Cell multiplicity is therefore either an input or an additional discrete search variable. Figure~\ref{fig:problem-definition} illustrates the task with gallium nitride.
 
Under a static-lattice approximation at pressure $p$, a common optimization target is
\begin{equation}\label{eq:enthalpy}
 X^*\in\operatorname*{arg\,min}_{X\in\mathcal X(\mathbf n,\mathcal C)}
 H(X;p),\qquad H(X;p)=E(X)+pV(X),
\end{equation}
where $\mathcal C$ specifies cell-size, symmetry, or other restrictions and $V$ is the cell volume.  Here $H$ is the enthalpy, the zero-temperature limit of the Gibbs free energy $G = E + pV - TS$, which determines phase stability at fixed pressure. At ambient pressure $pV$ is negligible and the target reduces to the energy $E$; at high pressure the $pV$ term favours denser structures and drives pressure-induced phase transitions.
This equation describes the optimization task: find the thermodynamically stable structure at pressure p. It does not imply that a finite stochastic search identifies the global minimum. Moreover, the reference structure in a benchmark need not be the minimizer of the selected approximate energy model \citep{CSP041,CSP046,CSP122,CSP129}.
 
\subsection{Related tasks that require separate comparisons}
Variable-composition search changes $\mathbf n$ as well as geometry. Comparing energies across different compositions then requires a consistent thermodynamic reference and competing-phase treatment; selecting the smallest raw total energy is not meaningful. Evolutionary phase searches and outer-loop composition optimization explicitly extend beyond fixed-composition CSP \citep{CSP080,CSP147,CSP196}. Similarly, prescribing a property while allowing composition to vary is inverse materials design, of which the structure-generation component can be relevant to CSP, but its success metric need not be structural recovery.
 
Molecular CSP typically starts from molecular identity or connectivity, sometimes with a supplied conformer. Its variables include molecular centers, orientations, cell geometry, and any flexible internal coordinates. Information about the space group, density, molecular conformation, or number of independent molecules can substantially change the difficulty \citep{CSP023,CSP031,CSP049,CSP063,CSP103}. Experimental cell dimensions, diffraction data, or related observations define an information-assisted task rather than prediction from composition or molecular identity alone \citep{CSP001}. Table~\ref{tab:tasks} summarizes the task definitions that should be kept separate when algorithms are compared.
 
\begin{table}[tb]
\centering\small
\caption{Task definitions that should be separated when comparing CSP algorithms. Additional inputs change the prediction problem.}\label{tab:tasks}
\begin{tabularx}{\linewidth}{@{}P{.19\linewidth}YY@{}}
\toprule Task & Supplied information & Principal output or question \\
\midrule
Fixed-composition CSP & Element counts or formula plus cell multiplicity; physical conditions & Candidate lattices and positions at the specified composition \\
Variable-composition search & Element pool, composition bounds, thermodynamic references & Structures and competing compositions \\
Molecular CSP & Molecular identity, optional conformers, packing bounds & Packings and, where searched, conformations \\
Observation-assisted prediction & Composition or molecular identity plus diffraction, cell, or other measurements & Structures consistent with the additional observations \\
Unconditional crystal generation & Learned prior and sampling configuration & New compositions and structures; not a prescribed formula \\
Structure reconstruction & A complete structure or a structure-derived latent code & Recovery of the supplied structural information \\
\bottomrule
\end{tabularx}
\end{table}
 
\subsection{Symmetry, equivalence, and physical ranking}
One special characteristic of CSP is the symmetry and geometric transformation related to crystal structures.
Different atom orders, origins, orientations, and lattice bases can describe the same crystal. Representations and structural matching criteria must account for these equivalences. Space-group and Wyckoff parameterizations reduce redundant degrees of freedom, but also constrain the accessible structures. A space group alone does not specify a structure: compatible site multiplicities, species assignments, and positional parameters remain to be selected \citep{CSP072,CSP107,CSP154,CSP177}. The distinction is particularly important when a benchmark supplies a symmetry label derived from the reference structure. Section~\ref{sec:structural-priors} examines how these priors enter construction, variation, exploration, and learned generation.
 
Symmetry and equivalence determine which candidates are distinct; a separate question is how distinct candidates should be ranked. At finite temperature the relevant ranking may involve Gibbs free energies rather than static enthalpies. Vibrational, configurational, electronic, magnetic, or nuclear-quantum contributions can matter depending on the system. Molecular polymorph studies and recent free-energy learning approaches address parts of this problem \citep{CSP089,CSP135,CSP181}. A harmonic stability check, a static convex-hull calculation, and a free-energy ranking therefore provide different information. None by itself establishes a synthesis pathway or the absence of alternative structures.
 
\section{A layered taxonomy of CSP algorithms}
\begin{figure}[!htb]
\centering
\begin{tikzpicture}[font=\sffamily\small,
box/.style={draw=teal!60!black,rounded corners=3pt,fill=teal!5,align=center,text width=42mm,minimum height=13mm,inner sep=3mm},
wide/.style={box,text width=145mm,fill=blue!4},
arr/.style={-Stealth,thick,draw=teal!70!black}]
\node[wide] (input) {Task and representation\\Formula / cell bounds / molecular identity / optional symmetry or observations};
\node[box,below=10mm of input.south,anchor=north] (b) {Random search};
\node[box,left=4mm of b] (a) {Construction\\Templates / enumeration};
\node[box,right=4mm of b] (c) {Evolutionary search};
\node[box,below=4mm of a] (d) {Swarm / differential evolution};
\node[box,below=4mm of b] (e) {Landscape exploration};
\node[box,below=4mm of c] (f) {Adaptive / ML-guided search};
\node[box,text width=145mm,below=4mm of e] (g) {Learned generation: diffusion, flows, autoregression, latent and hierarchical models};

\node[wide,below=10mm of g] (eval) {Local relaxation and selection\\Empirical potentials / electronic structure / machine-learning potentials};

\node[wide,below=7mm of eval] (out) {Distinct outcomes\\Structure recovery / energy ranking / thermodynamic and dynamical evidence};
\draw[arr] (eval.south)--(out.north);
\node[draw=teal!30,rounded corners,fit=(a)(b)(c)(d)(e)(f)(g),inner sep=3mm] (group) {};
\draw[arr] (input.south)--(group.north);
\draw[arr] (group.south)--(eval.north);
\end{tikzpicture}
\caption{A layered view of crystal structure prediction. The seven search families are practical categories. Relaxation and evaluation can be reused across search algorithm families; history or new training data can feed back into subsequent structural proposals. Besides fixed-composition CSP, we also cover molecular crystals and variable-composition CSP, which are related but independent tasks.}\label{fig:taxonomy}
\end{figure}
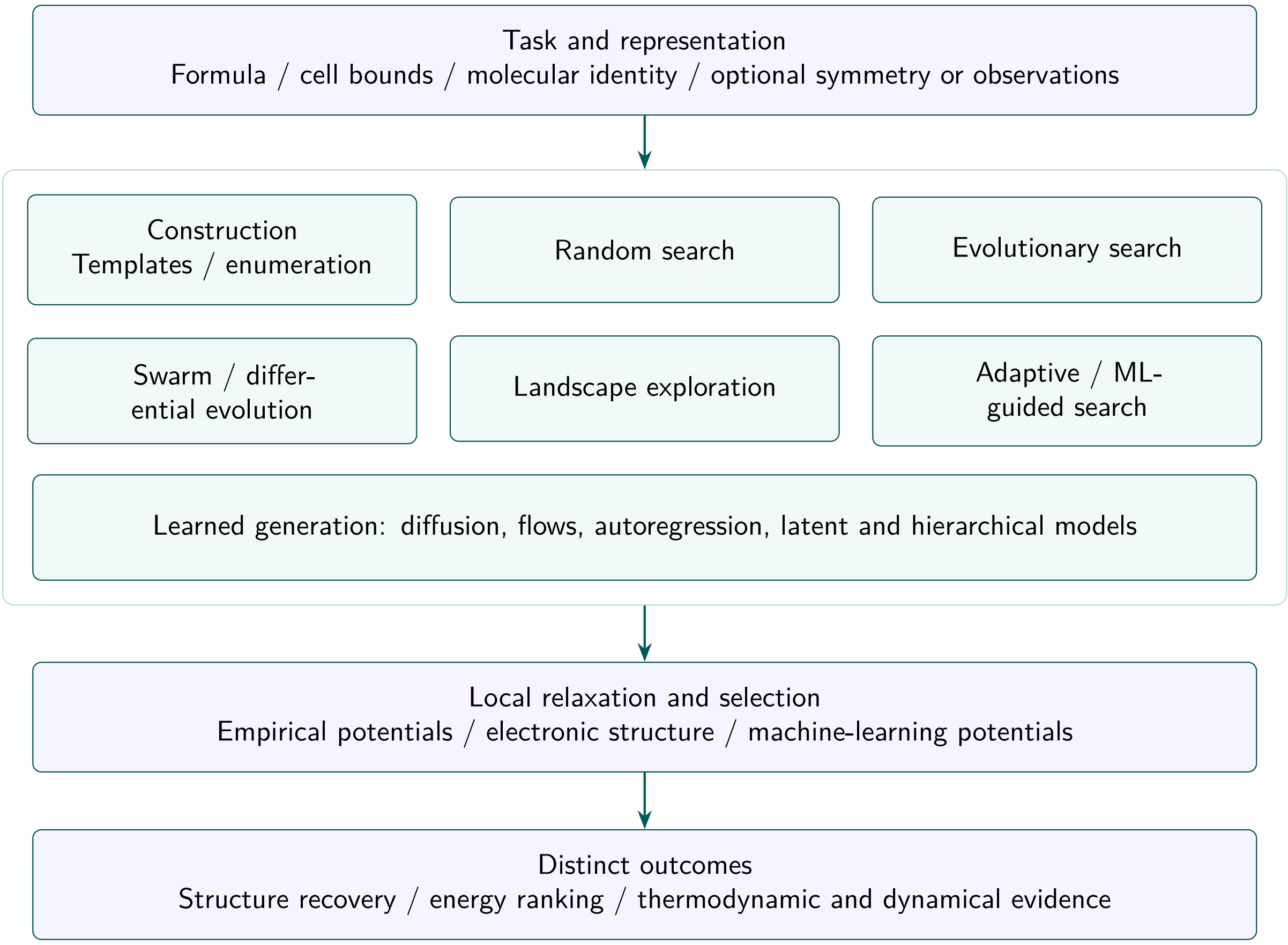

With the prediction task and its structural variables defined, a CSP workflow can be decomposed into representation, exploration, local relaxation, candidate selection, and final ranking or qualification. The taxonomy in Figure~\ref{fig:taxonomy} assigns primary categories to the exploration mechanism. It leaves the evaluator as a separate choice, because the same search can use an empirical potential, density-functional theory (DFT), or a machine learning (ML) potential. Hybrid algorithms occupy more than one category without requiring a new category for every combination.
 
Constructive CSP methods define candidates through templates, graphs, modules, occupations, or bounded discrete formulations. Random search draws starting structures at random within chemical and crystallographic bounds and relaxes each independently, whereas population methods build new candidates from information in earlier ones. Landscape methods explore transitions between basins. Adaptive methods learn during a search or allocate evaluations using a surrogate. Unlike constructive methods, whose candidates are defined by explicit rules, trained generators learn a proposal distribution from a training corpus and reuse it across new queries via sampling.
This division cuts across the usual split into ``traditional'' and ``AI'' methods: a learned potential can support an otherwise classical search, while a modern generative model can require substantial conventional relaxation. Table~\ref{tab:families} summarizes representative families and their main sources of restriction.
 
\begin{table}[tb]
\centering\small
\caption{Representative algorithm families and their main sources of restriction. Examples identify algorithmic lineages, not a performance ranking.}\label{tab:families}
\begin{tabularx}{\linewidth}{@{}P{.19\linewidth}P{.24\linewidth}Y@{}}
\toprule Family & Representative methods & Main source of search bias or limitation \\
\midrule
Construction & enumlib; GRINSP; TCSP; CSPML & Parent lattice, template library, topology, or discretization \\
Random & AIRSS; RandSpg; Genarris & Proposal distribution, chemical bounds, and available relaxations \\
Evolutionary & USPEX; XtalOpt; GAtor & Variation operators, diversity control, and fitness evaluator \\
Swarm / differential evolution & CALYPSO; SGO & Coordinate encoding and population dynamics \\
Landscape & Minima hopping; metadynamics; SSW; FUSE & Escape moves, collective variables, and starting basins \\
Adaptive / ML-guided & AGA; active-learning USPEX; CrySPY & Surrogate support, acquisition policy, and training cost \\
Learned generation & DiffCSP; CrystalFlow; uFlowCSP;  sequence models & Training distribution, representation, and conditioning contract \\
\bottomrule
\end{tabularx}
\end{table}
 
\subsection{Historical coverage of the surveyed literature}
\begin{figure}[!htb]
\centering
\includegraphics[width=\linewidth]{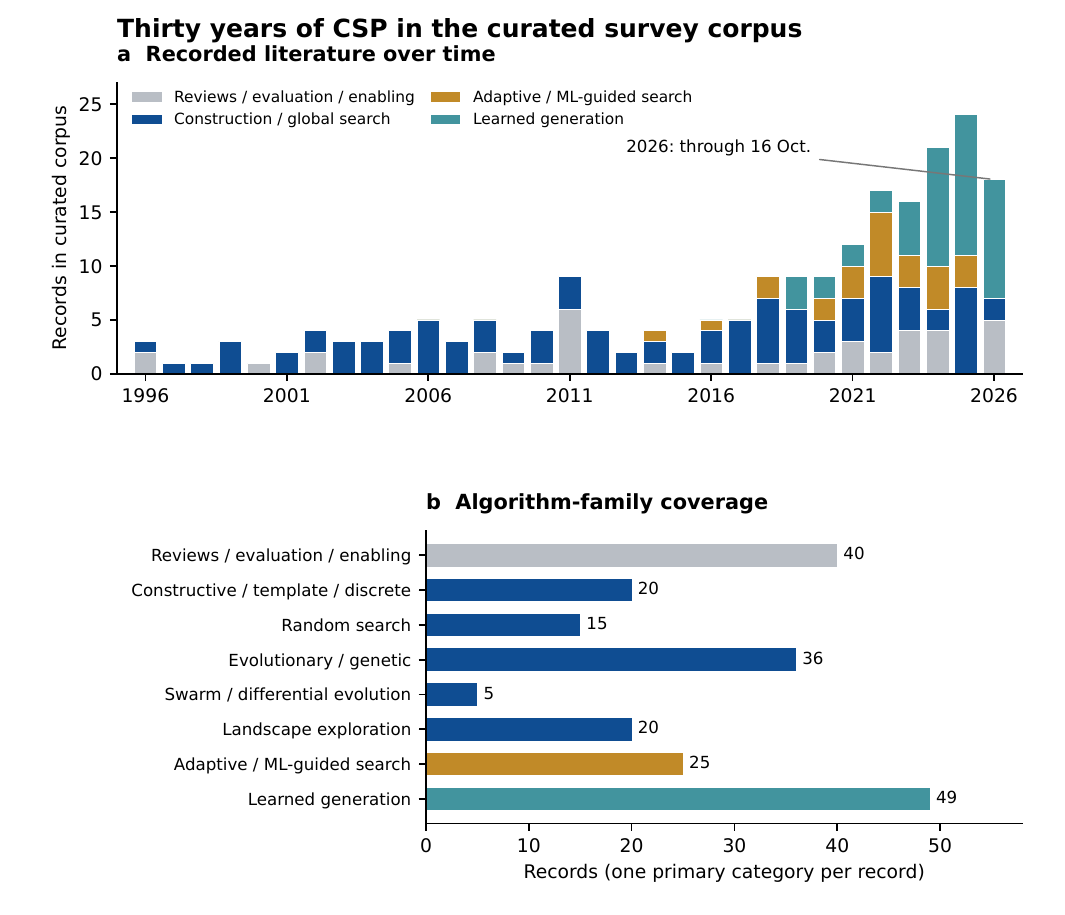}
\caption{Historical and categorical coverage of the curated survey corpus. (a) Annual counts grouped into four broad strands. (b) Counts in the eight primary categories, with one assignment per record. The underlying title, year, category, and source URL for each of the 210 records are retained with the figure data. Publication years follow the catalogue's bibliographic convention; conference or journal publication can follow an earlier preprint. The partial 2026 interval and targeted selection limit bibliometric interpretation.}\label{fig:literature-trends}
\end{figure}

Figure~\ref{fig:literature-trends} summarizes the 210 curated records shared by this survey and the companion repository as of 16 October 2026. Each record receives one primary category, so hybrid methods are counted once. The literature classification adds a category for reviews, evaluation, and enabling contributions to the seven proposal/search families.
The recent concentration of learned-generation papers complements the longer development of constructive and global-search CSP methods. These counts characterize the selection assembled for this review, including methodological, software, evaluation, and enabling contributions. They are not an exhaustive bibliometric census or a measure of community adoption.
 
\subsection{Representative algorithms and implementation resources}
Table~\ref{tab:algorithm-directory} brings together established global-search packages, template and learned-potential hybrids, and recent generative methods used in CSP comparisons. It provides a practical guide to their principal mechanisms and public implementation resources. Fixed-composition prediction and broader crystal generation are distinguished where the same model family exposes multiple modes; detailed methodological comparisons follow in the subsequent sections.
\begingroup
\small
\setlength{\tabcolsep}{4pt}
\renewcommand{\arraystretch}{1.13}
\begin{longtable}{@{}P{.22\linewidth}P{.15\linewidth}P{.47\linewidth}P{.10\linewidth}@{}}
\caption{Established CSP algorithms and representative recent generative methods. USPEX and CALYPSO lead the established global-search category. Other entries with supplied T46 results are ordered within their category by top-1 recovery and then top-20 recovery, followed by methods without comparable results. Paper links identify the method reference; code, project, archive, or access links identify implementation resources.}\label{tab:algorithm-directory}\\
\toprule
Method and citation & Family & Key features & Links \\
\midrule
\endfirsthead
\multicolumn{4}{l}{\textit{Table \thetable\ continued}}\\
\toprule
Method and citation & Family & Key features & Links \\
\midrule
\endhead
\midrule
\multicolumn{4}{r}{\textit{Continued on next page}}\\
\endfoot
\bottomrule
\endlastfoot
\multicolumn{4}{@{}l}{\textbf{Established global-search methods}}\\*
\addlinespace[2pt]
USPEX \citep{CSP027,CSP030} & Evolutionary & Heredity, mutation, local relaxation, and diversity control; later extensions add variable composition and structural priors. & \href{https://doi.org/10.1063/1.2210932}{Paper}\newline\href{https://uspex-team.org/documentation.html/uspex/downloads}{Access} \\
\addlinespace[3pt]
CALYPSO \citep{CSP043,CSP054} & Particle swarm & Population-based search with symmetry-guided initialization, local optimization, and structural duplicate control. & \href{https://doi.org/10.1103/PhysRevB.82.094116}{Paper}\newline\href{https://download.calypso.cn/home/}{Access} \\
\addlinespace[3pt]
MAGUS \citep{MAGUS2023} & Graph-assisted search & Evolutionary exploration using graph-derived structural units and machine-learning acceleration. & \href{https://doi.org/10.1093/nsr/nwad128}{Paper}\newline\href{https://zenodo.org/records/14730874}{Archive} \\
\addlinespace[3pt]
XtalOpt \citep{CSP053} & Evolutionary & Asynchronous population search; structural variation, duplicate filtering, and interfaces to external energy evaluators. & \href{https://doi.org/10.1016/j.cpc.2010.07.048}{Paper}\newline\href{https://github.com/xtalopt/XtalOpt}{Code} \\
\addlinespace[3pt]
AIRSS \citep{CSP028,CSP045} & Random search & Chemically sensible random cells, optional symmetry or molecular units, and repeated local relaxation. & \href{https://doi.org/10.1103/PhysRevLett.97.045504}{Paper}\newline\href{https://www.mtg.msm.cam.ac.uk/Codes/AIRSS}{Project} \\
\addlinespace[3pt]
Minima hopping \citep{CSP042} & Landscape search & Variable-cell search alternating molecular-dynamics escapes and local minimization; adaptive escape and acceptance rules. & \href{https://doi.org/10.1063/1.3512900}{Paper}\newline\href{https://gitlab.com/goedeckergroup/ase_mh}{Code} \\
\addlinespace[3pt]
SPAI / CALYPSO \citep{CSP108} & Symmetry search & Divide-and-conquer exploration of a symmetry hierarchy with adaptive subspace selection and symmetry-preserving construction. & \href{https://doi.org/10.1063/5.0074677}{Paper}\newline\href{https://download.calypso.cn/home/}{Access} \\
\addlinespace[3pt]
\multicolumn{4}{@{}l}{\textbf{Template methods and learned-potential evolutionary hybrids}}\\*
\addlinespace[2pt]
TCSPv3 \citep{Feng2026TCSP3,TCSPv3Software2026,CSP194} & Template & Coefficient-preserving, simultaneous element substitution with composition and cell validation; complementary template retrieval, soft oxidation-state and radius-compatibility scoring, structural deduplication, and optional ORB-v3 relaxation and ranking; separate data assets required. & \href{https://arxiv.org/abs/2610.34029}{Paper}\newline\href{https://arxiv.org/abs/2503.23183}{Lineage}\newline\href{https://github.com/usccolumbia/TCSPV3}{Code} \\
\addlinespace[3pt]
TCSP / TCSP 2.0 \citep{CSP123,CSP194} & Template & Composition-based retrieval and substitution; version 2.0 adds learned oxidation states, element embeddings, and chemistry heuristics. & \href{https://pubs.acs.org/doi/10.1021/acs.inorgchem.1c03879}{Paper}\newline\href{https://github.com/usccolumbia/TCSP}{Code} \\
\addlinespace[3pt]
CSPML \citep{CSP115} & Template + ML & Metric learning selects substitution templates for a target composition; evaluation depends on template-library overlap. & \href{https://www.sciencedirect.com/science/article/pii/S0927025622002555}{Paper}\newline\href{https://github.com/Minoru938/CSPML}{Code} \\
\addlinespace[3pt]
ParetoCSP \citep{CSP141} & Evolutionary + MLIP & Age--fitness Pareto search with M3GNet relaxation and energy evaluation; balances low energy and population diversity. & \href{https://www.oaepublish.com/articles/jmi.2023.33}{Paper}\newline\href{https://github.com/sadmanomee/ParetoCSP}{Code} \\
\addlinespace[3pt]
ParetoCSP2 \citep{CSP175} & Evolutionary + MLIP & Adaptive space-group diversity control extends Pareto search toward polymorph exploration. & \href{https://advanced.onlinelibrary.wiley.com/doi/full/10.1002/advs.202510792}{Paper}\newline\href{https://github.com/usccolumbia/ParetoCSP2}{Code} \\
\addlinespace[3pt]
\multicolumn{4}{@{}l}{\textbf{Diffusion, variational, and manifold generators}}\\*
\addlinespace[2pt]
Crystalite \citep{CSP184} & Diffusion Transformer & Lightweight geometric attention; separate fixed-composition CSP and de novo configurations, with subatomic tokens in the latter. & \href{https://arxiv.org/abs/2604.02270}{Paper}\newline\href{https://github.com/joshrosie/crystalite}{Code} \\
\addlinespace[3pt]
EquiCSP \citep{CSP144} & Diffusion & Composition-conditioned generation incorporating lattice-permutation and periodic-translation equivariance. & \href{https://proceedings.mlr.press/v235/lin24b.html}{Paper}\newline\href{https://github.com/EmperorJia/EquiCSP}{Code} \\
\addlinespace[3pt]
DiffCSP \citep{CSP128} & Diffusion & Fixed-composition joint generation of lattice and fractional coordinates with periodic equivariant modeling. & \href{https://arxiv.org/abs/2309.04475}{Paper}\newline\href{https://github.com/jiaor17/DiffCSP}{Code} \\
\addlinespace[3pt]
MatterGen \citep{CSP159,MatterGenCSPCode} & Diffusion & Joint chemistry/geometry generation and property adapters; a distinct released CSP configuration fixes supplied atom types. & \href{https://www.nature.com/articles/s41586-025-08628-5}{Paper}\newline\href{https://github.com/microsoft/mattergen}{Code} \\
\addlinespace[3pt]
cond-CDVAE \citep{CSP142} & Conditional VAE & Composition- and pressure-conditioned crystal generation with a diffusion decoder. & \href{https://www.nature.com/articles/s41524-024-01443-y}{Paper}\newline\href{https://github.com/ixsluo/cond-cdvae}{Code} \\
\addlinespace[3pt]
TGDMat \citep{TGDMat2025} & Text-guided diffusion & Joint diffusion of species, coordinates, and lattice with textual conditions; descriptions may supply symmetry beyond composition. & \href{https://arxiv.org/abs/2503.00522}{Paper}\newline\href{https://github.com/kdmsit/TGDMat}{Code} \\
\addlinespace[3pt]
GemsDiff \citep{CSP157} & Diffusion & Vector-field-oriented lattice and coordinate diffusion; released formula sampler fixes the atom inventory. & \href{https://ojs.aaai.org/index.php/AAAI/article/view/30224}{Paper}\newline\href{https://github.com/aklipf/gemsdiff}{Code} \\
\addlinespace[3pt]
CDVAE \citep{CSP113} & VAE + diffusion & Latent crystal generation with a score-based decoder; original de novo and reconstruction tasks differ from formula-only CSP. & \href{https://arxiv.org/abs/2110.06197}{Paper}\newline\href{https://github.com/txie-93/cdvae}{Code} \\
\addlinespace[3pt]
CrystalGRW \citep{CSP183} & Manifold diffusion & Geodesic random walks with crystallographic guidance; fixed-species code options are distinct from dedicated CSP training. & \href{https://www.nature.com/articles/s41598-026-62470-x}{Paper}\newline\href{https://github.com/trachote/crystalgrw}{Code} \\
\addlinespace[3pt]
SymmCD \citep{CSP178} & Symmetry diffusion & Generates asymmetric units and site symmetries, then expands the crystal; exact formula counts require additional control. & \href{https://arxiv.org/abs/2502.03638}{Paper}\newline\href{https://github.com/sibasmarak/SymmCD}{Code} \\
\addlinespace[3pt]
DiffCSP++ \citep{CSP154} & Symmetry diffusion & Space-group-compatible lattice and Wyckoff-coordinate parameterization constrains joint diffusion. & \href{https://arxiv.org/abs/2402.03992}{Paper}\newline\href{https://github.com/jiaor17/DiffCSP-PP}{Code} \\
\addlinespace[3pt]
\multicolumn{4}{@{}l}{\textbf{Flow-based generators}}\\*
\addlinespace[2pt]
OMatG \citep{CSP173} & Stochastic interpolants & Continuous geometry transport and discrete species modeling; CSP mode fixes species while de novo generation samples chemistry. & \href{https://proceedings.mlr.press/v267/hollmer25a.html}{Paper}\newline\href{https://github.com/FERMat-ML/OMatG}{Code} \\
\addlinespace[3pt]
uFlowCSP \citep{CSP195} & MeanFlow & Few-step average-velocity generation with chemical descriptors, formula embedding, canonical atom ordering, and auxiliary symmetry training; formula-only inference. & \href{https://arxiv.org/abs/2609.09799}{Paper}\newline\href{https://github.com/usccolumbia/uFlowCSP}{Code} \\
\addlinespace[3pt]
CrystalFlow \citep{CSP169} & Continuous flow & Equivariant conditional flow matching for lattice and atomic coordinates, including composition and pressure conditions. & \href{https://www.nature.com/articles/s41467-025-64364-4}{Paper}\newline\href{https://github.com/ixsluo/CrystalFlow}{Code} \\
\addlinespace[3pt]
FlowMM \citep{CSP146} & Riemannian flow & Flow matching on crystal-variable manifolds; dedicated known-composition CSP and de novo generation settings. & \href{https://proceedings.mlr.press/v235/miller24a.html}{Paper}\newline\href{https://github.com/facebookresearch/flowmm}{Code} \\
\addlinespace[3pt]
\multicolumn{4}{@{}l}{\textbf{Autoregressive and masked-token generators}}\\*
\addlinespace[2pt]
Uni-3DAR \citep{CSP179} & Autoregressive & Compressed spatial tokens and hierarchical 3D generation; CSP decoding constrains the remaining per-element atom inventory. & \href{https://arxiv.org/abs/2503.16278v1}{Paper}\newline\href{https://github.com/dptech-corp/Uni-3DAR}{Code} \\
\addlinespace[3pt]
MaskGXT \citep{MaskGXT2026} & Masked Transformer & Composition-conditioned iterative token unmasking, symmetry tokens, space-group-stratified sampling, and sub-bin coordinate refinement. & \href{https://arxiv.org/abs/2606.22866}{Paper}\newline\href{https://github.com/kiyoung98/MaskGXT}{Code} \\
\addlinespace[3pt]
CrystalFormer \citep{CSP177,CrystalFormerCSP2025} & Autoregressive & Space-group/Wyckoff token generation; CSP extension supports formula conditioning, with version-dependent full-cell count control. & \href{https://doi.org/10.1016/j.scib.2025.09.035}{Paper}\newline\href{https://github.com/deepmodeling/CrystalFormer}{Code} \\
\addlinespace[3pt]
\end{longtable}
\noindent\textit{Ordering.} The T46 ordering uses the supplied leaderboard snapshot \citep{CSPLeaderboardSnapshot2026}.

\section{CSP algorithms based on chemical construction, templates, and discrete search}
We begin with methods that construct candidates from existing structural information, including prototypes \cite{hu2026comprehensive}, connectivity patterns, chemical modules, and bounded occupational assignments. Figure~\ref{fig:construction-methods} summarizes these routes and the limit each one imposes.
 
\begin{figure}[!htb]
\centering
\begin{tikzpicture}[x=1mm,y=1mm,font=\sffamily\footnotesize,
ptitle/.style={font=\sffamily\footnotesize\bfseries,anchor=north west},
small/.style={font=\sffamily\scriptsize,text=black!65,align=center,execute at begin node={\hyphenpenalty=10000\exhyphenpenalty=10000}},
ex/.style={font=\sffamily\scriptsize,text=black!60,anchor=south west,align=left,execute at begin node={\hyphenpenalty=10000\exhyphenpenalty=10000}},
lim/.style={font=\sffamily\scriptsize,text=orange!55!black,anchor=south west,align=left,execute at begin node={\hyphenpenalty=10000\exhyphenpenalty=10000}},
arr/.style={-{Stealth[length=1.5mm]},thick,black!60},
A/.style={circle,fill=blue!55!black,draw=white,line width=.3pt,minimum size=2.4mm,inner sep=0},
B/.style={circle,fill=orange!85!black,draw=white,line width=.3pt,minimum size=1.9mm,inner sep=0},
C/.style={circle,fill=teal!65!black,draw=white,line width=.3pt,minimum size=2.4mm,inner sep=0},
Dd/.style={circle,fill=red!60!black,draw=white,line width=.3pt,minimum size=1.9mm,inner sep=0}]
\newcommand{\panel}[4]{\draw[draw=teal!60!black,rounded corners=3pt,fill=teal!3] (0,0) rectangle (#3,#4); \node[ptitle] at (1.5,#4-0.8) {#1\quad #2};}
\newcommand{\ucell}[3]{\begin{scope}[shift={(#1)}]\draw[black!45] (0,0) rectangle (8,8); \foreach \p in {(0,0),(8,0),(0,8),(8,8)} \node[#2] at \p {}; \node[#3] at (4,4) {};\end{scope}}

\begin{scope}[shift={(0,0)}]
\panel{a}{Template transfer}{52}{50}
\ucell{6,23}{A}{B} \node[small] at (10,19.5) {template};
\draw[arr] (16,27)--(22,27) node[midway,above,small] {swap};
\ucell{24,23}{C}{Dd} \node[small] at (28,19.5) {decorated};
\draw[arr] (34.5,27)--(40,27) node[midway,above,small] {relax};
\begin{scope}[shift={(41.5,23)}] \draw[black!45] (0,0)--(8.6,0.4)--(8.2,8.4)--(-0.4,7.8)--cycle; \foreach \p in {(0,0),(8.6,0.4),(-0.4,7.8),(8.2,8.4)} \node[C] at \p {}; \node[Dd] at (4.2,4.3) {}; \end{scope}
\node[small] at (45.5,19.5) {candidate};
\node[small] at (26,39.5) {select an analogue, substitute species,\\then refine the geometry};
\node[lim,text width=49mm] at (1.5,8.4) {cannot propose a structure type missing from the catalogue};
\node[ex,text width=49mm] at (1.5,0.8) {TCSP, TCSP 2.0, CSPML; data-mined ionic substitution};
\end{scope}

\begin{scope}[shift={(55,0)}]
\panel{b}{Ordering enumeration}{52}{50}
\begin{scope}[shift={(4,22)}]
 \draw[black!45] (0,0) grid[step=5] (10,10);
 \foreach \x in {0,5,10} \foreach \y in {0,5,10} \node[A,minimum size=1.8mm] at (\x,\y) {};
 \node[small] at (5,-3.4) {parent lattice};
\end{scope}
\draw[arr] (16,27)--(21,27);
\foreach \k/\cfg in {0/{1,0,0,1},1/{1,1,0,0},2/{0,1,1,0}}{
 \begin{scope}[shift={(23.5+\k*9.2,24)}]
  \draw[black!30] (0,0) grid[step=3] (6,6);
  \draw[black!70,dashed] (0,0) rectangle (6,6);
  \foreach \i in {0,1,2} \foreach \j in {0,1,2} {
    \pgfmathtruncatemacro\s{mod(\i,2)+2*mod(\j,2)}
    \pgfmathtruncatemacro\c{{\cfg}[\s]}
    \pgfmathtruncatemacro\img{(\i==2)||(\j==2)}
    \ifnum\img=1 \def\op{.35}\def\sz{1.2mm}\else\def\op{1}\def\sz{1.6mm}\fi
    \ifnum\c=1 \node[C,minimum size=\sz,opacity=\op] at (3*\i,3*\j) {};
    \else \node[B,minimum size=\sz,opacity=\op] at (3*\i,3*\j) {}; \fi }
 \end{scope}}
\draw[red!60!black,thick] (41.4,23.5)--(48.4,30.5); \draw[red!60!black,thick] (41.4,30.5)--(48.4,23.5);
\node[small] at (36,18.6) {supercell decorations;\\equivalents removed};
\node[small] at (26,39.5) {exhaustive within declared parent,\\cell-size and composition bounds};
\node[lim,text width=49mm] at (1.5,7.4) {cannot reach other lattices or reconstructive changes};
\node[ex,text width=49mm] at (1.5,0.8) {derivative-structure enumeration};
\end{scope}

\begin{scope}[shift={(110,0)}]
\panel{c}{Frameworks and modules}{52}{50}
\begin{scope}[shift={(6,22)}]
 \coordinate (n1) at (0,0); \coordinate (n2) at (6,2); \coordinate (n3) at (2,8); \coordinate (n4) at (9,9); \coordinate (n5) at (12,1);
 \foreach \a/\b in {n1/n2,n2/n3,n1/n3,n2/n4,n3/n4,n2/n5,n4/n5} \draw[black!55,thick] (\a)--(\b);
 \foreach \n in {n1,n2,n3,n4,n5} \node[A,minimum size=2mm] at (\n) {};
 \node[small] at (6,-3.4) {network};
\end{scope}
\begin{scope}[shift={(30,22)}]
 \draw[orange!70!black,fill=orange!15] (0,0)--(6,0)--(3,6)--cycle; \draw[orange!70!black,fill=orange!15] (7,2)--(13,2)--(10,8)--cycle;
 \draw[teal!60!black,fill=teal!12] (3,6)--(10,8)--(6.5,11)--cycle;
 \node[small] at (6.5,-3.4) {chemical modules};
\end{scope}
\node[small] at (26,39.5) {search over connectivity or reusable\\units, then embed in 3D and relax};
\node[lim,text width=49mm] at (1.5,7.4) {limited to the graph vocabulary or the module library};
\node[ex,text width=49mm] at (1.5,0.8) {framework enumeration, GRINSP; MC-EMMA, FUSE};
\end{scope}

\begin{scope}[shift={(0,-44)}]
\panel{d}{Symmetry-aware construction}{79}{40}
\node[draw=blue!55!black,fill=blue!8,rounded corners=2pt,inner sep=1.2mm,align=center,font=\sffamily\scriptsize] (sg) at (14,22) {space group\\$+$ Wyckoff sites};
\begin{scope}[shift={(40,22)}]
 \draw[teal!50!black,thick] (-8,-5)--(5,-5)--(8,5)--(-5,5)--cycle; \draw[black!40,dashed] (0,-6.5)--(0,6.5);
 \node[A] at (-3.5,1.8) {}; \node[A] at (3.5,1.8) {}; \node[B] at (-2,-2.2) {}; \node[B] at (2,-2.2) {};
\end{scope}
\draw[arr] (sg.east)--(30,22);
\node[small,anchor=west,text width=27mm,align=left] at (51,22) {fewer free coordinates; used to seed random, evolutionary and learned searches};
\node[lim,text width=75mm] at (1.5,7.4) {a too-narrow symmetry choice excludes the solution};
\node[ex,text width=75mm] at (1.5,0.8) {RandSpg, PyXtal};
\end{scope}

\begin{scope}[shift={(83,-44)}]
\panel{e}{Discretized optimization}{79}{40}
\begin{scope}[shift={(6,15)}]
 \foreach \x in {0,3,...,15} \foreach \y in {0,3,...,12} \fill[black!30] (\x,\y) circle (0.35);
 \foreach \p in {(3,3),(9,6),(15,12),(6,12)} \node[A,minimum size=2.1mm] at \p {};
 \foreach \p in {(12,0),(0,9)} \node[B] at \p {};
\end{scope}
\draw[arr] (24,21)--(30,21);
\node[draw=teal!60!black,fill=teal!10,rounded corners=2pt,inner sep=1.2mm,align=center,font=\sffamily\scriptsize] at (40,21) {integer program\\or Ising model};
\draw[arr] (50.5,21)--(56,21);
\node[small,anchor=west,text width=20mm,align=left] at (56.5,21) {continuous refinement (DFT)};
\node[lim,text width=75mm] at (1.5,7.4) {optimality holds only for the discretized problem};
\node[ex,text width=75mm] at (1.5,0.8) {integer-programming CSP; CRYSIM (GPU Ising optimizer)};
\end{scope}
\end{tikzpicture}
\caption{Construction, template, and discrete-search methods. (a) Template transfer substitutes species into a known crystal and relaxes the result. (b) Derivative-structure enumeration generates the symmetry-distinct decorations of a parent lattice. (c) Framework and module methods search over connectivity or reusable chemical units before embedding them in three dimensions. (d) Symmetry-aware constructors build crystals from a space group and compatible Wyckoff sites. (e) Discretized formulations solve an integer or Ising problem before continuous refinement. Orange text states the coverage limit of each route; representative methods are listed at the bottom of each panel.}
\label{fig:construction-methods}
\end{figure}
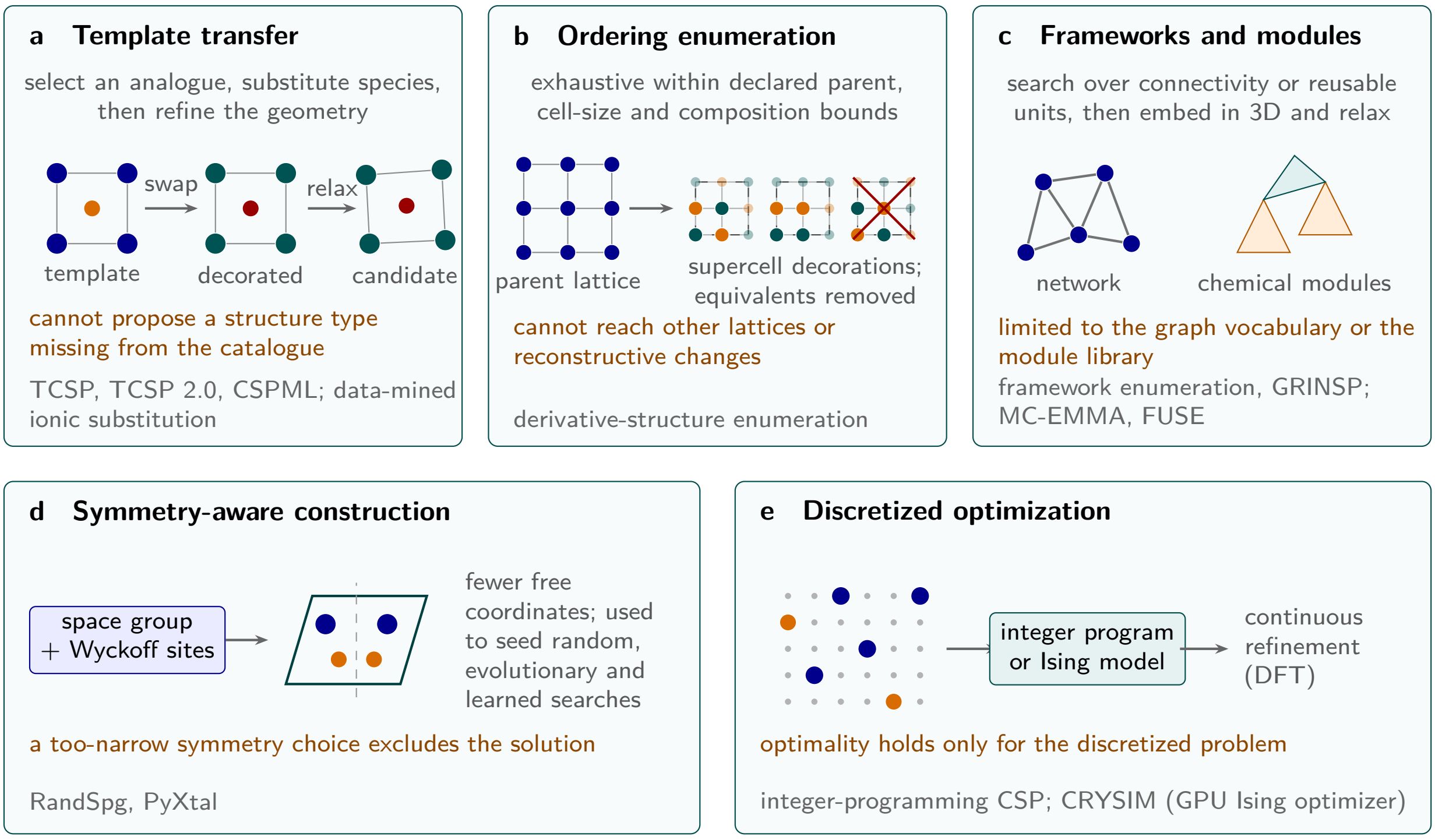
 
\subsection{Prototype transfer and elemental substitution}
Template methods begin with a structural hypothesis supplied by an existing crystal. They select chemically plausible analogues, replace species, and refine the resulting geometry. Data-mined correlations between known prototypes provided an early route to prioritizing structural candidates, while ionic-substitution models learned which replacements occur in known compounds \citep{CSP017,CSP029,CSP048}. More recent CSPML \citep{CSP115}, TCSP \citep{CSP123}, and TCSP~2.0 \citep{CSP194} studies use learned similarity or chemistry-informed rules to improve template selection. The {TCSPv3} algorithm \citep{Feng2026TCSP3} showed that substitution itself can be a dominant error source: an audit of TCSP~2.0 on T180 found that 20\% of its selected structures had the wrong composition, because element mappings ignored stoichiometric coefficients and sequential text substitution corrupted element symbols. Enforcing coefficient-preserving, simultaneous substitution with composition and cell validation, together with structural deduplication and machine-learning-potential ranking, raised the reported top-1 and top-20 match rates on T180 from 28.3\% and 45.6\% to 72.2\% and 90.6\%.
In molecular crystals, analogue-based generation similarly transfers packing information from related structures \citep{CSP067}.
 
The principal advantage of these template-based CSP algorithms is that much of the arrangement is inherited from a viable example. The main limitation follows from the same assumption: a search restricted to a template catalogue cannot directly propose a topology missing from that catalogue. Relaxation can substantially alter a decorated template, but it gives no systematic access to unrepresented structures. Evaluations should therefore report whether the reference structure, or an isostructural analogue, was available to the template selector, and separate near-template retrieval from success on a withheld prototype family.
 
Elemental substitution and occupational ordering solve related but different problems. Replacing a species on every equivalent site creates one decoration. Mixed substitutions introduce multiple possible assignments, and ordering can lower the parent symmetry. Derivative-structure enumeration explicitly generates symmetry-distinct supercells and decorations of a parent lattice; fixed-concentration enumeration imposes additional composition restrictions \citep{CSP035,CSP056}. Within declared parent, cell-size, species, and occupancy bounds the enumeration is exhaustive, but it cannot reach other lattices or reconstructive transformations.
 
\subsection{Graphs, modules, and symmetry-aware construction}
Framework and network methods move part of the problem from unconstrained coordinates to connectivity. Enumeration of periodic tetrahedral frameworks and systematic crystalline networks explores bounded graph classes before geometrical realization and energetic assessment \citep{CSP004,CSP006,CSP011,CSP019,CSP021}. GRINSP \citep{CSP024} constructs inorganic networks using polyhedral connectivity and geometric constraints. Such methods work well when coordination chemistry narrows the plausible space, and they reach only what their graph vocabulary and embedding rules allow.
 
Chemical modules provide an intermediate representation between full prototypes and individual atoms. MC-EMMA \citep{CSP070} and the Flexible Unit Structure Engine \citep{CSP083} use chemically motivated components as building blocks or search units. The resulting proposal space can retain useful local chemistry while allowing different larger-scale arrangements. The module library is still a prior, however: these methods recombine known units but cannot invent a missing coordination environment.
 
A third route constrains symmetry rather than connectivity or chemistry. Symmetry-aware constructors such as RandSpg \citep{CSP072} and PyXtal \citep{CSP107} generate structures consistent with requested space groups and compatible Wyckoff assignments. They initialize random, evolutionary, or learned searches and so act as components of those searches. Symmetry reduces the number of free coordinates and avoids some invalid configurations, but an overly narrow symmetry choice can exclude the solution. Search protocols must state whether symmetry is fixed, sampled, predicted, or allowed to change during relaxation.
 
\subsection{Bounded mathematical optimization}
A further approach discretizes the problem itself, which permits a more formal discussion of optimality. Integer-programming CSP \citep{CSP133} assigns atoms to a periodic discretization and couples the resulting solutions to continuous refinement. Its optimality guarantees hold for the discretized formulation; the continuous DFT relaxation that follows does not carry them over to arbitrary crystals. CRYSIM \citep{CSP176} explores another formulation, encoding crystallographic variables for a GPU-based Ising optimizer.
 
More generally, exhaustive enumeration and certified optimization are meaningful only within a specified domain and objective. Stochastic search has broader flexibility, but usually lacks a certificate that all lower-energy alternatives have been excluded. Claims of either kind hold only within the stated cell, representation, interaction model, and numerical bounds.
 
\section{Random structure search}
Stochastic methods give up such certificates in exchange for flexibility, and the simplest of them is random search, which generates starting structures by sampling within chemical and crystallographic bounds and then relaxes each toward a local minimum. AIRSS \citep{CSP028,CSP045} developed this strategy into a practical first-principles workflow, with an early silane study followed by a general methodological account. Its proposals are far from uniform over coordinates: useful ones impose chemically sensible separation distances, cell volumes, symmetry, or structural units. The proposal distribution and the relaxation basin volumes jointly determine which minima are encountered. Figure~\ref{fig:random-search} illustrates this dependence.
 
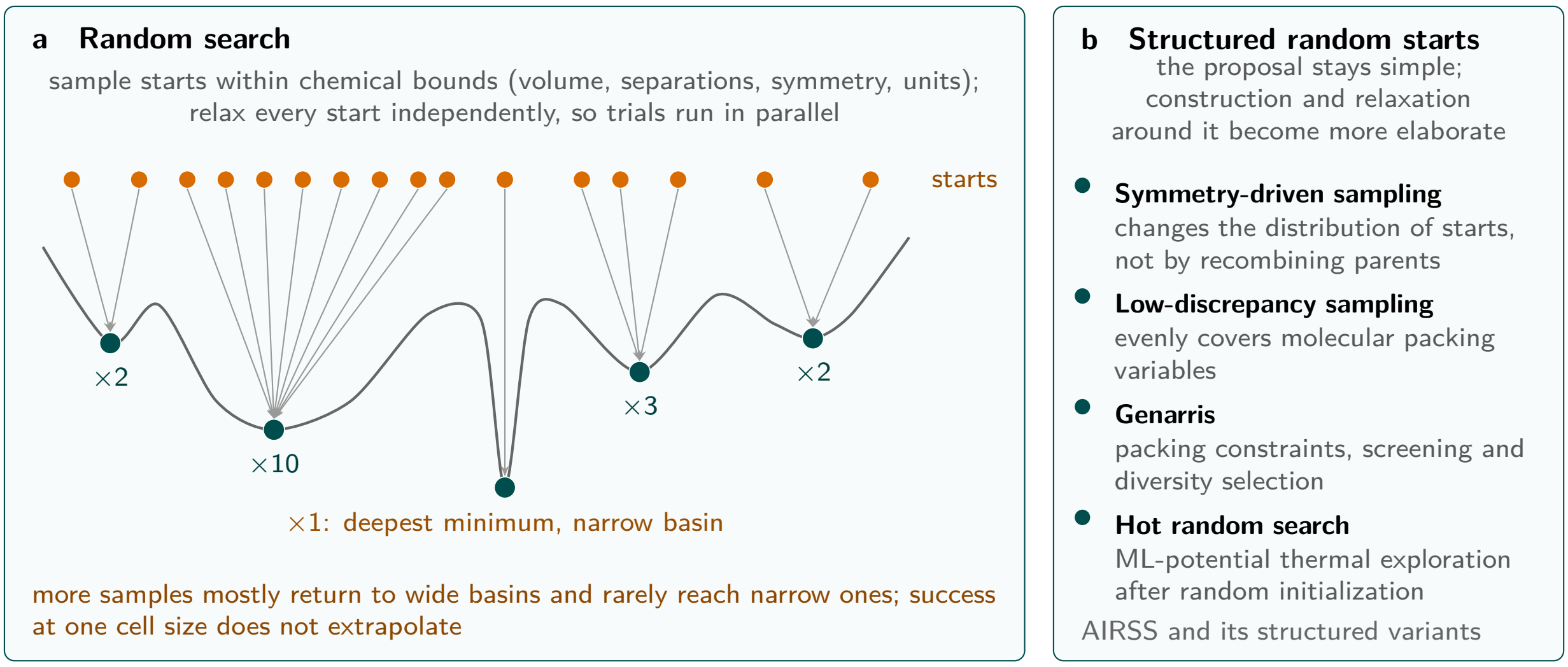
\begin{figure}[!htb]
\centering
\begin{tikzpicture}[x=1mm,y=1mm,font=\sffamily\footnotesize,
ptitle/.style={font=\sffamily\footnotesize\bfseries,anchor=north west},
small/.style={font=\sffamily\scriptsize,text=black!65,align=center,execute at begin node={\hyphenpenalty=10000\exhyphenpenalty=10000}},
ex/.style={font=\sffamily\scriptsize,text=black!60,anchor=south west,align=left,execute at begin node={\hyphenpenalty=10000\exhyphenpenalty=10000}},
lim/.style={font=\sffamily\scriptsize,text=orange!55!black,anchor=south west,align=left,execute at begin node={\hyphenpenalty=10000\exhyphenpenalty=10000}},
item/.style={font=\sffamily\scriptsize,anchor=north west,text width=46mm,align=left,execute at begin node={\hyphenpenalty=10000\exhyphenpenalty=10000}},
start/.style={circle,fill=orange!85!black,draw=white,line width=.3pt,inner sep=0,minimum size=1.8mm},
mini/.style={circle,fill=teal!60!black,draw=white,line width=.3pt,inner sep=0,minimum size=2.2mm}]

\draw[draw=teal!60!black,rounded corners=3pt,fill=teal!3] (0,0) rectangle (106,68);
\node[ptitle] at (1.5,67.2) {a\quad Random search};
\node[small] at (53,58.5) {sample starts within chemical bounds (volume, separations, symmetry, units);\\relax every start independently, so trials run in parallel};
\begin{scope}[shift={(0,12)}]
\draw[thick,black!60] plot[smooth,tension=0.5] coordinates {(4,31) (11,21) (16,25) (22,15) (28,12) (36,15) (44,24) (49.5,23.5) (52,6) (54.5,23.5) (58,25) (66,18) (74,26) (80,23) (84,21.5) (88,24) (94,32)};
\foreach \m/\x/\y in {1/11/21,2/28/12,3/52/6,4/66/18,5/84/21.5} \node[mini] (m\m) at (\x,\y) {};
\foreach \sx/\m in {7/1,14/1,19/2,23/2,27/2,31/2,35/2,39/2,43/2,46/2,52/3,60/4,64/4,70/4,79/5,90/5}{
  \node[start] (s\sx) at (\sx,38) {};
  \draw[-{Stealth[length=1.1mm]},black!40] (s\sx.south)--(m\m.north);}
\node[small,text=orange!60!black,anchor=west] at (95,38) {starts};
\foreach \x/\y/\n in {11/21/2,28/12/10,66/18/3,84/21.5/2} \node[small,text=teal!50!black] at (\x,\y-3.6) {$\times$\n};
\node[small,text=orange!60!black] at (52,2.2) {$\times$1: deepest minimum, narrow basin};
\end{scope}
\node[lim,text width=102mm] at (1.5,0.8) {more samples mostly return to wide basins and rarely reach narrow ones; success at one cell size does not extrapolate};

\begin{scope}[shift={(109,0)}]
\draw[draw=teal!60!black,rounded corners=3pt,fill=teal!3] (0,0) rectangle (53,68);
\node[ptitle] at (1.5,67.2) {b\quad Structured random starts};
\node[small,text width=49mm] at (26.5,58.5) {the proposal stays simple; construction and relaxation around it become more elaborate};
\foreach \y/\t/\d in {51/{Symmetry-driven sampling}/{changes the distribution of starts, not by recombining parents},39.5/{Low-discrepancy sampling}/{evenly covers molecular packing variables},28/{Genarris}/{packing constraints, screening and diversity selection},16.5/{Hot random search}/{ML-potential thermal exploration after random initialization}}{
  \fill[teal!60!black] (3,\y-1.6) circle (0.8);
  \node[item] at (5,\y) {\textbf{\t}\\[0.5pt]{\color{black!65}\d}};}
\node[ex,text width=49mm] at (1.5,0.8) {AIRSS and its structured variants};
\end{scope}
\end{tikzpicture}
\caption{Random structure search. (a) Starts sampled within chemical bounds relax independently into the basins of the energy landscape; the number of starts reaching each minimum (labels) follows the basin volumes, so a narrow basin can hold the lowest minimum yet be found only rarely. (b) Structured variants keep the simple proposal but change the starting distribution or the construction and relaxation around it.}
\label{fig:random-search}
\end{figure}
 
The random search method is straightforward to parallelize because independent trials need little communication. It also provides a strong diagnostic baseline: improvements can be assessed relative to chemically reasonable random starts under the same evaluator and relaxation budget. The difficulty is that additional sampling may repeatedly return to large, already known basins while missing narrow ones. Increasing structural complexity changes both the number of relevant degrees of freedom and the distribution of accessible basins, so success at one cell size cannot be extrapolated without measurement.
 
Random search can be structured without becoming evolutionary (Figure~\ref{fig:random-search}b). Symmetry-driven sampling \citep{CSP082}, for example, changes the distribution of initial configurations rather than recombining successful parents. In molecular CSP, low-discrepancy sampling \citep{CSP066} and the Genarris family \citep{CSP079,CSP096,CSP171} use packing-specific constraints and screening to organize the initial pool. Recent hot-random-search work \citep{CSP165} adds ML-potential-assisted thermal exploration to random initialization. In each case the proposal mechanism stays simple while the construction and relaxation stages around it become more sophisticated.
 
\section{Population-based global optimization}
\subsection{Evolutionary and genetic search}
Whereas independent random trials do not share information between candidates, evolutionary CSP maintains a population and uses evaluated candidates to construct new ones. Its basic ingredients are selection, variation, relaxation, and replacement, with some mechanism for retaining diversity. Early inorganic genetic algorithms \citep{CSP007} combined structural variation with energy minimization; periodic real-space formulations \citep{CSP026,CSP033,CSP037} subsequently developed operators for both atomic arrangements and cells. USPEX \citep{CSP027,CSP030} combined specialized evolutionary variation with ab initio or MLIP-based relaxation and established a widely used structure-search implementation. Figure~\ref{fig:population-search} summarizes the population schemes discussed in this section.
 
USPEX represents each candidate by its lattice and fractional coordinates and relaxes every offspring before selection, so the population evolves on locally optimized structures \citep{CSP027,CSP050}. New candidates come from heredity, which combines spatially coherent slabs of two parents, and from lattice, permutation, and soft-mode mutations, which follow low-frequency vibrational modes; a share of new random structures is added in each generation to preserve diversity. Real-space heredity can, however, destroy favourable local motifs when the parents differ in cell shape, density, or symmetry. The Latent Genetic Algorithm \citep{Zheng2026LatentGA} addresses this by recombining parents in the latent space of a pretrained universal interatomic potential: it interpolates the parents' latent representations and recovers an offspring by optimizing atomic positions and lattice vectors to match the interpolated target, without a trained decoder. In this preprint, the approach raises HfO$_2$ ground-state recovery from 20--35\% to 60--95\% and reduces the cost of a variable-supercell search over 16 perovskites about tenfold; these tests use the authors' own target systems rather than shared benchmark cohorts, and the cost of the inverse optimization should be counted when comparing search budgets. Structural fingerprints quantify the distance between candidates and are used to suppress near-duplicates and maintain niches in the population \citep{OganovValle2009Fingerprints}. Later developments introduced symmetric initialization and environment-sensitive variation operators, discussed in Section~\ref{sec:structural-priors} \citep{CSP057}. The method has been extended from fixed-composition bulk crystals to variable-composition searches that explore a whole chemical system in one run \citep{Oganov2010VariableComposition}, to surface reconstructions with variable stoichiometry \citep{Zhu2013USPEXSurfaces}, and to molecular crystals with rigid molecular units \citep{MolecularUSPEX2012}. The fitness need not be the enthalpy: hardness and other properties can be optimized directly \citep{Lyakhov2011Superhard}, and evolutionary metadynamics explores transformations between phases \citep{CSP055}. To reduce the cost of the underlying calculations, USPEX has been coupled to moment-tensor machine-learning potentials trained by active learning, which select new configurations for first-principles labelling as the search proceeds \citep{CSP085}.
 
USPEX is best known through its wide applications. It predicted the superconducting H$_3$S phase \citep{DiscDuan2014Pressurein} later found to superconduct at 203~K \citep{DiscDrozdov2015Convention}, the transparent high-pressure phase of sodium \citep{DiscMa2009Transparen}, ``forbidden'' sodium chlorides such as NaCl$_3$ \citep{DiscZhang2013Unexpected}, the stable helium compound Na$_2$He \citep{DiscDong2017A}, and superhard $\gamma$-boron \citep{DiscOganov2009Ionic}. In the audit of Section~\ref{sec:discoveries}, USPEX is the search method most often associated with experimentally examined predictions.

\begin{figure}[!htb]
\centering
\begin{tikzpicture}[x=1mm,y=1mm,font=\sffamily\footnotesize,
stage/.style={draw=teal!60!black,fill=teal!10,rounded corners=2pt,minimum height=15mm,text width=23mm,align=center,inner sep=1mm,execute at begin node={\hyphenpenalty=10000}},
small/.style={font=\sffamily\scriptsize,text=black!65,align=center,execute at begin node={\hyphenpenalty=10000\exhyphenpenalty=10000}},
ptitle/.style={font=\sffamily\footnotesize\bfseries,anchor=north west},
ex/.style={font=\sffamily\scriptsize,text=black!60,anchor=south west,align=left,execute at begin node={\hyphenpenalty=10000\exhyphenpenalty=10000}},
arr/.style={-{Stealth[length=1.6mm]},thick,black!70},
cand/.style={draw=orange!75!black,fill=orange!14,rounded corners=1pt,minimum size=4mm,inner sep=0}]

\draw[draw=teal!60!black,rounded corners=3pt,fill=teal!3] (0,0) rectangle (162,46);
\node[ptitle] at (1.5,45.2) {a\quad Evolutionary and genetic search};
\node[draw=orange!75!black,fill=orange!8,rounded corners=2pt,minimum width=22mm,minimum height=17mm] (pop) at (16,25) {};
\foreach \i/\dx/\dy in {1/-6/3,2/0/3,3/6/3,4/-6/-3.5,5/0/-3.5,6/6/-3.5} \node[cand] at (16+\dx,26+\dy) {};
\node[font=\sffamily\scriptsize\bfseries] at (16,18.2) {population};
\node[stage] (sel) at (46,25) {\textbf{Select}\\[1pt]{\scriptsize energy (fitness); age, niche or Pareto rank}};
\node[stage] (var) at (77,25) {\textbf{Vary}\\[1pt]{\scriptsize heredity (cut-and-splice), mutation, permutation, soft-mode}};
\node[stage] (rel) at (108,25) {\textbf{Relax}\\[1pt]{\scriptsize local minimization with DFT or an ML potential}};
\node[stage] (div) at (139,25) {\textbf{Keep diverse}\\[1pt]{\scriptsize fingerprints, duplicate removal, random injection, space-group balance}};
\draw[arr] (pop.east)--(sel.west); \draw[arr] (sel.east)--(var.west); \draw[arr] (var.east)--(rel.west); \draw[arr] (rel.east)--(div.west);
\draw[arr] (div.south) |- (90,12) -| node[pos=0.25,above,small] {next generation} (pop.south);
\node[ex,text width=158mm] at (2,1.2) {USPEX, XtalOpt and GASP: genetic search with crystal-specific operators.\quad GAtor: clustering niches.\quad ParetoCSP: age--fitness selection with M3GNet.\quad ParetoCSP2: space-group diversity.};

\begin{scope}[shift={(0,-52)}]
\draw[draw=teal!60!black,rounded corners=3pt,fill=teal!3] (0,0) rectangle (52,48);
\node[ptitle] at (1.5,47.2) {b\quad Particle swarm};
\begin{scope}[shift={(26,28)},scale=0.78]
 \foreach \r/\c in {17/8,13/14,9/22,5/34} \draw[blue!40!black!40,fill=blue!\c] (0,0) ellipse ({\r*1.15} and {\r*0.75});
 \fill[orange!85!black,draw=white,line width=.3pt] (1,2.4)--(2.9,0.5)--(1,-1.4)--(-0.9,0.5)--cycle;
 \node[font=\sffamily\scriptsize] at (4.5,-2.6) {$\mathbf g$};
 \foreach \x/\y/\px/\py in {-15/8/-9/5,13/-7/8/-3,-10/-9/-6/-4,15/9/9/4}{
   \fill[orange!85!black] (\x,\y) circle (0.9);
   \draw[-{Stealth[length=1.3mm]},teal!60!black,thick] (\x,\y)--($(\x,\y)!0.45!(1,0.5)$);
   \draw[black!45,densely dotted] (\x,\y)--(\px,\py); \fill[black!55] (\px,\py) circle (0.5);}
 \node[small] at (-15.5,11) {$\mathbf x_i$}; \node[small] at (-6.5,6.5) {$\mathbf p_i$};
\end{scope}
\node[small,anchor=south] at (26,9.6) {\tiny $\mathbf v_i\leftarrow\omega\mathbf v_i+c_1r_1(\mathbf p_i-\mathbf x_i)+c_2r_2(\mathbf g-\mathbf x_i)$};
\node[small,anchor=south] at (26,5.6) {personal best $\mathbf p_i$; swarm best $\mathbf g$};
\node[ex] at (1.5,0.8) {CALYPSO};
\end{scope}

\begin{scope}[shift={(55,-52)}]
\draw[draw=teal!60!black,rounded corners=3pt,fill=teal!3] (0,0) rectangle (52,48);
\node[ptitle] at (1.5,47.2) {c\quad Differential evolution};
\coordinate (xa) at (9,19); \coordinate (xb) at (40,38); \coordinate (xc) at (28,30); \coordinate (xi) at (43,17);
\coordinate (v) at ($(xa)+0.8*($(xb)-(xc)$)$);
\draw[-{Stealth[length=1.4mm]},black!55,thick] (xc)--(xb) node[midway,above left,small,inner sep=0.6pt] {$\mathbf x_b-\mathbf x_c$};
\draw[-{Stealth[length=1.4mm]},teal!60!black,thick] (xa)--(v) node[midway,above left,small,text=teal!50!black,inner sep=0.6pt] {$F(\mathbf x_b-\mathbf x_c)$};
\foreach \p/\l/\a in {xa/{$\mathbf x_a$}/below,xb/{$\mathbf x_b$}/above,xc/{$\mathbf x_c$}/left,xi/{target $\mathbf x_i$}/below} {\fill[orange!85!black] (\p) circle (0.9); \node[small,\a] at (\p) {\l};}
\fill[teal!60!black] (v) circle (0.9); \node[small,right,text=teal!50!black] at (v) {mutant $\mathbf v$};
\draw[-{Stealth[length=1.4mm]},black!45,dashed] (v) to[bend left=20] node[pos=0.62,above right,small,inner sep=0.6pt] {crossover} (xi);
\node[small,anchor=south] at (26,8.6) {keep the trial vector if its objective is lower};
\node[ex,text width=49mm] at (1.5,0.8) {SGO; MUSE combines it with annealing and basin hopping};
\end{scope}

\begin{scope}[shift={(110,-52)}]
\draw[draw=teal!60!black,rounded corners=3pt,fill=teal!3] (0,0) rectangle (52,48);
\node[ptitle] at (1.5,47.2) {d\quad Geometry-constrained};
\foreach \ox/\lab/\col/\cells in {6/{target map}/blue/{1/2,2/1,2/3,3/2,3/4,4/3,1/4,4/1,5/5},30/{candidate map}/orange/{1/2,2/1,2/3,3/2,4/4,1/4,4/1,5/3,3/5}}{
  \begin{scope}[shift={(\ox,22)}]
   \draw[black!35] (0,0) grid[step=3] (15,15);
   \foreach \i/\j in \cells \fill[\col!55!black!70] ({(\i-1)*3+0.3},{15-\j*3+0.3}) rectangle ({\i*3-0.3},{15-(\j-1)*3-0.3});
   \node[small] at (7.5,-2.6) {\lab};
  \end{scope}}
\node[font=\sffamily\small] at (25.5,29.5) {$\approx$};
\node[small,anchor=south] at (26,9.4) {objective: agreement with a supplied\\contact map or distance matrix};
\node[ex,text width=49mm] at (1.5,0.8) {CMCrystal, DMCrystal,\\CMCrystalHS, CMCrystalMOO};
\end{scope}
\end{tikzpicture}
\caption{Population-based global optimization. (a) The evolutionary loop shared by genetic CSP codes: candidates are selected by energy (and, in some methods, by age, niche, or Pareto rank), varied by crystal-specific operators, relaxed, and filtered for diversity before the next generation. (b) Particle-swarm optimization moves each candidate toward its own best position $\mathbf p_i$ and the best position of the swarm $\mathbf g$. (c) Differential evolution builds a mutant from the scaled difference of two candidates and accepts the trial vector after crossover only if it improves the objective. (d) Geometry-constrained optimization replaces the energy by agreement with a predicted contact map or distance matrix. Representative methods are listed at the bottom of each panel.}
\label{fig:population-search}
\end{figure}
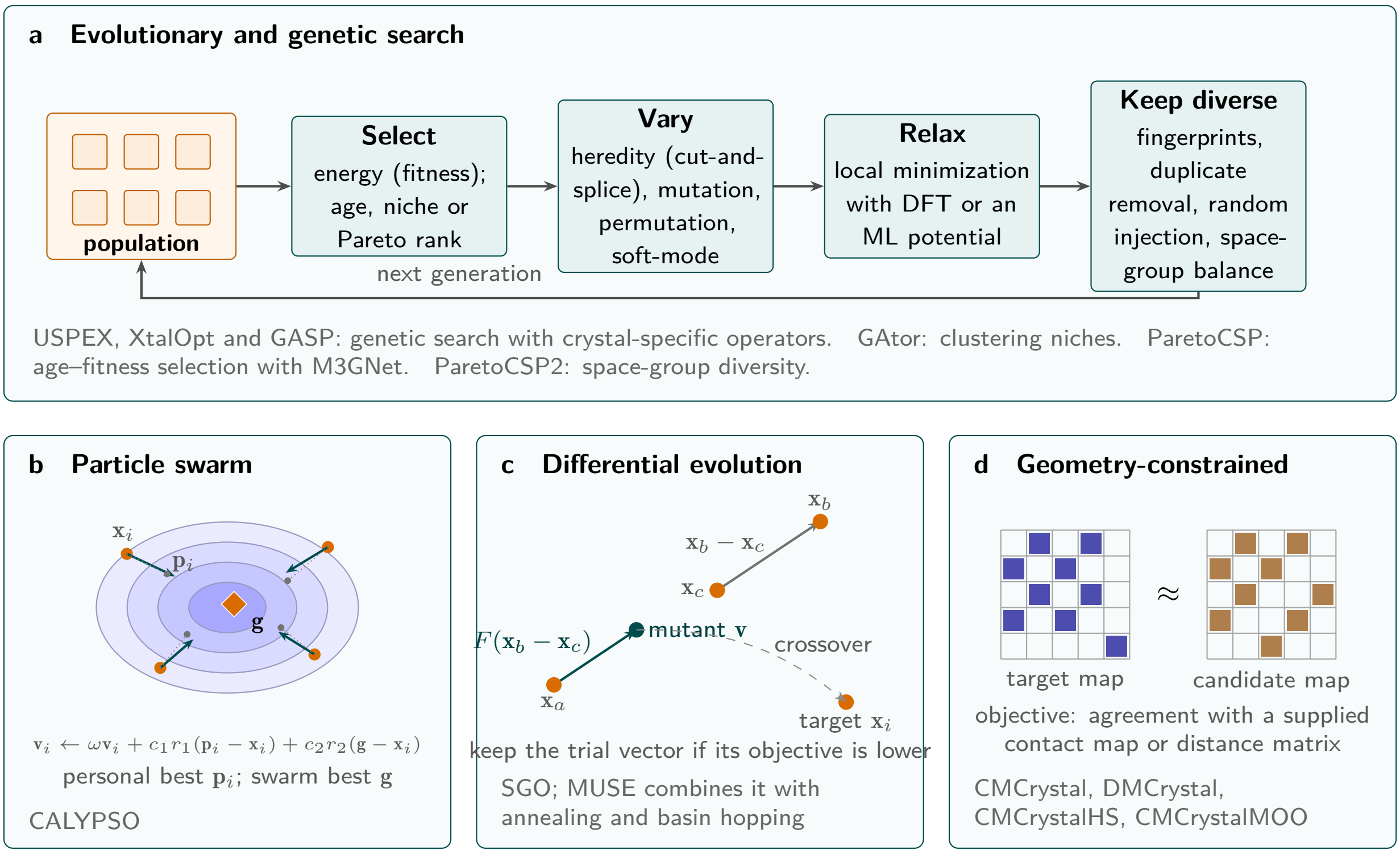
 
XtalOpt \citep{CSP053,CSP052,CSP069,CSP074,CSP084,CSP091} provides an independent open implementation of evolutionary CSP, with subsequent software releases extending its capabilities. Later work incorporates multiple objectives and variable-composition Pareto search \citep{CSP158,CSP196}. GASP \citep{CSP080} provides another implementation supporting structure and phase prediction. Such releases do not define new optimization paradigms, but a benchmark should state which one it ran.
 
The effectiveness of an evolutionary search depends on the compatibility between its representation and its operators. Recombining arbitrary coordinate arrays may create severe overlaps or destroy useful motifs. Crystal-specific operators seek to preserve useful structural fragments while changing composition-compatible arrangements, lattice parameters, or collective distortions. Selection alone also tends to concentrate the population around a small set of basins. Fingerprints, duplicate rejection, random injection, and other diversity mechanisms are therefore central algorithmic choices \citep{CSP044,CSP050,CSP057}.
 
Population diversity can also be organized explicitly by niches or competing objectives. GAtor \citep{CSP076,CSP078} develops evolutionary search for molecular crystals with clustering-based niches. ParetoCSP \citep{CSP141} combines an NSGA-III-based multiobjective genetic algorithm with the M3GNet interatomic potential, treating genotypic age and predicted energy as competing selection criteria. Age--fitness selection helps maintain exploration while the learned potential makes candidate evaluation and relaxation more affordable.
 
ParetoCSP2 \citep{CSP175} extends this search direction to polymorph prediction through adaptive space-group diversity control, together with an age--fitness evolutionary mechanism, improved population initialization, and iterative structural relaxation. Controlling the representation of space groups discourages the population from concentrating in a single symmetry class. Polymorph coverage thus becomes an explicit search objective, although space-group diversity is only a proxy for it, since different structures can share a space group. Because a lower-energy population can lose distinct alternatives, lowest-energy discovery, diversity retention, and recovery of multiple forms are best reported separately.
 
\subsection{Particle swarms and differential evolution}
Particle-swarm CSP uses a population whose updates depend on individual and collective search information. CALYPSO \citep{CSP043,CSP054} introduced this strategy for crystal prediction and combined it with structural construction, local optimization, and duplicate control. The quality of the search depends on the swarm rule and equally on how cells and positions are encoded, how invalid updates are repaired, and how new diversity enters the population.
 
CALYPSO has since grown from a bulk-crystal optimizer into a general structure-search platform \citep{Wang2015CALYPSOReview,Wang2016CALYPSOApplications,Tong2019CALYPSOMethodology}. Its current releases share a common core: symmetry-constrained generation of initial and replacement structures across the space groups, a fraction of randomly generated structures injected each generation to maintain diversity, bond-characterization-matrix fingerprints that remove near-duplicates before expensive relaxation, and interfaces to external first-principles and force-field codes for local optimization. Around this core, dedicated modes address isolated clusters, two-dimensional layers, reconstructed surfaces and interfaces \citep{Wang2015CALYPSOReview,Gao2022CALYPSO2D}, and objective functions can target properties such as hardness in addition to enthalpy \citep{Zhang2019CALYPSOSuperhard}. The most recent developments target the cost of the underlying energy evaluations. Machine-learning potentials can be trained in advance to replace DFT or learned on the fly during the search, which reproduced the experimental B$_{36}$ and B$_{40}$ clusters and proposed a putative B$_{84}$ ground state at a cost several orders of magnitude below a fully DFT-based search \citep{Tong2018CALYPSOML}. A concurrent-learning scheme now couples swarm exploration, quantum-mechanical labelling of the most representative structures, and iterative Deep Potential training, yielding compact training sets for multicomponent systems such as Mg--Al alloys and Li--La--H superhydrides \citep{Wang2024ConcurrentCSP}. The symmetry-orientated divide-and-conquer strategy, SPAI, is discussed in Section~\ref{sec:structural-priors}.
 
CALYPSO's influence is clearest in its range of applications. It has been used most prominently to predict compressed hydrogen-rich superconductors, including the hydrogen sulfide and lanthanum hydride phases whose high critical temperatures were subsequently confirmed experimentally \citep{Cui2019CALYPSOHydrides}. Other application areas include superhard materials \citep{Zhang2019CALYPSOSuperhard}, two-dimensional materials and surfaces \citep{Gao2022CALYPSO2D}, isolated clusters \citep{Wang2015CALYPSOReview}, and high-pressure phases of planetary and geoscience materials \citep{Hermann2019CALYPSOGeoscience}. This breadth has made CALYPSO, alongside USPEX, a standard general-purpose CSP package and a common reference point for newer search and learned-generation methods.
 
Differential evolution is a related population scheme that constructs variations from differences between candidate parameter vectors. SGO \citep{CSP073} applies this family of methods to atomistic global optimization, including crystal structures. Multiobjective differential evolution \citep{CSP064} also appears in property-oriented materials design. Although differential evolution is itself evolutionary in a broad mathematical sense, separating it from crystal-specific genetic recombination is useful for comparing implementations. Other implementations combine several of these strategies; MUSE \citep{CSP060}, for example, couples evolutionary search with annealing and basin hopping.
 
No population rule removes dependence on the energy evaluator. The comparison by Ji and co-workers \citep{CSP041} between genetic and minima-hopping searches shows why comparisons must specify composition, system size, and supplied information.
 
\subsection{Geometry-constrained global optimization}\label{sec:geometry-search}
Global optimizers can reconstruct structures by matching geometric descriptors instead of minimizing an energy model. CMCrystal \citep{CMCrystal2021} formulates reconstruction as contact-map matching under supplied crystallographic information and compares six global optimization algorithms. DMCrystal \citep{DMCrystal2020} uses a genetic algorithm to fit pairwise atomic distances, with space-group and lattice information constraining reconstruction. Distances provide more geometric information than binary contacts, but both approaches depend on the quality and completeness of the supplied descriptors. Their tests show that such constraints are useful and a composition-only predictor would also have to infer them.
 
CMCrystalHS \citep{CMCrystalHS2021} addresses high-symmetry structures by using PyXtal to generate and filter candidates with compatible Wyckoff assignments, then applying differential evolution to the free positional parameters with contact-map agreement as the objective. This couples a discrete assignment problem to continuous coordinate optimization. CMCrystalMOO \citep{CMCrystalMOO2022} instead combines contact-map agreement, individual age, and coordination-number agreement in a multiobjective genetic search. The age objective discourages premature convergence, while coordination constraints address local environments extending beyond contacts within the chosen cell. ParetoCSP later carried the age objective over to energy search with a neural potential.
 
\section{Energy-landscape exploration}
While population methods organize information across candidate structures, landscape methods emphasize connected sequences of configurations or local minima. Annealing changes acceptance behavior through a temperature-like parameter. Basin-hopping-style methods combine perturbation and local minimization so that acceptance operates on relaxed states rather than arbitrary high-energy geometries. Their performance depends on the move distribution and on the balance between revisiting useful basins and escaping already explored regions. Earlier landscape studies motivated the use of such searches for identifying promising solids and connecting them to synthesis considerations \citep{CSP002,CSP012,CSP032}. Figure~\ref{fig:landscape-methods} compares how these methods move across a landscape.
 
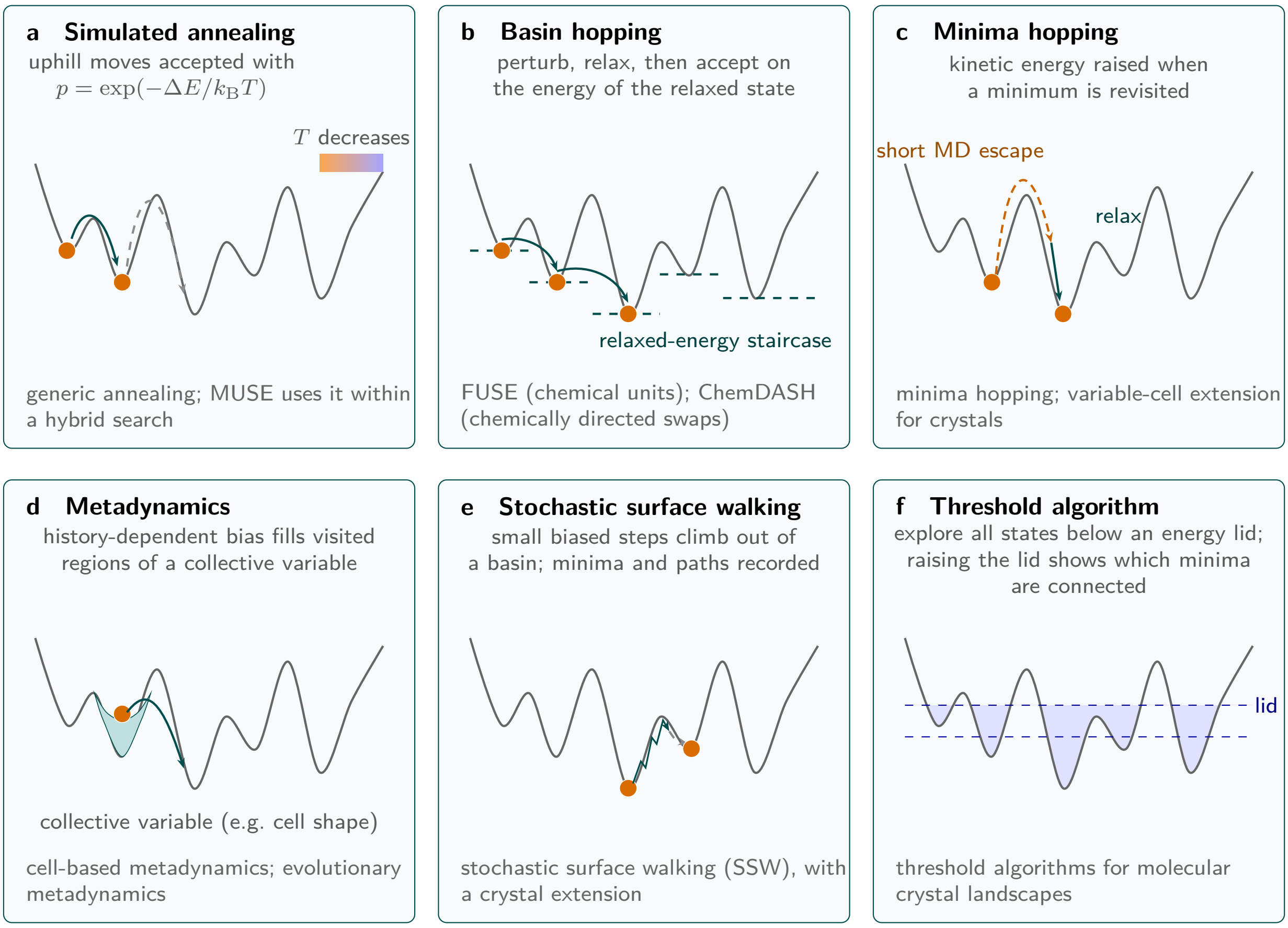
\begin{figure}[!htb]
\centering
\begin{tikzpicture}[x=1mm,y=1mm,font=\sffamily\footnotesize,
ptitle/.style={font=\sffamily\footnotesize\bfseries,anchor=north west},
small/.style={font=\sffamily\scriptsize,text=black!65,align=center,execute at begin node={\hyphenpenalty=10000\exhyphenpenalty=10000}},
ex/.style={font=\sffamily\scriptsize,text=black!60,anchor=south west,align=left,text width=49mm,execute at begin node={\hyphenpenalty=10000\exhyphenpenalty=10000}},
walk/.style={-{Stealth[length=1.3mm]},thick,teal!60!black},
dot/.style={circle,fill=orange!85!black,draw=white,line width=.3pt,inner sep=0,minimum size=2.2mm}]

\newcommand{\cspland}{\draw[thick,black!60] plot[smooth,tension=0.55] coordinates {(4,32) (8,21) (11.5,25) (15,17) (19.5,28) (24,13) (28,22) (32,18) (36,29) (40,15) (44,24) (48,31)};}
\newcommand{\panel}[2]{\draw[draw=teal!60!black,rounded corners=3pt,fill=teal!3] (0,0) rectangle (52,56); \node[ptitle] at (1.5,55.2) {#1\quad #2};}

\begin{scope}[shift={(0,0)}]
\panel{a}{Simulated annealing}
\begin{scope}[shift={(0,4)}]
\cspland
\node[dot] at (8,21) {}; \node[dot] at (15,17) {};
\draw[walk] (8.6,22.5) .. controls (10,27) and (12.5,26.5) .. (14.4,19);
\draw[walk,dashed,black!45] (15.6,19) .. controls (17,30) and (20,31) .. (22.8,15.5);
\node[small] at (20,43) {uphill moves accepted with\\$p=\exp(-\Delta E/k_\mathrm{B}T)$};
\shade[left color=orange!70,right color=blue!35] (40,31) rectangle (48,33.2); \node[small] at (44,35.4) {$T$ decreases};
\end{scope}
\node[ex] at (1.5,0.8) {generic annealing; MUSE uses it within a hybrid search};
\end{scope}

\begin{scope}[shift={(55,0)}]
\panel{b}{Basin hopping}
\begin{scope}[shift={(0,4)}]
\cspland
\foreach \a/\b/\e in {4/11.5/21,11.5/19.5/17,19.5/28/13,28/36/18,36/48/15} \draw[teal!60!black,dashed,line width=.8pt] (\a,\e)--(\b,\e);
\foreach \p in {(8,21),(15,17),(24,13)} \node[dot] at \p {};
\draw[walk] (8,22.4) to[bend left=40] (15,18.4); \draw[walk] (15,18.4) to[bend left=40] (24,14.4);
\node[small] at (26,43) {perturb, relax, then accept on\\the energy of the relaxed state};
\node[small,text=teal!50!black] at (35,9.5) {relaxed-energy staircase};
\end{scope}
\node[ex] at (1.5,0.8) {FUSE (chemical units); ChemDASH (chemically directed swaps)};
\end{scope}

\begin{scope}[shift={(110,0)}]
\panel{c}{Minima hopping}
\begin{scope}[shift={(0,4)}]
\cspland
\node[dot] at (15,17) {}; \node[dot] at (24,13) {};
\draw[-{Stealth[length=1.3mm]},thick,orange!80!black,densely dashed] (15.5,18) .. controls (17,33) and (20,33) .. (22.5,22);
\draw[walk] (22.5,22)--(23.6,14.6);
\node[small,text=orange!60!black] at (11,33.5) {short MD escape};
\node[small,text=teal!50!black] at (31,25.5) {relax};
\node[small] at (26,43) {kinetic energy raised when\\a minimum is revisited};
\end{scope}
\node[ex] at (1.5,0.8) {minima hopping; variable-cell extension for crystals};
\end{scope}

\begin{scope}[shift={(0,-60)}]
\panel{d}{Metadynamics}
\begin{scope}[shift={(0,4)}]
\cspland
\fill[teal!25,draw=teal!60!black,line width=.3pt] plot[smooth,tension=0.55] coordinates {(11.5,25) (13,22.6) (15,21.6) (17,22.6) (18.6,25)} -- plot[smooth,tension=0.55] coordinates {(18.6,25) (17.4,21.6) (15,17) (12.6,21.6) (11.5,25)} -- cycle;
\node[dot] at (15,22.4) {};
\draw[walk] (15.6,22.6) .. controls (17.5,25) and (19.5,26) .. (22.8,15.5);
\node[small] at (26,43) {history-dependent bias fills visited\\regions of a collective variable};
\node[small] at (26,8.6) {collective variable (e.g.\ cell shape)};
\end{scope}
\node[ex] at (1.5,0.8) {cell-based metadynamics; evolutionary metadynamics};
\end{scope}

\begin{scope}[shift={(55,-60)}]
\panel{e}{Stochastic surface walking}
\begin{scope}[shift={(0,4)}]
\cspland
\node[dot] at (24,13) {};
\draw[walk,line width=.6pt] (24.6,13.6)--(25.6,16.4)--(26.2,15.6)--(27,19.4)--(27.6,18.6)--(28.4,21.6)--(29.2,20.2);
\draw[walk,dashed,black!45] (29.2,20.2) .. controls (30,19) .. (31.6,18.6);
\node[dot] at (32,18) {};
\node[small] at (26,43) {small biased steps climb out of\\a basin; minima and paths recorded};
\end{scope}
\node[ex] at (1.5,0.8) {stochastic surface walking (SSW), with a crystal extension};
\end{scope}

\begin{scope}[shift={(110,-60)}]
\panel{f}{Threshold algorithm}
\begin{scope}[shift={(0,4)}]
\begin{scope}
\clip (4,0) rectangle (48,23.5);
\fill[blue!12] plot[smooth,tension=0.55] coordinates {(4,32) (8,21) (11.5,25) (15,17) (19.5,28) (24,13) (28,22) (32,18) (36,29) (40,15) (44,24) (48,31)} -- (48,23.5) -- (4,23.5) -- cycle;
\end{scope}
\cspland
\draw[blue!60!black,dashed] (4,23.5)--(48,23.5) node[right,small,text=blue!50!black,inner sep=0.5pt] {lid};
\draw[blue!60!black,dashed] (4,19.5)--(48,19.5);
\node[small] at (26,42) {explore all states below an energy lid;\\raising the lid shows which minima\\are connected};
\end{scope}
\node[ex] at (1.5,0.8) {threshold algorithms for molecular crystal landscapes};
\end{scope}
\end{tikzpicture}
\caption{Energy-landscape exploration on a schematic one-dimensional landscape. (a) Simulated annealing accepts uphill moves with a Boltzmann probability at a decreasing temperature. (b) Basin hopping perturbs and relaxes, so that acceptance acts on the energies of relaxed states (dashed staircase). (c) Minima hopping escapes a minimum by a short molecular-dynamics trajectory and then relaxes, raising the kinetic energy when a minimum is revisited. (d) Metadynamics adds a history-dependent bias that fills visited regions of a collective variable. (e) Stochastic surface walking climbs out of a basin by small biased displacements, recording minima and pathways. (f) Threshold algorithms explore all states below an energy lid, revealing which minima are connected. Representative methods are listed at the bottom of each panel.}
\label{fig:landscape-methods}
\end{figure}
 
Minima hopping \citep{CSP020,CSP042} combines dynamical escape trajectories with local relaxation. The original methodological paper treats complex molecular systems; the variable-cell extension explicitly addresses crystalline enthalpy landscapes. This distinction matters when assigning historical priority and when transferring a cluster optimizer to periodic CSP. Changes in lattice degrees of freedom, pressure treatment, and permitted cell shapes are part of the crystal algorithm.
 
Metadynamics uses history-dependent bias in selected collective variables to encourage movement away from visited regions. Cell-based formulations \citep{CSP018,CSP025} adapt this idea to crystalline transformations. Its reach depends on the collective variables: a useful cell deformation may expose a transition, whereas a transformation requiring an unrepresented internal rearrangement may remain difficult. Evolutionary metadynamics \citep{CSP055} explicitly connects this landscape strategy to population-based structural variation.
 
Stochastic surface walking \citep{CSP058,CSP062} combines biased displacements and acceptance steps to explore minima and pathways, with a separate extension to crystalline lattice and atomic variables. FUSE \citep{CSP083} uses flexible chemical units within basin-hopping exploration, and ChemDASH \citep{CSP093,CSP119} introduces chemically directed moves. Molecular threshold algorithms \citep{CSP116,CSP136} examine connected regions of a crystal-energy landscape and help distinguish closely related minima from more separated forms. A catalogue of local minima thus carries information beyond its lowest-energy member, although connectivity under a model is not a kinetic description of crystallization.
\section{Symmetry, topology, and chemical knowledge in CSP}\label{sec:structural-priors}
Across the search families discussed above, performance depends on both the exploration mechanism and the structural knowledge encoded in candidates and moves. This section examines symmetry, topology, and chemical constraints as ingredients shared across those families. Structural knowledge can reduce the number of independent variables, suppress implausible configurations, preserve useful fragments, remove equivalent candidates, or allocate effort among competing structural families. Table~\ref{tab:structural-priors} compares these roles across representative methods. Figure~\ref{fig:structural-priors} shows where each kind of knowledge enters a search and how strongly it constrains the result.
 
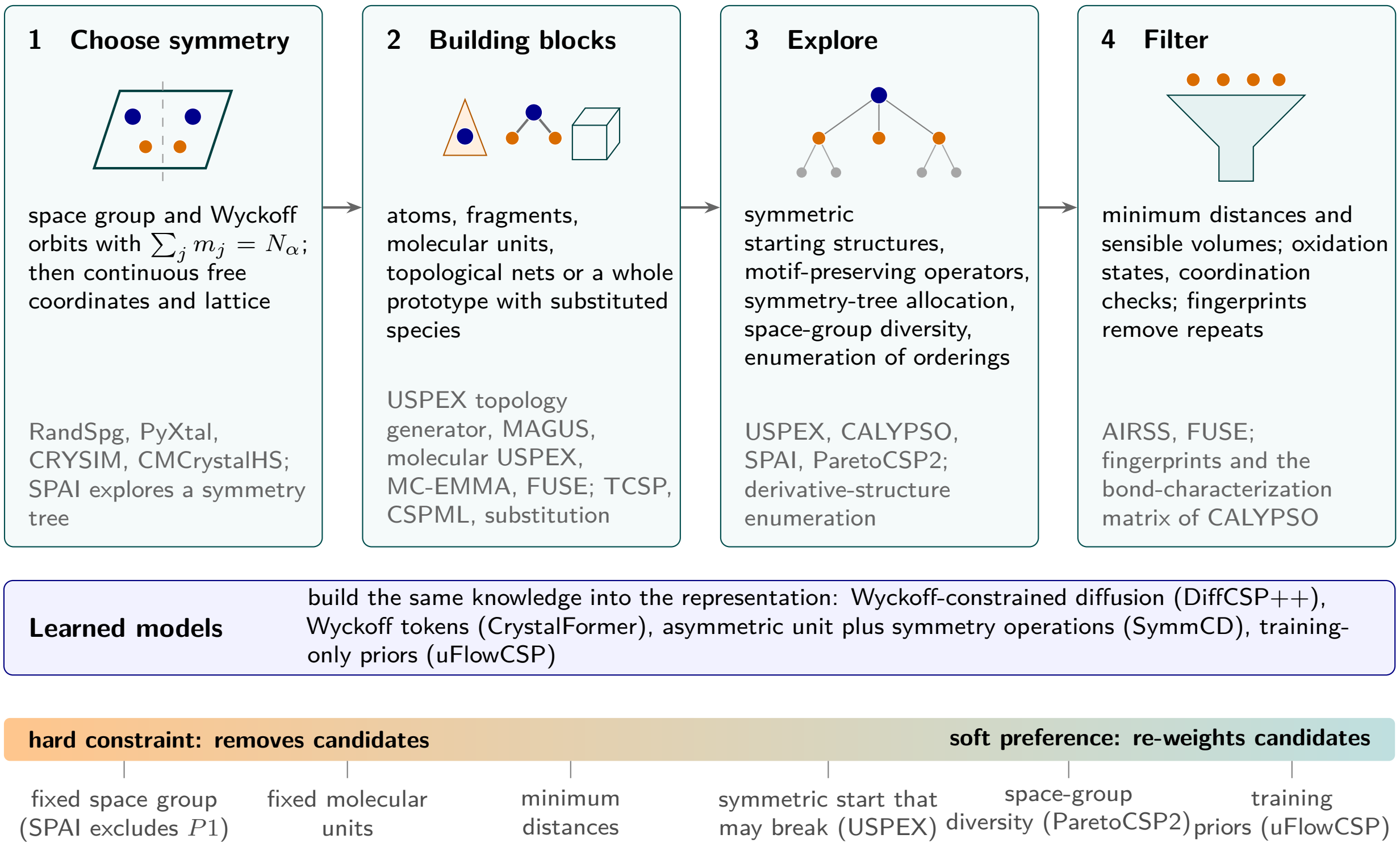
\begin{figure}[!htb]
\centering
\begin{tikzpicture}[x=1mm,y=1mm,font=\sffamily\footnotesize,
stage/.style={draw=teal!60!black,fill=teal!4,rounded corners=3pt,minimum width=37mm,minimum height=63mm,anchor=north west},
stitle/.style={font=\sffamily\footnotesize\bfseries,anchor=north west,text width=35mm},
desc/.style={font=\sffamily\scriptsize,anchor=north west,text width=33.5mm,align=left,execute at begin node={\hyphenpenalty=10000\exhyphenpenalty=10000}},
ex/.style={font=\sffamily\scriptsize,text=black!60,anchor=south west,text width=33.5mm,align=left,execute at begin node={\hyphenpenalty=10000\exhyphenpenalty=10000}},
arr/.style={-{Stealth[length=1.8mm]},thick,black!60},
atomA/.style={circle,fill=blue!55!black,draw=white,line width=.3pt,minimum size=2.0mm,inner sep=0},
atomB/.style={circle,fill=orange!85!black,draw=white,line width=.3pt,minimum size=1.6mm,inner sep=0}]

\foreach \i/\x in {1/0,2/41.7,3/83.4,4/125} \node[stage] (s\i) at (\x,0) {};
\foreach \i/\j in {1/2,2/3,3/4} \draw[arr] ($(s\i.east)+(0,8)$)--($(s\j.west)+(0,8)$);

\node[stitle] at (1.5,-1.5) {1\quad Choose symmetry};
\begin{scope}[shift={(18.5,-14.5)}]
 \draw[teal!50!black,thick] (-8,-4.5)--(5,-4.5)--(8,4.5)--(-5,4.5)--cycle;
 \draw[black!40,dashed] (0,-6)--(0,6);
 \node[atomA] at (-3.5,1.5) {}; \node[atomA] at (3.5,1.5) {}; \node[atomB] at (-2,-2) {}; \node[atomB] at (2,-2) {};
\end{scope}
\node[desc] at (1.5,-22) {space group and Wyckoff orbits with $\sum_j m_j=N_\alpha$; then continuous free coordinates and lattice};
\node[ex] at (1.5,-62) {RandSpg, PyXtal, CRYSIM, CMCrystalHS; SPAI explores a symmetry tree};

\node[stitle] at (43.2,-1.5) {2\quad Building blocks};
\begin{scope}[shift={(60.2,-14.5)}]
 \draw[orange!70!black,fill=orange!12] (-9,-3)--(-4,-3)--(-6.5,3.5)--cycle; \node[atomA] at (-6.5,-0.8) {};
 \draw[black!55,thick] (-1,-1)--(1.5,2)--(4,-1); \node[atomB] at (-1,-1) {}; \node[atomB] at (4,-1) {}; \node[atomA] at (1.5,2) {};
 \draw[teal!50!black] (6,-3.5) rectangle (10,0.5); \draw[teal!50!black] (6,0.5)--(7.5,2.5)--(11.5,2.5)--(10,0.5); \draw[teal!50!black] (11.5,2.5)--(11.5,-1.5)--(10,-3.5);
\end{scope}
\node[desc] at (43.2,-22) {atoms, fragments, molecular units, topological nets or a whole prototype with substituted species};
\node[ex] at (43.2,-62) {USPEX topology generator, MAGUS, molecular USPEX, MC-EMMA, FUSE; TCSP, CSPML, substitution};

\node[stitle] at (84.9,-1.5) {3\quad Explore};
\begin{scope}[shift={(101.9,-14.5)}]
 \node[atomA] (r) at (0,4) {};
 \foreach \dx/\n in {-7/a,0/b,7/c} {\node[atomB] (\n) at (\dx,-1) {}; \draw[black!50] (r)--(\n);}
 \foreach \p/\dx in {a/-2,a/2,c/-2,c/2} \draw[black!35] (\p)--++(\dx,-4) node[circle,fill=black!35,inner sep=0,minimum size=1.2mm] {};
\end{scope}
\node[desc] at (84.9,-22) {symmetric starting structures, motif-preserving operators, symmetry-tree allocation, space-group diversity, enumeration of orderings};
\node[ex] at (84.9,-62) {USPEX, CALYPSO, SPAI, ParetoCSP2; derivative-structure enumeration};

\node[stitle] at (126.5,-1.5) {4\quad Filter};
\begin{scope}[shift={(143.5,-15.5)}]
 \draw[teal!50!black,fill=teal!10] (-8,5)--(8,5)--(2,-1)--(2,-5)--(-2,-5)--(-2,-1)--cycle;
 \foreach \dx in {-5,-1.5,2,5} \node[atomB] at (\dx,6.8) {};
\end{scope}
\node[desc] at (126.5,-22) {minimum distances and sensible volumes; oxidation states, coordination checks; fingerprints remove repeats};
\node[ex] at (126.5,-62) {AIRSS, FUSE; fingerprints and the bond-characterization matrix of CALYPSO};

\draw[draw=blue!50!black,fill=blue!5,rounded corners=3pt] (0,-67) rectangle (162,-78);
\node[font=\sffamily\footnotesize\bfseries,anchor=west] at (1.5,-72.5) {Learned models};
\node[font=\sffamily\scriptsize,anchor=west,text width=126mm,align=left] at (34,-72.5) {build the same knowledge into the representation: Wyckoff-constrained diffusion (DiffCSP++), Wyckoff tokens (CrystalFormer), asymmetric unit plus symmetry operations (SymmCD), training-only priors (uFlowCSP)};

\shade[left color=orange!45,right color=teal!25,rounded corners=2pt] (0,-83) rectangle (162,-88);
\node[font=\sffamily\scriptsize\bfseries,anchor=west] at (1.5,-85.5) {hard constraint: removes candidates};
\node[font=\sffamily\scriptsize\bfseries,anchor=east] at (160.5,-85.5) {soft preference: re-weights candidates};
\foreach \x/\t in {14/{fixed space group\\(SPAI excludes $P1$)},40/{fixed molecular\\units},66/{minimum\\distances},96/{symmetric start that\\may break (USPEX)},124/{space-group\\diversity (ParetoCSP2)},150/{training\\priors (uFlowCSP)}}{
  \draw[black!50] (\x,-88)--(\x,-90);
  \node[font=\sffamily\scriptsize,anchor=north,align=center,text=black!70] at (\x,-90) {\t};}
\end{tikzpicture}
\caption{Where structural knowledge enters a CSP search. Symmetry can define the search variables (1), chemical knowledge the building blocks (2), and both can shape exploration (3) and filtering (4); representative methods are listed at the bottom of each stage. Learned generators build the same knowledge into their representation. The bar orders uses of this knowledge from hard constraints, which remove candidates and can exclude the target, to soft preferences, which only change how candidates are sampled or ranked.}
\label{fig:structural-priors}
\end{figure}
 
\subsection{Space groups and Wyckoff positions as search variables}
A space-group label specifies allowed symmetry operations, but does not determine the occupied sites, their chemical identities, or the remaining positional and lattice parameters. A useful description separates the discrete choice of space group and Wyckoff assignment from the continuous optimization within that assignment. For a fully occupied, ordered structure in a consistent cell setting, composition compatibility requires
\begin{equation}
  \sum_{j:s_j=\alpha} m_j = N_\alpha,
  \label{eq:wyckoff-composition}
\end{equation}
where orbit $j$ has multiplicity $m_j$, species $s_j$, and $N_\alpha$ is the required count of species $\alpha$ in that cell. The remaining search variables are the free coordinates of the occupied orbits and the symmetry-compatible lattice parameters. Multiplicity balance is necessary but insufficient: distinct orbits must not collide, special sites cannot be multiply occupied, and distance constraints must also be satisfied. A reduced formula therefore does not alone specify a feasible assignment; the number of formula units and the cell convention matter.
 
RandSpg \citep{CSP072} constructs random crystals through composition-compatible Wyckoff assignments and interatomic-distance checks. PyXtal \citep{CSP107} generalizes symmetry-based construction to atomic and molecular systems, including special positions and compatibility between molecular and site symmetry. Both are reused as constructors by other optimizers; CMCrystalHS \citep{CMCrystalHS2021}, for example, combines Wyckoff-based construction with differential evolution over free coordinates to match a supplied contact map. CRYSIM \citep{CSP176} explicitly separates space-group, Wyckoff-combination, and independent-coordinate variables in its optimization encoding. Such encodings make explicit the mixed discrete--continuous structure of the problem, which a Cartesian description of every atom hides.
 
\subsection{Symmetry in USPEX and CALYPSO: from initialization to structured exploration}
USPEX's methodological developments \citep{CSP057} show how symmetry and local chemical order can improve evolutionary search together. Random symmetric structures provide compact initial configurations and replenish population diversity. Environment-sensitive variation operators favor useful structural fragments, while soft-mutation introduces collective displacements associated with soft modes. The initial space group is not fixed, however: variation operators and subsequent relaxation can change symmetry, with small perturbations allowing symmetry breaking. Its crystallographic prior therefore guides candidate construction without permanently restricting every descendant to the same symmetry subspace.
 
CALYPSO \citep{CSP043,CSP054} likewise combines its particle-swarm update with symmetry-constrained generation, injection of random structures, and structural comparison. Its bond-characterization matrix is used to identify similar structures, helping avoid repeated exploration of equivalent or closely related candidates. Symmetry and duplicate control serve different purposes: the former organizes candidate construction, whereas the latter limits redundancy after candidates have been produced.
The CALYPSO team's symmetry-orientated divide-and-conquer method, SPAI \citep{CSP108}, makes symmetry part of the exploration strategy itself. A symmetry tree organizes space-group subspaces, compatible Wyckoff combinations, and configurations within those combinations. Artificial-bee-colony-inspired scout, onlooker, and their employed roles distribute exploration across space groups, alternative site assignments, and positional variations, respectively. Energy-dependent role changes concentrate effort in promising subspaces while renewing exploration. The division is thus over crystallographic choices; the material itself is not partitioned. The reported scheme excludes $P1$ candidates and preserves symmetry during evolution; this restriction helps explain both its efficiency for the tested complex symmetric structures and its limited coverage of unrestricted CSP.
 
\subsection{Topology, fragments, and modularity (reusable chemical building blocks)}
Topology describes connectivity rather than a particular set of bond lengths and angles. It is distinct from space-group symmetry: a symmetry label does not uniquely identify a bonding network, and a network can admit distorted embeddings with different geometric symmetries. The USPEX-associated topology-based generator \citep{USPEXTopology2019} selects nets from a topological database and converts suitable nets into stoichiometry-compatible crystal candidates. These candidates can seed evolutionary search or support a broader form of structural data mining. The approach transfers connectivity patterns beyond a single geometrical template, although its coverage still depends on the available nets and their realizable chemical decorations.
 
MAGUS \citep{MAGUS2023} (2023) uses graph-based decomposition to identify structural units that can be retained and rearranged during evolutionary search, alongside on-the-fly learning of interatomic potentials. Its later symmetry-principle development \citep{CSP170} combines a space-group miner, fragment reorganization, and symmetry-preserving mutation. This couples global crystallographic information to local fragments extracted from structures encountered during the search. Its structural knowledge is therefore updated during exploration instead of being fixed before the calculation begins.
 
Chemically defined units (e.g. polyhedra), modules or motifs \cite{aksenov2026crystal} provide another route to dimensional reduction. Constrained molecular USPEX \citep{MolecularUSPEX2012} treats molecular units as whole objects during candidate construction and variation, reducing the need to rediscover their internal bonding. MC-EMMA \citep{CSP070} uses chemically informed modular construction to explore complex inorganic phase fields. FUSE \citep{CSP083} constructs crystals from cation--anion submodules, stacking rules, and coordination checks, then varies their assembly within a basin-hopping search. Its approximate lattice choices do not impose a full internal space-group symmetry. A motif-based scheme \citep{Zhu2017Motifs} describes each structural motif, whatever its size, by a position and an orientation with five degrees of freedom, which greatly reduces the search space; applied to the cathode material LiFePO$_4$, it found many new low-energy structures. Polyhedra can serve as units in the same way: Yokoyama et al.\ \citep{Yokoyama2024Polyhedra} represent the shape and connectivity of space-filling polyhedra as a dual periodic graph and generate the crystal from the standard realization of that graph, recovering face-centred cubic, hexagonal close-packed, and body-centred cubic structures as a first demonstration. Chemical organization can therefore constrain a search without enforcing a space group, at the price of a representational limit: a fixed molecule or restricted motif vocabulary can miss dissociation, unusual coordination, or a reconstructive transformation unless the allowed moves can change those units.
 
\subsection{Prototypes, substitutions, and cation ordering}
A prototype supplies a stronger prior than a space-group label because it also supplies a pattern of occupied sites and a structural arrangement. Data-mined ionic substitution \citep{CSP048} learns which species can replace one another while retaining a known structure type. Template-based CSP methods, including TCSP \citep{CSP123} and CSPML \citep{CSP115}, use related structural information to propose candidates for a target composition. This works when chemistry changes more readily than the underlying framework, but local relaxation rarely escapes an incorrect parent topology.
 
Substitution and ordering are separate choices: choosing a set of replacement elements does not specify which symmetry-related sites they occupy in an ordered supercell. Derivative-structure enumeration \citep{CSP035,CSP056} uses the symmetry of a parent lattice to remove equivalent occupational arrangements and generate distinct decorations within specified cell and composition bounds. It can therefore support searches over cation, alloy, or vacancy ordering. Individual decorations may lower the parent symmetry or enlarge the repeat cell. Here symmetry only removes redundant arrangements; the final crystal need not keep the parent's space group.
 
\subsection{Chemical plausibility and domain-specific constraints}
AIRSS \citep{CSP045} shows how much physical knowledge random search can incorporate through sensible volumes, minimum separations, symmetry, and structural units. Such bounds prevent costly relaxation of obviously unsuitable starting configurations. Oxidation-state compatibility, ionic sizes, coordination preferences, and likely bonding environments can similarly prioritize substitutions or assemblies in ionic systems. FUSE's coordination checks \citep{CSP083} make this use of chemistry explicit. These rules are domain-dependent: an ambient-pressure coordination preference can wrongly exclude high-pressure phases, and an ionic description can mislead for metallic or strongly covalent compounds.
 
A hard restriction removes candidates from the admissible space, whereas a soft preference changes their sampling probability or score. Minimum-distance cutoffs, fixed molecular identities, required topologies, and preferred space groups can constrain a search in different ways. The appropriate strength depends on what is known about the target. Experimental cell or symmetry information can legitimately narrow an observation-assisted prediction, whereas using the reference structure's space group in a nominally formula-only test supplies additional information.
 
\subsection{Structural knowledge in learned search and generation}
 
\begin{table}[!htb]
\centering\small
\caption{How structural knowledge enters CSP. The same prior can be used at several stages; restrictions refer to the stated representation or search setting.}\label{tab:structural-priors}
\begin{tabularx}{\linewidth}{@{}P{.23\linewidth}Y Y@{}}
\toprule Method or component & Knowledge and algorithmic role & Main coverage consideration \\
\midrule
USPEX \citep{CSP057} & Symmetric initialization; environment-sensitive variation & Initial symmetry can change during evolution and relaxation. \\
CALYPSO \citep{CSP054} & Symmetric generation; bond-based duplicate control & Similarity filtering and symmetry sampling are separate choices. \\
SPAI / CALYPSO \citep{CSP108} & Hierarchical space-group, site-assignment, and coordinate exploration & Reported search excludes $P1$ candidates. \\
RandSpg / PyXtal \citep{CSP072,CSP107} & Compatible Wyckoff orbits and site constraints & A constructor supplies candidates, not an energy-based search by itself. \\
Topology-based USPEX generator \citep{USPEXTopology2019} & Database nets supply candidate connectivity & Coverage depends on nets and chemical decorations. \\
MAGUS \citep{MAGUS2023,CSP170} & Graph fragments, mined symmetry, and constrained variation & Fragment and symmetry choices bias accessible candidates. \\
Molecular USPEX; FUSE \citep{MolecularUSPEX2012,CSP083} & Molecular or chemical units reduce independent variables & Fixed units can exclude alternative bonding patterns. \\
TCSP; ionic substitution \citep{CSP123,CSP048} & Known prototypes and substitution preferences initialize structures & Missing parent topologies remain difficult to discover. \\
Derivative enumeration \citep{CSP035,CSP056} & Parent symmetry removes equivalent site decorations & Completeness is bounded by parent lattice and supercell size. \\
DiffCSP++; CrystalFormer; SymmCD \citep{CSP154,CSP177,CSP178} & Symmetry enters generative variables and sampling & Conditioning information and learned coverage must be specified. \\
uFlowCSP \citep{CSP195} & Electronegativity, radius, period/group descriptors and stoichiometry embedding; canonical atom ordering; Wyckoff grouping and crystal-system supervision during training & Formula-only inference uses chemical priors; training-only symmetry information does not impose a target space group at generation. \\
ParetoCSP2 \citep{CSP175} & Space-group diversity guides population management & Distinct polymorphs can share a symmetry label. \\
\bottomrule
\end{tabularx}
\end{table}
 
Learned CSP methods increasingly incorporate the same crystallographic variables. DiffCSP++ \citep{CSP154} constrains lattice and fractional-coordinate diffusion using space-group-compatible lattice representations and Wyckoff positions. CrystalFormer \citep{CSP177} generates sequences describing symmetry-inequivalent sites, including their Wyckoff labels, elements, and coordinates, before reconstructing the full crystal. SymmCD \citep{CSP178} learns a joint distribution over asymmetric-unit information and the transformations needed to generate the complete structure. These designs build crystallography into the representation and sampling procedure instead of leaving the network to infer it from unconstrained coordinates. uFlowCSP \citep{CSP195} incorporates chemical descriptors and symmetry-informed training priors while retaining formula-only inference; its flow-based formulation is discussed in Section~\ref{sec:flow-models}.
 
Building crystallographic symmetry into a model in this way should not be confused with rotational, translational, or permutation equivariance. Equivariance makes predictions transform consistently when an equivalent input description changes; it does not, by itself, require a generated crystal to occupy a particular space group or Wyckoff assignment. Likewise, the ability to specify a space group does not establish exact formula conditioning: site multiplicities and species counts must still satisfy Eq.~\ref{eq:wyckoff-composition}. The source of symmetry information---user-supplied, sampled, predicted, or inherited from a template---remains part of the prediction task.
 
ParetoCSP2 \citep{CSP175} provides a complementary use of symmetry by controlling space-group diversity during evolutionary optimization. It uses crystallographic labels to manage population coverage without committing the search to one label. Across these methods, structural knowledge has moved from fixed construction rules toward adaptive allocation and learned distributions, but the trade-off is unchanged: symmetry, chemical motifs, and topology improve efficiency by concentrating exploration, but can also remove the structure being sought when imposed too narrowly.

Figure~\ref{fig:algorithm-features} compares the roles of these priors across selected methods. A learned proposal can coexist with explicit chemistry or crystallography, while template and evolutionary methods can use learned evaluators without becoming neural generators. Exact full-cell composition is shown separately because a formula ratio, symmetry label, and atom inventory impose different constraints.
 
\begin{figure}[!htb]
\centering
\includegraphics[width=\linewidth]{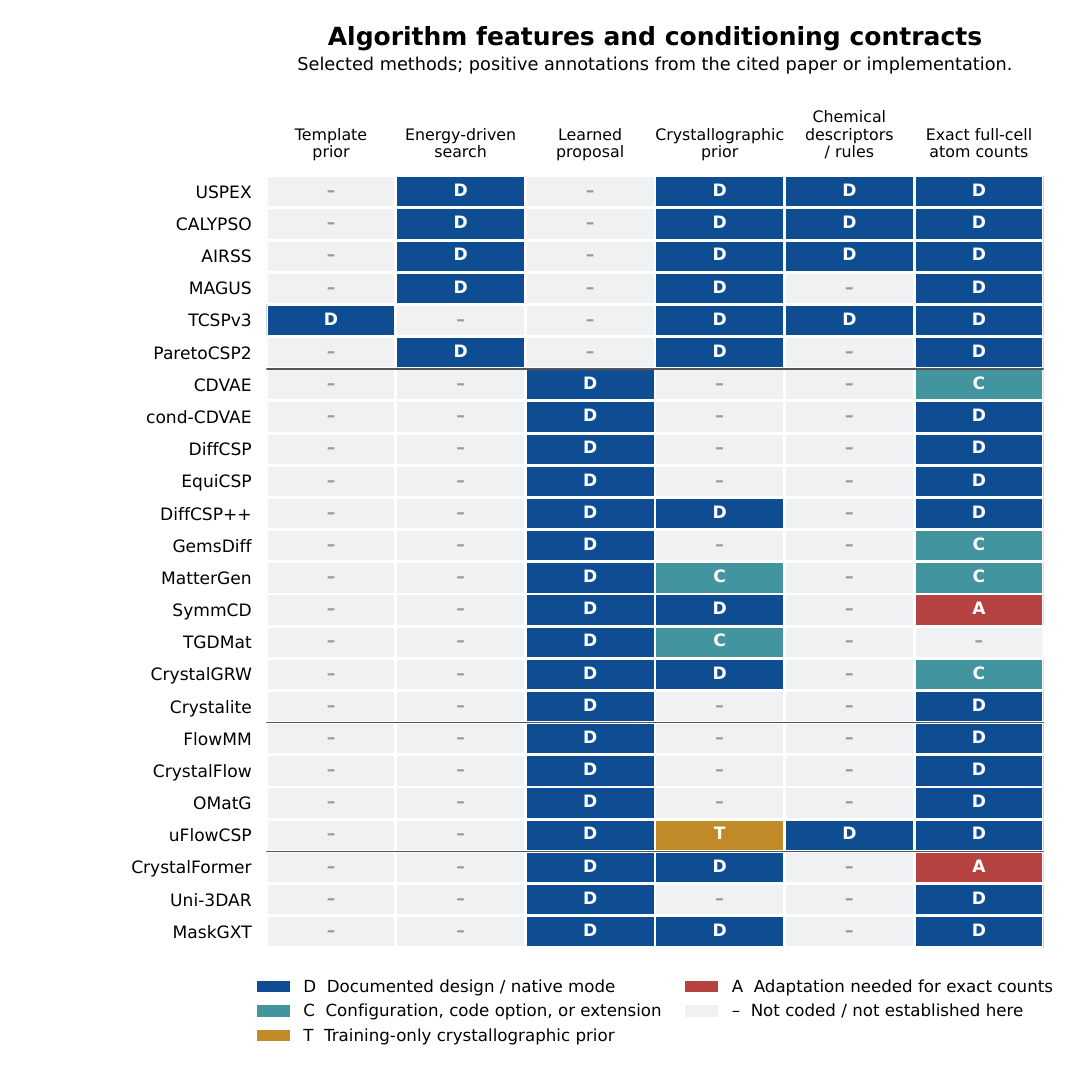}
\caption{Feature comparison of 24 representative CSP algorithms, based on the method references in Table~\ref{tab:algorithm-directory} and the accompanying source-linked feature data. D denotes a documented design feature or native mode; C denotes an implementation option, configuration, or extension; T denotes a training-only crystallographic prior; A identifies an exact-count requirement needing adaptation. A dash means the feature was not coded or established in this comparison, rather than proof of absence. Crystallographic priors include space groups, Wyckoff sites, point groups, or inherited template symmetry; geometric equivariance alone does not receive this label. Chemical descriptors/rules refer to explicit chemical information beyond a learned atomic-number embedding. Energy-driven search means target-specific objective exploration, not routine post-generation relaxation. Exact counts refer to the full-cell elemental inventory, not only reduced stoichiometry.}\label{fig:algorithm-features}
\end{figure}
\section{Machine-learning-assisted and adaptive search}
Structural knowledge can be specified explicitly or learned from data. We now examine how learned representations, evaluators, structural descriptors, and search policies enter the prediction workflow. Figure~\ref{fig:ml-assisted} shows where each enters a search loop.
 
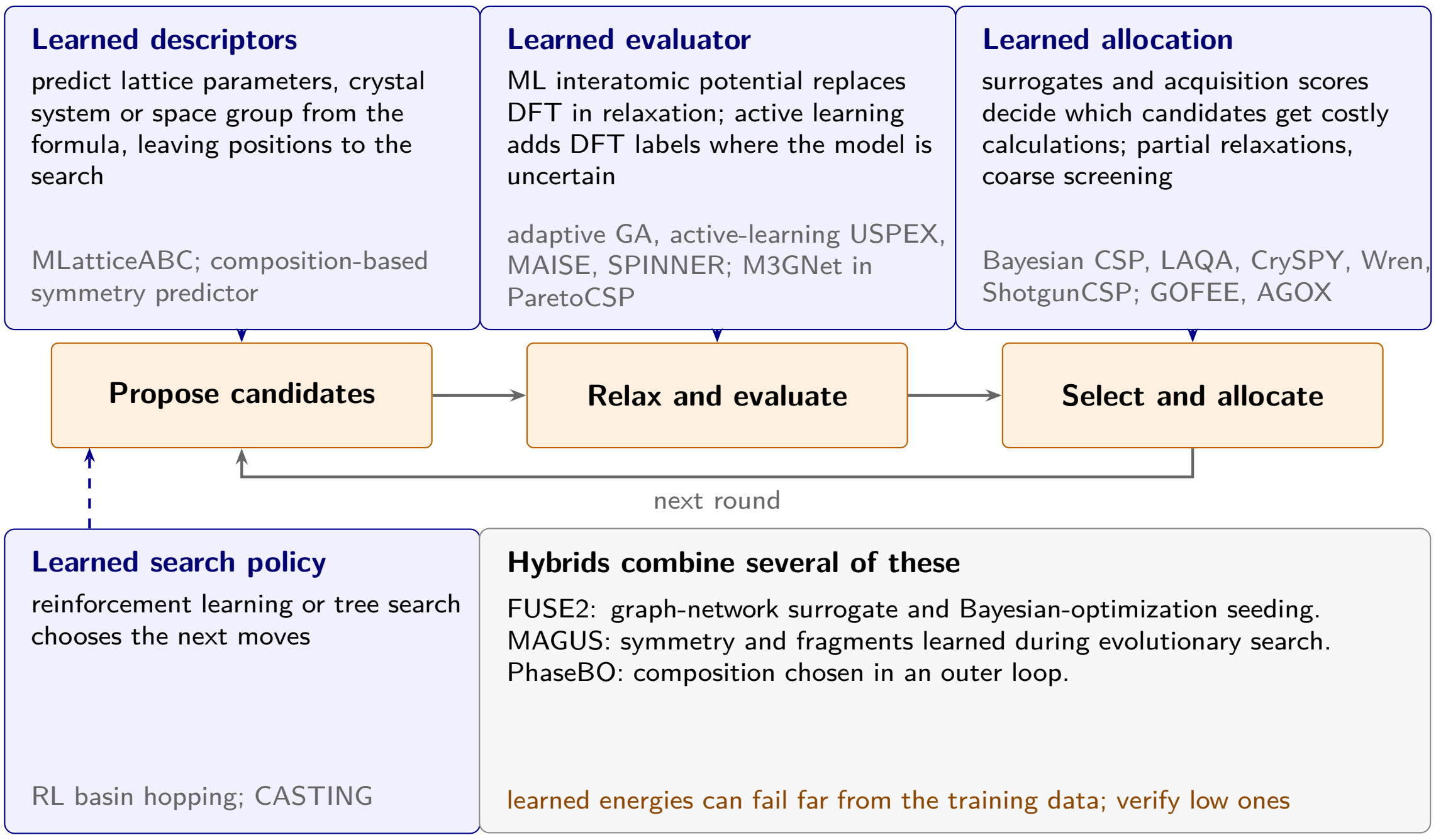
\begin{figure}[!htb]
\centering
\begin{tikzpicture}[x=1mm,y=1mm,font=\sffamily\footnotesize,
loopbox/.style={draw=orange!75!black,fill=orange!12,rounded corners=2pt,minimum width=40mm,minimum height=11mm,align=center,font=\sffamily\footnotesize\bfseries},
mod/.style={draw=blue!55!black,fill=blue!6,rounded corners=3pt,minimum width=50mm,minimum height=34mm,anchor=north west},
mtitle/.style={font=\sffamily\footnotesize\bfseries,anchor=north west,text=blue!45!black},
mdesc/.style={font=\sffamily\scriptsize,anchor=north west,text width=47mm,align=left,execute at begin node={\hyphenpenalty=10000\exhyphenpenalty=10000}},
ex/.style={font=\sffamily\scriptsize,text=black!60,anchor=south west,text width=47mm,align=left,execute at begin node={\hyphenpenalty=10000\exhyphenpenalty=10000}},
arr/.style={-{Stealth[length=1.7mm]},thick,black!60},
plug/.style={-{Stealth[length=1.5mm]},thick,blue!55!black,dashed}]

\node[loopbox] (p) at (31,0) {Propose candidates};
\node[loopbox] (e) at (81,0) {Relax and evaluate};
\node[loopbox] (s) at (131,0) {Select and allocate};
\draw[arr] (p.east)--(e.west); \draw[arr] (e.east)--(s.west);
\draw[arr] (s.south) -- ++(0,-3) -| node[pos=0.25,below,font=\sffamily\scriptsize,text=black!60] {next round} (p.south);

\node[mod] (m1) at (6,41) {};
\node[mtitle] at (7.5,40) {Learned descriptors};
\node[mdesc] at (7.5,35.5) {predict lattice parameters, crystal system or space group from the formula, leaving positions to the search};
\node[ex] at (7.5,8.0) {MLatticeABC; composition-based symmetry predictor};
\draw[plug] (31,7)--(p.north);

\node[mod] (m2) at (56,41) {};
\node[mtitle] at (57.5,40) {Learned evaluator};
\node[mdesc] at (57.5,35.5) {ML interatomic potential replaces DFT in relaxation; active learning adds DFT labels where the model is uncertain};
\node[ex] at (57.5,8.0) {adaptive GA, active-learning USPEX, MAISE, SPINNER; M3GNet in ParetoCSP};
\draw[plug] (81,7)--(e.north);

\node[mod] (m3) at (106,41) {};
\node[mtitle] at (107.5,40) {Learned allocation};
\node[mdesc] at (107.5,35.5) {surrogates and acquisition scores decide which candidates get costly calculations; partial relaxations, coarse screening};
\node[ex] at (107.5,8.0) {Bayesian CSP, LAQA, CrySPY, Wren, ShotgunCSP; GOFEE, AGOX};
\draw[plug] (131,7)--(s.north);

\node[mod,minimum height=32mm] (m4) at (6,-14) {};
\node[mtitle] at (7.5,-15) {Learned search policy};
\node[mdesc] at (7.5,-19.5) {reinforcement learning or tree search chooses the next moves};
\node[ex] at (7.5,-45) {RL basin hopping; CASTING};
\draw[plug] (15,-14)--(15,-5.5);

\draw[draw=black!45,fill=black!3,rounded corners=3pt] (56,-14) rectangle (156,-46);
\node[font=\sffamily\footnotesize\bfseries,anchor=north west] at (57.5,-15) {Hybrids combine several of these};
\node[font=\sffamily\scriptsize,anchor=north west,text width=96mm,align=left] at (57.5,-20) {FUSE2: graph-network surrogate and Bayesian-optimization seeding.\\MAGUS: symmetry and fragments learned during evolutionary search.\\PhaseBO: composition chosen in an outer loop.};
\node[font=\sffamily\scriptsize,anchor=south west,text width=96mm,align=left,text=orange!55!black] at (57.5,-45.4) {learned energies can fail far from the training data; verify low ones};
\end{tikzpicture}
\caption{Where learning enters a structure search. A classical loop proposes candidates, relaxes and evaluates them, and selects which to keep or compute next. Learned descriptors supply cell or symmetry hypotheses to the proposal step; learned interatomic potentials replace first-principles relaxation, with active learning adding reference labels; surrogates and acquisition scores allocate expensive calculations; learned policies choose search moves. Hybrids combine several of these components. Representative methods are listed in each box.}
\label{fig:ml-assisted}
\end{figure}
 
\subsection{Representing crystals for learning}\label{sec:representations}
Every learned component of a CSP workflow first converts a crystal into a model input, and that choice limits what the model can learn and what it can output (Figure~\ref{fig:representations}a). The full-cell description $X=(\mathbf A,\mathbf F,\mathbf L)$ is used directly by diffusion and flow models such as DiffCSP~\citep{CSP128}, which act on fractional coordinates and the lattice. A symmetry-reduced description keeps only the space group, the occupied Wyckoff orbits, and their free parameters; it underlies PyXtal~\citep{CSP107}, Wren~\citep{CSP120}, and CrystalFormer~\citep{CSP177}.
 
A crystal graph treats atoms as nodes and neighbours within a cutoff, including periodic images, as edges labelled by distance. Graphs were introduced for crystals by CGCNN~\citep{Xie2018CGCNN} and are the input of universal interatomic potentials such as M3GNet~\citep{CSP109} and CHGNet~\citep{CSP127}. Text and token sequences, such as the CIF text of CrystaLLM~\citep{CSP140} and the SLICES string~\citep{CSP126}, let language models generate crystals.
 
Other representations separate structural description from geometric reconstruction. Contact maps and distance matrices are intermediate targets of AlphaCrystal~\citep{CSP131} and AlphaCrystal-II~\citep{CSP139}, from which structures are reconstructed by optimization (Section~\ref{sec:geometry-search}). Voxel grids and invertible feature vectors were used by early generators such as iMatGen~\citep{CSP088} and FTCP~\citep{CSP111}.
 
These representations differ in which invariances they build in (atom permutation, translation, rotation, and the choice of unit cell), in whether they can be decoded back to a unique crystal, and in how the composition and cell size are fixed. A graph is invariant to atom order and orientation but cannot by itself be decoded into coordinates, which is why graphs serve mainly as inputs to energy and property models. Full-cell and symmetry-reduced descriptions decode directly into structures and are therefore used for generation, with symmetry-reduced descriptions trading generality for fewer variables. Sequence and voxel representations require learned or rule-based decoding, which is where validity and composition errors enter.
 
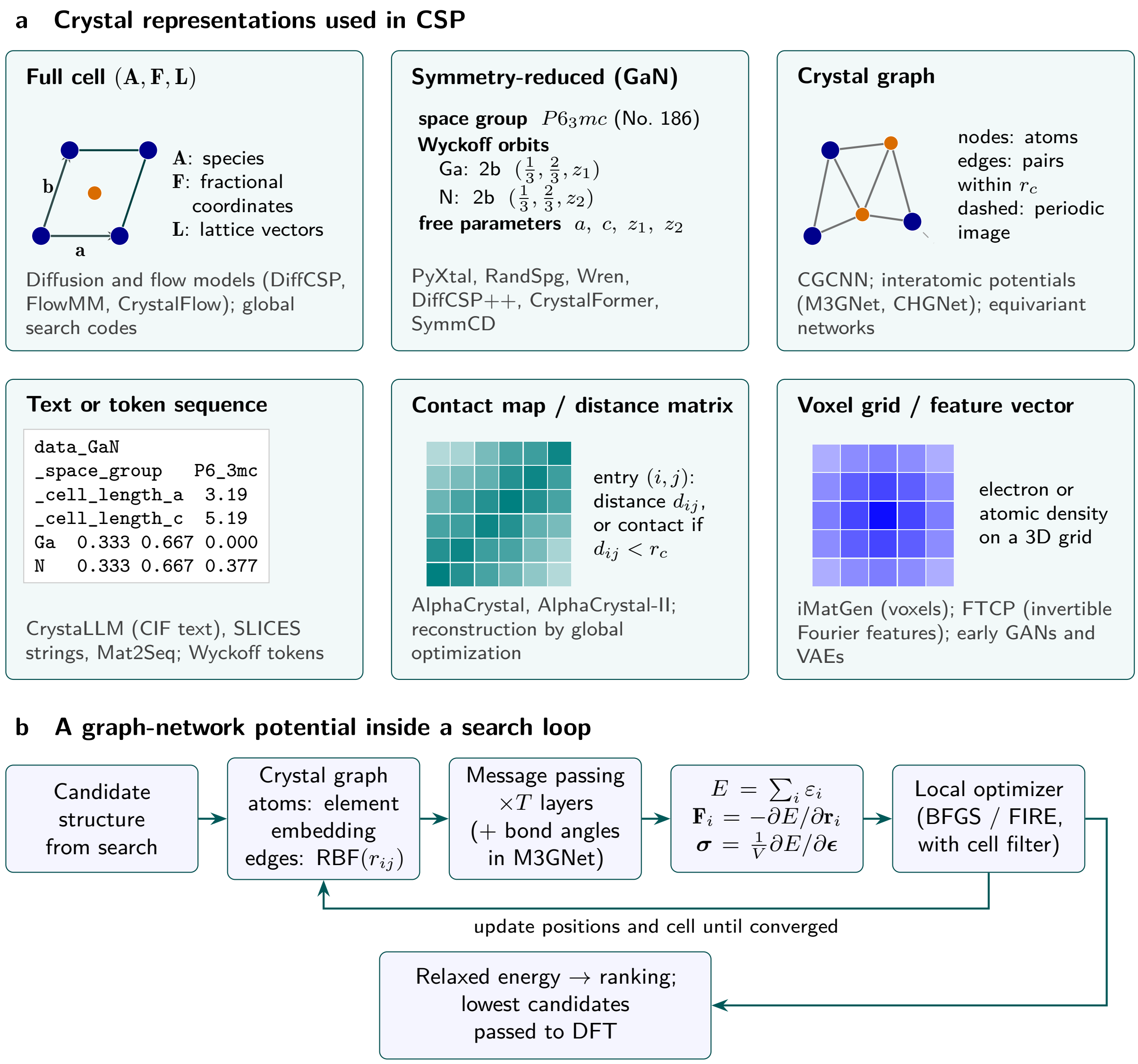
\begin{figure}[!htb]
\centering
\begin{tikzpicture}[font=\sffamily\footnotesize,
panel/.style={draw=teal!60!black,rounded corners=3pt,fill=teal!5,minimum width=50mm,minimum height=42mm},
ptitle/.style={font=\sffamily\footnotesize\bfseries,anchor=north west},
use/.style={font=\sffamily\scriptsize,anchor=south west,text width=46mm,align=left,text=black!75,execute at begin node={\hyphenpenalty=10000}},
lab/.style={font=\sffamily\scriptsize,align=left},
atomA/.style={circle,fill=blue!55!black,draw=white,line width=.3pt,minimum size=2.6mm,inner sep=0},
atomB/.style={circle,fill=orange!85!black,draw=white,line width=.3pt,minimum size=2.0mm,inner sep=0},
stp/.style={draw=teal!60!black,rounded corners=3pt,fill=blue!4,align=center,text width=24.5mm,minimum height=15mm,inner sep=1.2mm},
arr/.style={-Stealth,thick,draw=teal!70!black}]

\node[font=\sffamily\small\bfseries,anchor=west] at (0,48mm) {a\quad Crystal representations used in CSP};

\node[panel,anchor=north west] (p1) at (0,44mm) {};
\node[ptitle] at ($(p1.north west)+(1.5mm,-1mm)$) {Full cell $(\mathbf A,\mathbf F,\mathbf L)$};
\begin{scope}[shift={($(p1.north west)+(5mm,-26mm)$)}]
  \draw[thick,teal!50!black] (0,0)--(11mm,0)--(15mm,12mm)--(4mm,12mm)--cycle;
  \draw[-Stealth,black!70] (0,0)--(11mm,0);
  \draw[-Stealth,black!70] (0,0)--(4mm,12mm);
  \node[lab] at (5.5mm,-2.2mm) {$\mathbf a$};
  \node[lab] at (1mm,7mm) {$\mathbf b$};
  \node[atomA] at (0,0) {}; \node[atomA] at (11mm,0) {}; \node[atomA] at (15mm,12mm) {}; \node[atomA] at (4mm,12mm) {};
  \node[atomB] at (7.5mm,6mm) {};
  \node[lab,anchor=west] at (17mm,6mm) {$\mathbf A$: species\\$\mathbf F$: fractional\\\quad coordinates\\$\mathbf L$: lattice vectors};
\end{scope}
\node[use] at ($(p1.south west)+(1.5mm,1mm)$) {Diffusion and flow models (DiffCSP, FlowMM, CrystalFlow); global search codes};

\node[panel,anchor=north west] (p2) at (54mm,44mm) {};
\node[ptitle] at ($(p2.north west)+(1.5mm,-1mm)$) {Symmetry-reduced (GaN)};
\node[lab,anchor=north west,text width=45mm] at ($(p2.north west)+(2.5mm,-7mm)$) {%
\textbf{space group} \ $P6_3mc$ (No.\ 186)\\[1pt]
\textbf{Wyckoff orbits}\\
\quad Ga: 2b \ $(\tfrac13,\tfrac23,z_1)$\\
\quad N: \ 2b \ $(\tfrac13,\tfrac23,z_2)$\\[1pt]
\textbf{free parameters} \ $a,\ c,\ z_1,\ z_2$};
\node[use] at ($(p2.south west)+(1.5mm,1mm)$) {PyXtal, RandSpg, Wren, DiffCSP++, CrystalFormer, SymmCD};

\node[panel,anchor=north west] (p3) at (108mm,44mm) {};
\node[ptitle] at ($(p3.north west)+(1.5mm,-1mm)$) {Crystal graph};
\begin{scope}[shift={($(p3.north west)+(5mm,-26mm)$)}]
  \coordinate (n1) at (0,0); \coordinate (n2) at (7mm,3mm); \coordinate (n3) at (2.5mm,12mm);
  \coordinate (n4) at (11mm,13mm); \coordinate (n5) at (14mm,2mm);
  \foreach \a/\b in {n1/n2,n1/n3,n2/n3,n2/n4,n3/n4,n2/n5,n4/n5} \draw[black!55,thick] (\a)--(\b);
  \draw[black!40,dashed] (n5)--++(3mm,-3mm);
  \foreach \n in {n1,n3,n5} \node[atomA] at (\n) {};
  \foreach \n in {n2,n4} \node[atomB] at (\n) {};
  \node[lab,anchor=west] at (19mm,7mm) {nodes: atoms\\edges: pairs\\within $r_c$\\dashed: periodic\\image};
\end{scope}
\node[use] at ($(p3.south west)+(1.5mm,1mm)$) {CGCNN; interatomic potentials (M3GNet, CHGNet); equivariant networks};

\node[panel,anchor=north west] (p4) at (0,-2mm) {};
\node[ptitle] at ($(p4.north west)+(1.5mm,-1mm)$) {Text or token sequence};
\node[font=\ttfamily\scriptsize,align=left,anchor=north west,fill=white,draw=black!20,inner sep=1.5mm] at ($(p4.north west)+(2.5mm,-7mm)$) {%
data\_GaN\\
\_space\_group\ \ \ P6\_3mc\\
\_cell\_length\_a\ \ 3.19\\
\_cell\_length\_c\ \ 5.19\\
Ga\ \ 0.333\ 0.667\ 0.000\\
N\ \ \ 0.333\ 0.667\ 0.377};
\node[use] at ($(p4.south west)+(1.5mm,1mm)$) {CrystaLLM (CIF text), SLICES strings, Mat2Seq; Wyckoff tokens};

\node[panel,anchor=north west] (p5) at (54mm,-2mm) {};
\node[ptitle] at ($(p5.north west)+(1.5mm,-1mm)$) {Contact map / distance matrix};
\begin{scope}[shift={($(p5.north west)+(5mm,-29mm)$)}]
  \foreach \i in {0,...,5} \foreach \j in {0,...,5} {
    \pgfmathsetmacro{\v}{int(100-14*abs(\i-\j)-5*mod(\i*\j,3))}
    \fill[teal!\v!white] (\i*3.4mm,\j*3.4mm) rectangle ++(3.2mm,3.2mm);}
  \node[lab,anchor=west] at (22mm,10mm) {entry $(i,j)$:\\distance $d_{ij}$,\\or contact if\\$d_{ij}<r_c$};
\end{scope}
\node[use] at ($(p5.south west)+(1.5mm,1mm)$) {AlphaCrystal, \mbox{AlphaCrystal-II}; reconstruction by global optimization};

\node[panel,anchor=north west] (p6) at (108mm,-2mm) {};
\node[ptitle] at ($(p6.north west)+(1.5mm,-1mm)$) {Voxel grid / feature vector};
\begin{scope}[shift={($(p6.north west)+(5mm,-29mm)$)}]
  \foreach \i in {0,...,4} \foreach \j in {0,...,4} {
    \pgfmathsetmacro{\v}{int(max(5,95-22*sqrt((\i-2)^2+(\j-2)^2)))}
    \fill[blue!\v!white] (\i*4mm,\j*4mm) rectangle ++(3.8mm,3.8mm);}
  \node[lab,anchor=west] at (22mm,10mm) {electron or\\atomic density\\on a 3D grid};
\end{scope}
\node[use] at ($(p6.south west)+(1.5mm,1mm)$) {iMatGen (voxels); FTCP (invertible Fourier features); early GANs and VAEs};

\node[font=\sffamily\small\bfseries,anchor=west] at (0,-51mm) {b\quad A graph-network potential inside a search loop};
\node[stp,anchor=north west] (s1) at (0,-56mm) {Candidate\\structure\\from search};
\node[stp,right=4mm of s1] (s2) {Crystal graph\\atoms: element\\embedding\\edges: RBF$(r_{ij})$};
\node[stp,right=4mm of s2] (s3) {Message passing\\$\times T$ layers\\(+ bond angles\\in M3GNet)};
\node[stp,right=4mm of s3] (s4) {$E=\sum_i\varepsilon_i$\\$\mathbf F_i=-\partial E/\partial\mathbf r_i$\\$\boldsymbol\sigma=\tfrac{1}{V}\partial E/\partial\boldsymbol\epsilon$};
\node[stp,right=4mm of s4] (s5) {Local optimizer\\(BFGS / FIRE,\\with cell filter)};
\foreach \a/\b in {s1/s2,s2/s3,s3/s4,s4/s5} \draw[arr] (\a)--(\b);
\draw[arr] (s5.south) |- ++(0,-5mm) -| node[pos=.25,below,font=\sffamily\scriptsize]{update positions and cell until converged} (s2.south);
\node[stp,text width=44mm,below=10mm of s3] (s6) {Relaxed energy $\rightarrow$ ranking;\\lowest candidates passed to DFT};
\draw[arr] (s5.east) -- ++(3mm,0) |- (s6.east);
\end{tikzpicture}
\caption{Crystal representations and their role in CSP. (a) Six representations used by CSP methods, illustrated for wurtzite GaN, with examples of methods that use each. (b) How a graph-network interatomic potential is used inside a search: each candidate is converted to a crystal graph, message passing produces per-atom energies, and forces and stresses obtained as derivatives of the total energy drive a standard local optimizer; the relaxed energies rank candidates before the lowest are passed to DFT.}\label{fig:representations}
\end{figure}
 
\subsection{Learning the evaluator}
An ML interatomic potential approximates the energy and, where trained, forces and stresses used during structural relaxation. It lowers the cost of evaluating candidates but leaves open how new candidates are selected. The adaptive genetic algorithm \citep{CSP059} iteratively fitted potentials to first-principles data for use in evolutionary exploration. Active-learning potentials (MLIP) coupled to USPEX \citep{CSP085} subsequently made the evaluator update an explicit part of the search loop. MAISE \citep{CSP106} and SPINNER \citep{CSP110} provide further combinations of learned interactions and structural exploration.
 
Most current universal potentials are graph neural networks, and their role in a search follows from how they are built (Figure~\ref{fig:representations}b). Each atom starts from a learned embedding of its element, and each edge carries the interatomic distance expanded in radial basis functions; M3GNet~\citep{CSP109} adds three-body terms through bond angles, and CHGNet~\citep{CSP127} adds magnetic moments as a proxy for charge state. Several rounds of message passing update each atom's features from its neighbours, and the total energy is a sum of per-atom contributions. Forces and stresses are obtained as derivatives of this energy with respect to atomic positions and cell strain, so they are consistent with the energy by construction. The potential can therefore drive standard local optimizers, such as BFGS or FIRE with a filter for cell degrees of freedom, in place of DFT. ParetoCSP~\citep{CSP141}, for example, relaxes every candidate of its genetic search with M3GNet, so that thousands of candidates can be relaxed and ranked and only the lowest-energy structures are passed to DFT.
 
Training-set coverage is especially consequential in global search because trial structures need not resemble equilibrium crystals. Disordered-structure training \citep{CSP099} addresses this coverage problem from a different direction. General pretrained models such as M3GNet \citep{CSP109} and CHGNet \citep{CSP127} can provide reusable evaluators, but their energies can be unreliable for unusual chemistries, high pressures, or strongly distorted geometries, and a search may actively drift toward regions where the model is wrong. Low predicted energies therefore need independent checking.
 
Active learning places this issue inside a feedback loop: structures judged informative or insufficiently supported receive more accurate calculations, and the model is updated. Related approaches occur in both inorganic searches and molecular crystal-energy landscapes \citep{CSP085,CSP150,CSP162,CSP190}. The relevant comparison includes the cost of acquiring labels and training the potential, not only the speed of its final inference. Uncertainty indicators help select calculations, but their calibration has to be checked.
 
\subsection{Learning structural descriptors}
Learning can also supply intermediate crystallographic variables for candidate construction. MLatticeABC \citep{MLatticeABC2021} uses random-forest regression with composition descriptors and additional features, including atom-count information, to predict lattice parameters. The composition-based symmetry predictor of Li and co-workers \citep{CompositionSymmetry2021} uses enhanced descriptors and random forests to predict crystal systems and space groups from formulas. These predictors supply candidate cell or symmetry information and leave atomic positions and site assignments to a later search. They can narrow that search, although a composition with several polymorphs needs more than one cell or symmetry hypothesis, and descriptor accuracy says little about complete-structure recovery.
 
\subsection{Learning which calculation to perform}
Bayesian optimization and surrogate-guided search allocate a limited calculation budget by combining predicted quality with information about uncertainty or unexplored regions. Bayesian CSP \citep{CSP075}, graph-network optimization \citep{CSP100}, and symmetry-relaxation approaches \citep{CSP114} illustrate distinct uses of this principle. The learned object may be a global energy model, a ranking function over a candidate library, or an acquisition score. Their inputs and optimization domains differ accordingly.
 
LAQA \citep{CSP077,CSP118} changes resource allocation at a finer level by using partially completed local optimizations to decide which candidates warrant further relaxation. CrySPY \citep{CSP102} exposes several search strategies within a modular framework, including random, evolutionary, Bayesian, and look-ahead approaches. Hybrid Bayesian/evolutionary search \citep{CSP117} combines proposal mechanisms directly. GOFEE \citep{CSP095} and AGOX \citep{CSP112} provide adjacent atomistic optimization frameworks useful for understanding surrogate-driven exploration, although their original demonstrations focused largely on clusters and surfaces.
 
Coarse structural representations offer another means of reducing cost. Wren \citep{CSP120} predicts energies from coordinate-free crystallographic descriptions, allowing a search over symmetry-based candidates before detailed coordinates are evaluated. ShotgunCSP \citep{CSP153} instead emphasizes rapid screening of generated libraries using learned formation energies. Both still need a realizable geometry and a final relaxation.
 
\subsection{Learning search policies and composing methods}
Reinforcement learning can change the distribution of moves used during a CSP run, as in work on basin-hopping search policies \citep{CSP137}. Tree-search approaches such as CASTING \citep{CSP124} explore a different policy and planning formulation. These methods learn how to explore, not a distribution over complete crystals, so their results depend heavily on the reward definition, exploration cost, and evaluator. Reinforcement learning can also act on a generator rather than a search policy. CrystalGRPO \citep{CSP182} post-trains flow-based CSP models with a reward that combines a MACE-predicted energy with a StructureMatcher recovery score, since predicted energy alone does not identify the reference polymorph. It offers one mode that favours single-draw (top-1) recovery and another that regularizes towards the pretrained trajectories and uses a coverage-aware advantage to keep the candidate diversity needed for top-20 recovery. Its reward still relies on the reference structure and a learned potential, so the same cautions about reward design and evaluator dependence apply (Section~\ref{sec:generative-models}).
 
Hybrid methods combine these ingredients in different ways. FUSE2 \citep{CSP148} uses graph-network energy-surrogate and Bayesian-optimization seeding within a flexible-unit search. MAGUS \citep{CSP170} learns useful symmetry and structural fragments during evolutionary exploration. PhaseBO \citep{CSP147} places composition selection outside the structure/energy calculation. Their gains come from different components, which a generic label such as ``AI-assisted CSP'' does not identify.
 
\section{Deep generative models for crystal structure prediction}\label{sec:generative-models}
The preceding methods use learning to evaluate candidates, supply intermediate descriptors, or guide exploration. Deep generative models also learn to propose structural configurations from a training corpus, shifting attention to the distribution of candidates itself. Figure~\ref{fig:generative-mechanisms} summarizes the five generative mechanisms that current CSP models build on.
 
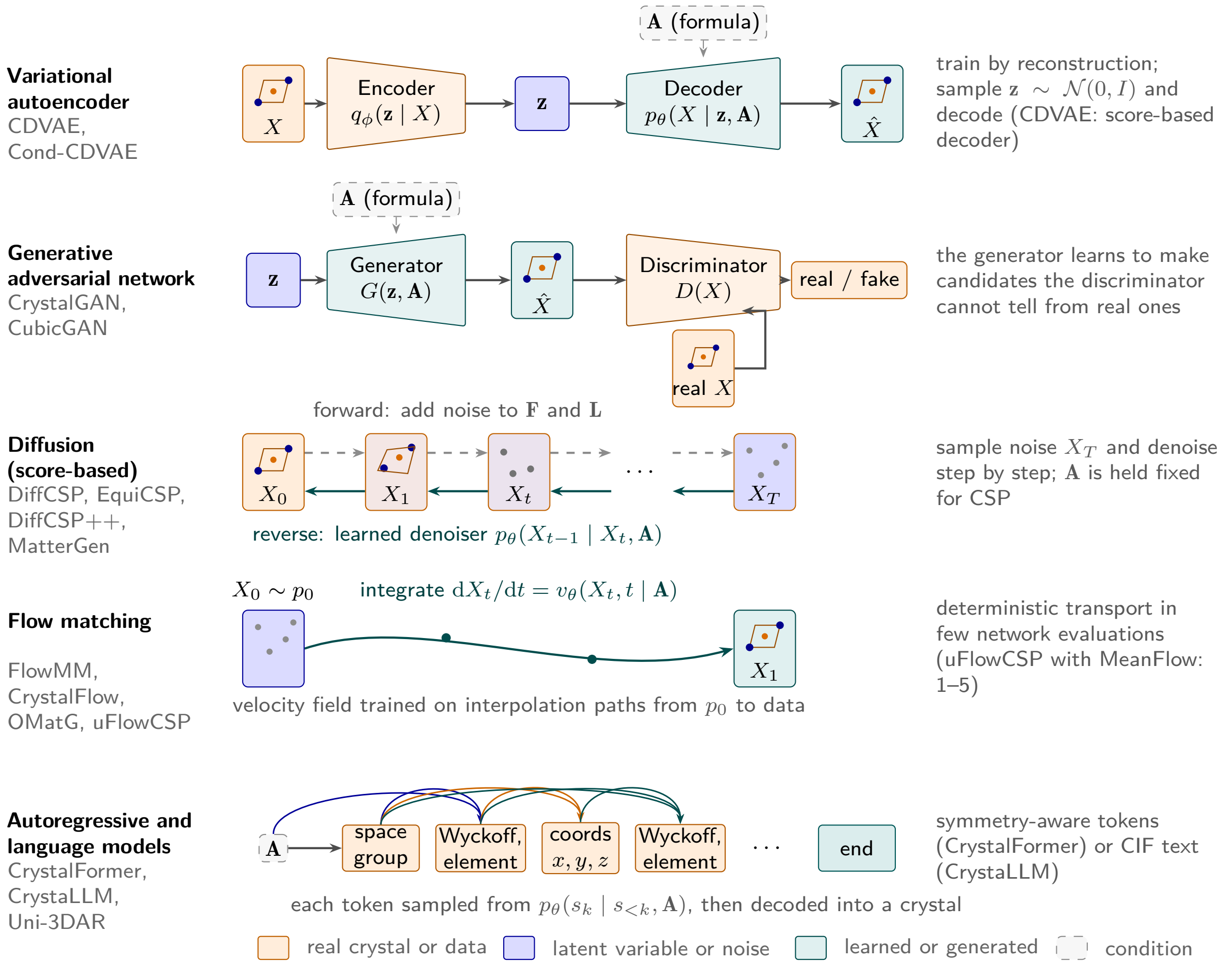
\begin{figure}[!htb]
\centering
\begin{tikzpicture}[x=1mm,y=1mm,font=\sffamily\footnotesize,
data/.style={draw=orange!75!black,fill=orange!14,rounded corners=2pt,minimum width=8mm,minimum height=10mm,inner sep=0.5mm,align=center},
gen/.style={draw=teal!60!black,fill=teal!12,rounded corners=2pt,minimum width=8mm,minimum height=10mm,inner sep=0.5mm,align=center},
lat/.style={draw=blue!60!black,fill=blue!14,rounded corners=2pt,minimum width=7mm,minimum height=7mm,inner sep=0.5mm},
noise/.style={draw=blue!60!black,fill=blue!14,rounded corners=2pt,minimum width=8mm,minimum height=10mm,inner sep=0.5mm,align=center},
cond/.style={draw=black!45,dashed,fill=black!3,rounded corners=2pt,inner sep=0.8mm,font=\sffamily\scriptsize},
tok/.style={rounded corners=1.5pt,minimum width=10mm,minimum height=6mm,inner sep=0.4mm,font=\sffamily\scriptsize,align=center},
arr/.style={-{Stealth[length=1.6mm]},thick,black!70},
carr/.style={-{Stealth[length=1.4mm]},black!45,dashed},
rowlab/.style={anchor=north west,align=left,text width=28mm,font=\sffamily\scriptsize\bfseries,execute at begin node={\hyphenpenalty=10000\exhyphenpenalty=10000}},
ex/.style={font=\sffamily\scriptsize,text=black!60,align=left,execute at begin node={\hyphenpenalty=10000\exhyphenpenalty=10000}},
note/.style={font=\sffamily\scriptsize,text=black!65,align=center,execute at begin node={\hyphenpenalty=10000\exhyphenpenalty=10000}}]

\newcommand{\xtal}[2]{%
 \begin{scope}[shift={(#1)},scale=#2]
  \draw[orange!60!black,line width=.4pt] (-2.2,-1.6)--(1.4,-1.6)--(2.2,1.6)--(-1.4,1.6)--cycle;
  \fill[blue!55!black] (-2.2,-1.6) circle (0.55); \fill[blue!55!black] (2.2,1.6) circle (0.55);
  \fill[orange!85!black] (0,0) circle (0.45);
 \end{scope}}
\newcommand{\net}[5]{%
 \draw[draw=#4!60!black,fill=#4!10,rounded corners=1pt] ($(#1)+(-#2/2,#3/2)$)--($(#1)+(#2/2,#3/2-2.2)$)--($(#1)+(#2/2,-#3/2+2.2)$)--($(#1)+(-#2/2,-#3/2)$)--cycle;
 \node[align=center,font=\sffamily\scriptsize] at (#1) {#5};}
\newcommand{\netr}[5]{%
 \draw[draw=#4!60!black,fill=#4!10,rounded corners=1pt] ($(#1)+(-#2/2,#3/2-2.2)$)--($(#1)+(#2/2,#3/2)$)--($(#1)+(#2/2,-#3/2)$)--($(#1)+(-#2/2,-#3/2+2.2)$)--cycle;
 \node[align=center,font=\sffamily\scriptsize] at (#1) {#5};}

\def\y{0}
\node[rowlab] at (0,\y+6) {Variational\\autoencoder};
\node[ex,anchor=north west,text width=28mm] at (0,\y-0.5) {CDVAE, Cond-CDVAE};
\node[data] (x1) at (36,\y) {}; \xtal{36,\y+1.6}{0.9}; \node[font=\sffamily\scriptsize] at (36,\y-3) {$X$};
\net{52,\y}{18}{12}{orange}{Encoder\\$q_\phi(\mathbf z\mid X)$}
\node[lat] (z1) at (71,\y) {$\mathbf z$};
\netr{92,\y}{20}{12}{teal}{Decoder\\$p_\theta(X\mid\mathbf z,\mathbf A)$}
\node[gen] (o1) at (114,\y) {}; \xtal{114,\y+1.6}{0.9}; \node[font=\sffamily\scriptsize] at (114,\y-3) {$\hat X$};
\draw[arr] (x1.east)--(43,\y); \draw[arr] (61,\y)--(z1.west); \draw[arr] (z1.east)--(82,\y); \draw[arr] (102,\y)--(o1.west);
\node[cond] (c1) at (92,\y+10.5) {$\mathbf A$ (formula)}; \draw[carr] (c1.south)--(92,\y+6);
\node[note,anchor=west,text width=38mm,align=left] at (121,\y) {train by reconstruction; sample $\mathbf z\sim\mathcal N(0,I)$ and decode (CDVAE: score-based decoder)};

\def\y{-23}
\node[rowlab] at (0,\y+6) {Generative\\adversarial network};
\node[ex,anchor=north west,text width=28mm] at (0,\y-0.5) {CrystalGAN, CubicGAN};
\node[lat] (z2) at (36,\y) {$\mathbf z$};
\netr{52,\y}{18}{12}{teal}{Generator\\$G(\mathbf z,\mathbf A)$}
\node[gen] (o2) at (71,\y) {}; \xtal{71,\y+1.6}{0.9}; \node[font=\sffamily\scriptsize] at (71,\y-3) {$\hat X$};
\net{92,\y}{20}{12}{orange}{Discriminator\\$D(X)$}
\node[data] (r2) at (92,\y-11.5) {}; \xtal{92,\y-10}{0.75}; \node[font=\sffamily\scriptsize] at (92,\y-14) {real $X$};
\node[draw=orange!75!black,fill=orange!14,rounded corners=2pt,inner sep=1mm,font=\sffamily\scriptsize] (d2) at (111,\y) {real / fake};
\draw[arr] (z2.east)--(43,\y); \draw[arr] (61,\y)--(o2.west); \draw[arr] (o2.east)--(82,\y); \draw[arr] (r2.east)--++(4,0)|-(97,\y-4); \draw[arr] (102,\y)--(d2.west);
\node[cond] (c2) at (52,\y+10.5) {$\mathbf A$ (formula)}; \draw[carr] (c2.south)--(52,\y+6);
\node[note,anchor=west,text width=38mm,align=left] at (121,\y) {the generator learns to make candidates the discriminator cannot tell from real ones};

\def\y{-48}
\node[rowlab] at (0,\y+6) {Diffusion\\(score-based)};
\node[ex,anchor=north west,text width=28mm] at (0,\y-0.5) {DiffCSP, EquiCSP,\\DiffCSP++, MatterGen};
\foreach \i/\xx/\op/\lab in {0/36/0/{$X_0$},1/52/30/{$X_1$},2/68/55/{$X_t$},3/100/100/{$X_T$}}{
  \pgfmathsetmacro\o{\op}
  \node[rounded corners=2pt,minimum width=8mm,minimum height=10mm,draw=orange!75!black,fill=orange!14] (d\i) at (\xx,\y) {};
  \fill[blue!14,rounded corners=2pt,opacity=\o/100] ($(d\i.south west)+(.15,.15)$) rectangle ($(d\i.north east)-(.15,.15)$);
  \node[font=\sffamily\scriptsize] at (\xx,\y-3) {\lab};}
\xtal{36,\y+1.6}{0.9}
\begin{scope}[shift={(52,\y+1.6)}] \fill[blue!55!black] (-1.6,-1.1) circle (0.5); \fill[blue!55!black] (2.0,1.6) circle (0.5); \fill[orange!85!black] (0.4,0.3) circle (0.42); \draw[orange!60!black,line width=.4pt] (-2.4,-1.4)--(1.5,-1.8)--(2.4,1.7)--(-1.2,1.3)--cycle; \end{scope}
\foreach \p in {(66,\y+2.6),(69.5,\y+0.4),(67.2,\y-0.2)} \fill[black!55] \p circle (0.45);
\foreach \p in {(98,\y+2.8),(101.5,\y+1.2),(99.5,\y-0.4),(102.6,\y+3.4)} \fill[black!45] \p circle (0.4);
\node at (84,\y) {\textbf{$\cdots$}};
\foreach \a/\b in {d0/d1,d1/d2} {\draw[arr,dashed,black!45] ($(\a.east)+(0,2.5)$)--($(\b.west)+(0,2.5)$); \draw[arr,teal!60!black] ($(\b.west)+(0,-2.5)$)--($(\a.east)+(0,-2.5)$);}
\draw[arr,dashed,black!45] ($(d2.east)+(0,2.5)$)--(80,\y+2.5); \draw[arr,dashed,black!45] (88,\y+2.5)--($(d3.west)+(0,2.5)$);
\draw[arr,teal!60!black] ($(d3.west)+(0,-2.5)$)--(88,\y-2.5); \draw[arr,teal!60!black] (80,\y-2.5)--($(d2.east)+(0,-2.5)$);
\node[note] at (60,\y+8.2) {forward: add noise to $\mathbf F$ and $\mathbf L$};
\node[note,text=teal!50!black] at (60,\y-8.2) {reverse: learned denoiser $p_\theta(X_{t-1}\mid X_t,\mathbf A)$};
\node[note,anchor=west,text width=38mm,align=left] at (121,\y) {sample noise $X_T$ and denoise step by step; $\mathbf A$ is held fixed for CSP};

\def\y{-71}
\node[rowlab] at (0,\y+6) {Flow matching};
\node[ex,anchor=north west,text width=28mm] at (0,\y-0.5) {FlowMM, CrystalFlow,\\OMatG, uFlowCSP};
\node[noise] (f0) at (36,\y) {}; \foreach \p in {(34,\y+2.8),(37.6,\y+1.2),(35.5,\y-0.4),(38.6,\y+3.4)} \fill[black!45] \p circle (0.4);
\node[font=\sffamily\scriptsize] at (36,\y+7.5) {$X_0\sim p_0$};
\node[gen] (f1) at (100,\y) {}; \xtal{100,\y+1.6}{0.9}; \node[font=\sffamily\scriptsize] at (100,\y-3) {$X_1$};
\draw[thick,teal!60!black,-{Stealth[length=1.8mm]}] (f0.east) .. controls (55,\y+5) and (78,\y-5) .. (f1.west);
\foreach \t/\xx/\yy in {0.33/58.5/1.4,0.66/77.5/-1.4} \fill[teal!60!black] (\xx,\y+\yy) circle (0.6);
\node[note,text=teal!50!black] at (68,\y+7.5) {integrate $\mathrm{d}X_t/\mathrm{d}t=v_\theta(X_t,t\mid\mathbf A)$};
\node[note] at (68,\y-7.5) {velocity field trained on interpolation paths from $p_0$ to data};
\node[note,anchor=west,text width=38mm,align=left] at (121,\y) {deterministic transport in few network evaluations (uFlowCSP with MeanFlow: 1--5)};

\def\y{-97}
\node[rowlab] at (0,\y+6) {Autoregressive and\\language models};
\node[ex,anchor=north west,text width=28mm] at (0,\y-0.5) {CrystalFormer, CrystaLLM,\\Uni-3DAR};
\node[cond] (c5) at (36,\y) {$\mathbf A$};
\foreach \i/\xx/\t/\c in {1/50/{space\\group}/orange,2/63/{Wyckoff,\\element}/orange,3/76/{coords\\$x,y,z$}/orange,4/89/{Wyckoff,\\element}/orange,5/112/{end}/teal}{
  \node[tok,draw=\c!70!black,fill=\c!14] (t\i) at (\xx,\y) {\t};}
\node at (100.5,\y) {\textbf{$\cdots$}};
\foreach \a/\b/\col in {c5/t2/blue!60!black,t1/t2/blue!60!black,t1/t3/orange!80!black,t2/t3/orange!80!black,t1/t4/teal!60!black,t2/t4/teal!60!black,t3/t4/teal!60!black}
  \draw[-{Stealth[length=1.3mm]},\col,line width=.55pt] (\a.north) .. controls +(0,6) and +(0,6) .. (\b.north);
\draw[arr] (c5.east)--(t1.west);
\node[note] at (82,\y-7.5) {each token sampled from $p_\theta(s_k\mid s_{<k},\mathbf A)$, then decoded into a crystal};
\node[note,anchor=west,text width=38mm,align=left] at (121,\y) {symmetry-aware tokens (CrystalFormer) or CIF text (CrystaLLM)};

\def\y{-110}
\node[draw=orange!75!black,fill=orange!14,rounded corners=1pt,minimum width=4mm,minimum height=3mm] at (36,\y) {}; \node[ex,anchor=west] at (39,\y) {real crystal or data};
\node[draw=blue!60!black,fill=blue!14,rounded corners=1pt,minimum width=4mm,minimum height=3mm] at (68,\y) {}; \node[ex,anchor=west] at (71,\y) {latent variable or noise};
\node[draw=teal!60!black,fill=teal!12,rounded corners=1pt,minimum width=4mm,minimum height=3mm] at (106,\y) {}; \node[ex,anchor=west] at (109,\y) {learned or generated};
\node[cond,minimum width=4mm,minimum height=3mm] at (140,\y) {}; \node[ex,anchor=west] at (143,\y) {condition};
\end{tikzpicture}
\caption{Generative mechanisms used for crystal structure prediction. Each row shows how a model family turns noise or a latent variable into a crystal $X=(\mathbf A,\mathbf F,\mathbf L)$, with representative CSP models on the left. In CSP the atom types $\mathbf A$ (the formula) are supplied as a condition (dashed boxes) and only the fractional coordinates $\mathbf F$ and lattice $\mathbf L$ are generated. Variational autoencoders decode a sampled latent vector; generative adversarial networks train a generator against a discriminator; diffusion models reverse a gradual noising of $\mathbf F$ and $\mathbf L$; flow matching integrates a learned velocity field from a base distribution to the data; autoregressive and language models generate a crystal token by token, each token conditioned on the previous ones. Orange: real crystals; blue: latent variables or noise; teal: learned networks and generated crystals.}
\label{fig:generative-mechanisms}
\end{figure}
 
\subsection{From reconstruction to a conditional proposal distribution}
A trained generative CSP model approximates a distribution over structures compatible with specified inputs. For a fixed atom multiset, a useful schematic expression is
\begin{equation}\label{eq:conditional}
 (\mathbf F,\mathbf L)\sim p_\theta(\mathbf F,\mathbf L\mid\mathbf A,\mathbf c),
\end{equation}
where $\mathbf c$ denotes additional conditions such as pressure or symmetry. In de novo generation, the model also samples $\mathbf A$ or an equivalent composition representation. In reconstruction, the conditioning information includes the original structure or an encoding of it. The three tasks can share an architecture while answering different questions.
 
Deep generative models of crystals predate diffusion-based generators. CrystalGAN \citep{CSP086} (2018 preprint; 2019 proceedings) used adversarial domain translation to propose ternary hydrides from binary ones. iMatGen \citep{CSP088} (2019) sampled vanadium oxides from the latent space of a variational autoencoder built on an invertible image representation of the crystal. Kim et al. \citep{CSP097} (2020) conditioned a point-cloud GAN on composition and generated Mg--Mn--O structures for specified compositions, and CubicGAN \citep{CSP105} (2021) extended adversarial generation to large-scale sampling of cubic crystals conditioned on elements and space group. Related adversarial, density-grid, and latent-variable formulations appeared over the same period \citep{CSP087,CSP092,CSP101}. These representations made it possible to learn structural regularities, but a valid decoded crystal, a faithful reconstruction, and a prescribed-formula prediction are separate achievements. FTCP \citep{CSP111} and other representation-based inverse-design approaches further connected latent variables to properties. Whether such a model performs CSP depends on what is supplied and held fixed at inference.
 
CDVAE \citep{CSP113} (2021 preprint; ICLR 2022) followed this line by combining a variational latent representation with a periodic, equivariant score-based crystal decoder. CDVAE is one of the earliest successful deep generative models for crystal generation and prediction, laying the foundation for many follow-up improvements \cite{DiffCSPData,CSP144}. While its released code includes a low-level fixed-species option for CSP~\citep{CDVAECode}, the published model was not evaluated as a formula-only CSP workflow. SyMat \citep{CSP138} likewise separates some composition/lattice information from coordinate generation. A coordinate model that receives a supplied lattice solves a narrower task than one that predicts both the lattice and the coordinates.
 
\subsection{Diffusion models}
Diffusion-based CSP learns to reverse a noise process on structural variables, adapting denoising diffusion and score-based generative models \citep{SohlDickstein2015Diffusion,Ho2020DDPM,Song2021ScoreSDE} to periodic crystals, as first done in CDVAE \citep{CSP113}. In its forward corruption process, a diffusion model adds known noise to a clean sample to create a noised sample and clean sample pair to train a noise predictor, which is then used in the reverse/inference model to denoise a noised sample into a denoised sample iteratively. These two processes
must respect the geometry of crystal periodic coordinates and the chosen lattice representation. Fractional coordinates avoid treating periodic images as unrelated Cartesian positions, while equivariant networks seek consistent behavior under changes of orientation and atom ordering. DiffCSP \citep{CSP128} formulates joint generation of lattice and atomic coordinates conditioned on the supplied species. EquiCSP \citep{CSP144} develops related symmetry and periodicity considerations.
 
There are several distinct ways to add crystallographic knowledge. DiffCSP++ \citep{CSP154} constrains generation using space-group and site information, while SymmCD \citep{CSP178} uses asymmetric-unit and site-symmetry structure in its generative representation. The benefits of a reduced parameter space must be balanced against the source of the symmetry information. A correct space group supplied from the target is privileged information; one predicted from composition can be wrong; one sampled independently consumes part of the candidate budget. However, such space-group-conditioned CSP has a unique advantage to ensure coverage of structural diversity, which may benefit structure search in polymorph structure prediction and on-the-hull structure search given a chemical system. This important genre of CSP problems are recently investigated by Shibayama et al.\ \citep{Shibayama2026DiversityGA}, in which they combine the universal neural network potential PFP with a genetic algorithm that uses hull-informed filtering, an aging mechanism, and niching across stoichiometries, and report larger convex hulls with fewer trials than symmetry-aware random structure generation.
 
Diffusion can also model composition itself, which changes the task: MatterGen \citep{CSP159} jointly models composition and geometry for inorganic materials generation. Its release also exposes a distinct CSP configuration that fixes supplied atom types \citep{MatterGenCSPCode}. The paper's chemistry and property adapters do not by themselves preserve an exact formula: a base model trained to change species becomes a CSP model only when its inference configuration, training objective, checkpoint, and atom-count handling all fix the composition.
 
Pressure-conditioned Cond-CDVAE \citep{CSP142} extends the conditioning task beyond composition. Other diffusion-related approaches alter the decoder, vector field, latent representation, or dynamical formulation, including DP-CDVAE \citep{CSP143}, GemsDiff \citep{CSP157}, kinetic Langevin diffusion \citep{CSP172}, and Crystalite \citep{CSP184}. Diffusion is thus a family of formulations, and a shorter sampling trajectory, a more expressive backbone, and stronger conditioning are different contributions that call for different comparisons.
 
\subsection{Flow matching and related continuous models}\label{sec:flow-models}
Flow models learn a transformation from a base distribution to a distribution of crystal structures. In a schematic continuous formulation,
\begin{equation}\label{eq:flow}
 \frac{dX_t}{dt}=v_\theta(X_t,t;\mathbf c),
\end{equation}
where $X_t$ denotes the structural variables being evolved and the vector field is defined consistently with their geometry. Equation~\eqref{eq:flow} is a conceptual description; discrete species need their own treatment or must be held fixed.
 
FlowMM \citep{CSP146} develops Riemannian flow matching for both known-composition CSP and de novo generation. CrystalFlow \citep{CSP169} provides another flow-based approach with fixed-composition and pressure-conditioned settings. Open Materials Generation \citep{CSP173} uses stochastic interpolants. These approaches differ in their interpolation paths, structural parameterizations, and training objectives.
 
uFlowCSP \citep{CSP195} combines MeanFlow average-velocity generation \citep{Geng2025MeanFlow} with explicit chemical and crystallographic domain knowledge. Its Transformer receives four elemental descriptors---Pauling electronegativity, atomic radius, periodic-table period, and group---alongside learned element embeddings. A global embedding of element fractions encodes stoichiometry. Canonical atom ordering uses electronegativity; training preprocessing additionally groups symmetry-related atoms and orders them by Wyckoff letter. A coarse seven-class crystal-system token supplies auxiliary symmetry supervision during training and is replaced by a null token at inference. At inference, the model requires the supplied full-cell elemental inventory but no target space group. The preprint reports generation in one to five network evaluations.
 
In uFlowCSP, domain knowledge enters through token construction, global conditioning, and auxiliary training, providing a softer use of structural priors than the explicit symmetry constraints discussed in Section~\ref{sec:structural-priors}. Since the cost of a complete workflow also depends on the number of candidates, energy ranking, and relaxation, comparisons with other methods need more than the number of network evaluations.
 
Bayesian flow networks and manifold random walks provide additional generative formulations. CrysBFN \citep{CSP160} treats periodic structural variables within a Bayesian-flow approach, and CrystalGRW \citep{CSP183} uses geodesic random walks. Their ``flows'' and random walks describe sampling from a learned data distribution, not physical dynamics or a Boltzmann distribution.
 
\subsection{Autoregressive, language, and hierarchical models}
Sequence-based approaches turn crystal generation into structured prediction over tokens. SLICES \citep{CSP126} encodes a crystal through a graph-derived string, while CrystaLLM \citep{CSP140} models crystallographic text and Mat2Seq \citep{CSP149} uses an invariant tokenization. Their attractions include the reuse of sequence models and the possibility of expressing conditions as prefixes or tokens. Valid text, however, does not guarantee a valid, correctly conditioned, low-energy geometry, so decoding and geometric refinement remain part of the algorithm.
 
CrystalFormer \citep{CSP177} uses space-group and Wyckoff information in an autoregressive representation. A separate CrystalFormer-CSP preprint \citep{CrystalFormerCSP2025} addresses composition-given prediction. Formula-conditioned interfaces must also be interpreted at the level of the particular version and output filter \citep{CrystalFormerCode}: matching a reduced stoichiometric ratio is different from preserving a prescribed number of each atom in the cell. FlowLLM \citep{CSP145} combines a language-model base distribution with flow refinement. A fixed-species refinement stage can preserve the composition it receives without ensuring that the preceding language model generated the requested inventory.
 
Hierarchical prediction can also proceed through structural descriptors rather than complete sequences, connecting learned constraints to the reconstruction algorithms in Section~\ref{sec:geometry-search}. CRYSPNet \citep{CSP094} predicts crystallographic descriptors, and AlphaCrystal \citep{CSP131} and AlphaCrystal-II \citep{CSP139} use contact or distance information followed by structure reconstruction. NextCrystal \citep{CSP189} combines symmetry proposals, constrained assignments, and diffusion in a recent preprint. Splitting CSP into subproblems in this way means that errors propagate between stages, and evaluations should measure the full chain.
 
Other recent CSP approaches include energy-based generation, reward-proportional sequential construction, and unified spatial-token models \citep{CSP130,CSP163,CSP166,CSP179,CSP187}. They broaden the available proposal mechanisms, but their task scope varies: some produce only coarse crystallographic specifications, some jointly generate chemistry and geometry, and some expose separate CSP modes.
 
\subsection{What exact formula conditioning means}
For a requested unit-cell inventory $\mathbf n$, exact count preservation requires
\begin{equation}\label{eq:counts}
 \sum_{i=1}^{N}\mathbb{1}[a_i=s_k]=n_k\qquad\text{for every element }s_k.
\end{equation}
A formula embedding may encourage this relation without enforcing it. A sampler that fixes the supplied species tensor enforces counts by construction, while a post-filter only rejects mismatches after sampling. A reduced-formula filter enforces ratios but may admit several cell multiplicities.
 
We examined the papers and released code of 42 atomic generative models and hybrid workflows for this property. Twenty-one inspected modes preserve exact cell counts, including the hybrid FUSE2 workflow and some low-level code options. For nineteen, exact count preservation is not established and adaptation may be required; two document only unconditional sampling. These classifications are based on documentation and code inspection, not on running the models; Table~\ref{tab:formula} lists them model by model.
 
\begin{table}[tb]
\centering\small
\caption{How representative generative models handle a supplied formula (papers and code as of 16 October 2026).
}\label{tab:formula}
\begin{tabularx}{\linewidth}{@{}P{.20\linewidth}P{.24\linewidth}Y@{}}
\toprule Model / workflow & Formula-count handling & Interpretation\\
\midrule
DiffCSP; EquiCSP & Fixed supplied species & Formula-given lattice and coordinate generation \citep{CSP128,CSP144} \\
FlowMM; CrystalFlow & Separate CSP mode & Distinguish fixed-composition configuration from de novo generation \citep{CSP146,CSP169} \\
uFlowCSP \citep{CSP195} & Fixed supplied species and atom counts & Formula-only inference; crystal-system supervision is auxiliary during training, with cell multiplicity supplied by the query \\
MatterGen & Separate CSP configuration & Chemistry/property adapters alone do not establish exact counts \citep{CSP159,MatterGenCSPCode} \\
CDVAE; CrystalGRW & Released low-level option & Interface support is narrower evidence than a dedicated CSP evaluation \citep{CSP113,CSP183,CDVAECode,CrystalGRWCode} \\
CrystaLLM; Mat2Seq & Prompt / condition & Formula-related input exists; strict preservation is not established \citep{CSP140,CSP149} \\
CrystalFormer & Version-dependent CSP extension & Reduced-ratio filtering differs from prescribed cell counts \citep{CSP177,CrystalFormerCSP2025,CrystalFormerCode} \\
FlowLLM & Fixed-species refinement & End-to-end counts also depend on the preceding sequence stage \citep{CSP145} \\
iMatGen; DP-CDVAE & Unconditional sampling documented & Reconstruction or de novo generation does not demonstrate formula-only prediction \citep{CSP088,CSP143} \\
\bottomrule
\end{tabularx}
\end{table}
 
\subsection{Guidance, post-training, and learned ranking}
A generative proposal can be followed by energy ranking, relaxation, resampling, or additional search. DAO \citep{CSP193} pairs a composition-conditioned generator with an energy-related model. CrystalGRPO \citep{CSP182} studies reinforcement post-training for flow-based CSP in a preprint. Such steps can improve recovery or ranking, but they add evaluator calls and training exposure that comparisons should count, and a reward validated with the same surrogate that defined it is not an independent test.
 
Complete workflows and their components therefore both need to be compared. A generator evaluated before relaxation tests its proposal quality; the same generator after common relaxation tests the quality of the basins it reaches; additional learned reranking tests candidate selection. Reporting only the best of these stages conceals where an advantage originates. The same holds for classical initialization followed by ML relaxation and for learned generation followed by DFT.
\section{Molecular-crystal structure prediction}
\subsection{Packing and conformational search}
The learned proposal mechanisms discussed above also appear in molecular CSP, where molecular identity, packing, and conformation define a distinct set of structural variables. The molecular graph constrains intramolecular bonding, while packing introduces intermolecular translations, orientations, and cell degrees of freedom. Flexibility adds conformational variables that can couple strongly to packing. UPACK \citep{CSP008}, conformation-family Monte Carlo \citep{CSP010}, molecular genetic algorithms \citep{CSP015,CSP016}, and the CrystalPredictor/CrystalOptimizer lineage \citep{CSP023,CSP031,CSP049} illustrate the progression from relatively rigid building blocks to coupled conformational and packing exploration. Figure~\ref{fig:molecular-csp} summarizes the stages of molecular CSP discussed in this section.
 
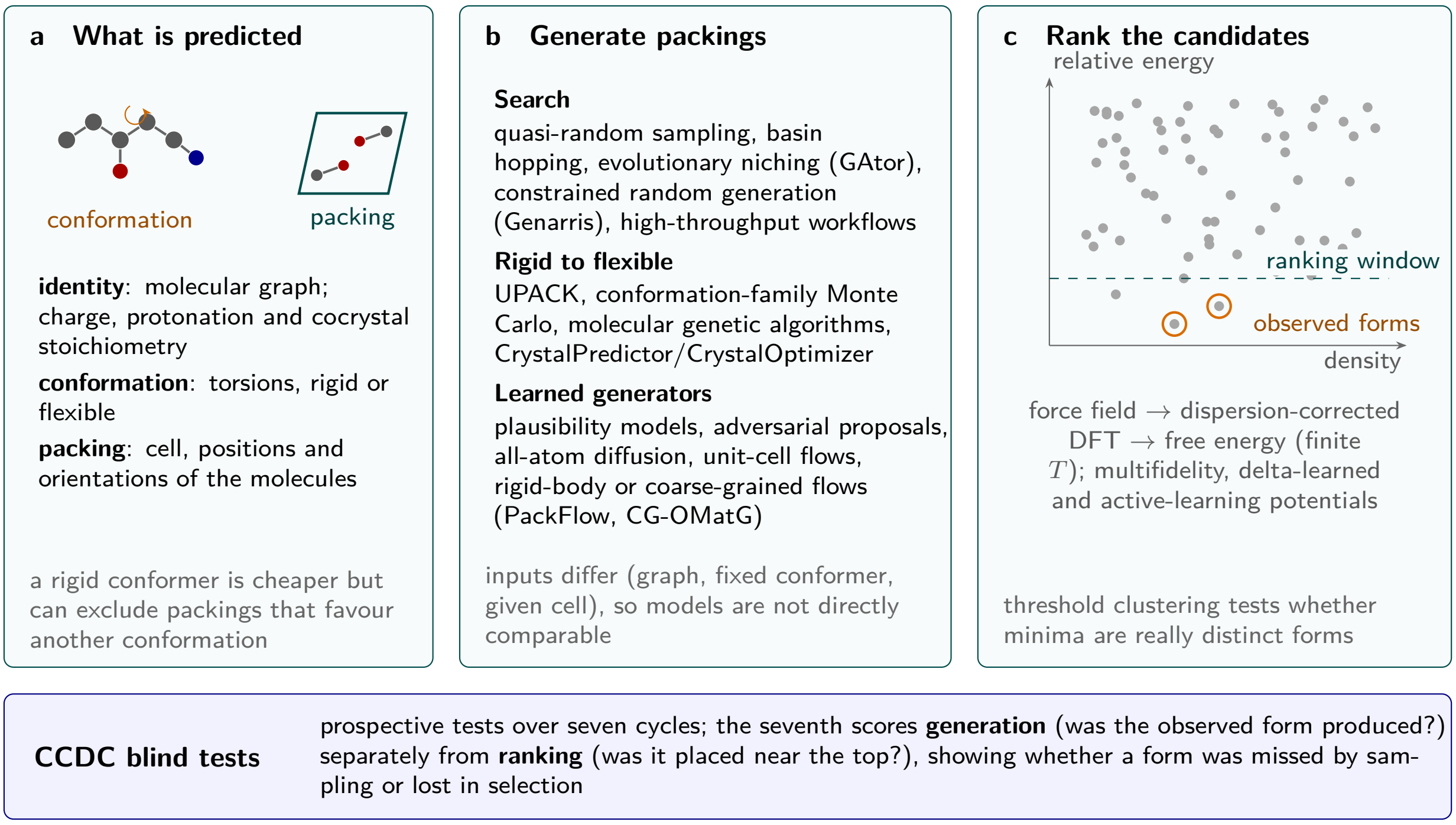
\begin{figure}[!htb]
\centering
\begin{tikzpicture}[x=1mm,y=1mm,font=\sffamily\footnotesize,
ptitle/.style={font=\sffamily\footnotesize\bfseries,anchor=north west},
small/.style={font=\sffamily\scriptsize,text=black!65,align=center,execute at begin node={\hyphenpenalty=10000\exhyphenpenalty=10000}},
item/.style={font=\sffamily\scriptsize,anchor=north west,align=left,execute at begin node={\hyphenpenalty=10000\exhyphenpenalty=10000}},
ex/.style={font=\sffamily\scriptsize,text=black!60,anchor=south west,align=left,execute at begin node={\hyphenpenalty=10000\exhyphenpenalty=10000}},
arr/.style={-{Stealth[length=1.7mm]},thick,black!60},
at1/.style={circle,fill=black!65,draw=white,line width=.3pt,inner sep=0,minimum size=2.0mm},
at2/.style={circle,fill=red!65!black,draw=white,line width=.3pt,inner sep=0,minimum size=1.8mm},
at3/.style={circle,fill=blue!55!black,draw=white,line width=.3pt,inner sep=0,minimum size=1.8mm}]
\newcommand{\panel}[3]{\draw[draw=teal!60!black,rounded corners=3pt,fill=teal!3] (0,0) rectangle (#3,74); \node[ptitle] at (1.5,73.2) {#1\quad #2};}

\begin{scope}[shift={(0,0)}]
\panel{a}{What is predicted}{48}
\begin{scope}[shift={(13,59)}]
 \draw[black!60,thick] (-6,0)--(-3,2)--(0,0)--(3,2)--(6,0); \draw[black!60,thick] (0,0)--(0,-3.5); \draw[black!60,thick] (6,0)--(8.5,-2);
 \node[at1] at (-6,0) {}; \node[at1] at (-3,2) {}; \node[at1] at (0,0) {}; \node[at1] at (3,2) {}; \node[at1] at (6,0) {};
 \node[at2] at (0,-3.5) {}; \node[at3] at (8.5,-2) {};
 \draw[-{Stealth[length=1.1mm]},orange!80!black] (1.1,3.8) arc[start angle=120,end angle=400,radius=1.1];
\end{scope}
\begin{scope}[shift={(33,53)}]
 \draw[teal!60!black,thick] (0,0)--(10,0)--(12,9)--(2,9)--cycle;
 \begin{scope}[shift={(3.5,2.6)},rotate=20] \draw[black!60,thick] (-1.6,0)--(1.6,0); \node[at1,minimum size=1.4mm] at (-1.6,0) {}; \node[at2,minimum size=1.3mm] at (1.6,0) {}; \end{scope}
 \begin{scope}[shift={(8.3,6.4)},rotate=200] \draw[black!60,thick] (-1.6,0)--(1.6,0); \node[at1,minimum size=1.4mm] at (-1.6,0) {}; \node[at2,minimum size=1.3mm] at (1.6,0) {}; \end{scope}
\end{scope}
\node[small,text=orange!60!black] at (13,50.2) {conformation};
\node[small,text=teal!50!black] at (39,50.2) {packing};
\node[item,text width=45mm] at (2.5,45) {\textbf{identity}: molecular graph; charge, protonation and cocrystal stoichiometry\\[2pt]\textbf{conformation}: torsions, rigid or flexible\\[2pt]\textbf{packing}: cell, positions and orientations of the molecules};
\node[ex,text width=45mm] at (1.5,0.8) {a rigid conformer is cheaper but can exclude packings that favour another conformation};
\end{scope}

\begin{scope}[shift={(51,0)}]
\panel{b}{Generate packings}{55}
\node[item,text width=51mm] at (2.5,66) {\textbf{Search}\\[1pt]
quasi-random sampling, basin hopping, evolutionary niching (GAtor), constrained random generation (Genarris), high-throughput workflows\\[3pt]
\textbf{Rigid to flexible}\\[1pt]
UPACK, conformation-family Monte Carlo, molecular genetic algorithms, CrystalPredictor/CrystalOptimizer\\[3pt]
\textbf{Learned generators}\\[1pt]
plausibility models, adversarial proposals, all-atom diffusion, unit-cell flows, rigid-body or coarse-grained flows (PackFlow, CG-OMatG)};
\node[ex,text width=51mm] at (1.5,0.8) {inputs differ (graph, fixed conformer, given cell), so models are not directly comparable};
\end{scope}

\begin{scope}[shift={(109,0)}]
\panel{c}{Rank the candidates}{53}
\begin{scope}[shift={(8,36)}]
 \draw[-{Stealth[length=1.2mm]},black!60] (0,0)--(40,0) node[below,small,inner sep=1.5pt,anchor=north east] {density};
 \draw[-{Stealth[length=1.2mm]},black!60] (0,0)--(0,30) node[above right,small,inner sep=1pt,anchor=south west] {relative energy};
 \pgfmathsetseed{7}
 \foreach \i in {1,...,60}{\pgfmathsetmacro\xx{4+rnd*33}\pgfmathsetmacro\yy{3.5+rnd^0.6*24}\fill[black!35] (\xx,\yy) circle (0.55);}
 \foreach \p in {(14,2.4),(19,4.4)} \fill[black!35] \p circle (0.55);
 \draw[orange!85!black,thick] (14,2.4) circle (1.3); \draw[orange!85!black,thick] (19,4.4) circle (1.3);
 \draw[teal!60!black,dashed] (0,7.5)--(38,7.5); \node[small,text=teal!50!black,anchor=west,fill=teal!3,inner sep=0.5pt] at (24,9.3) {ranking window};
 \node[small,text=orange!60!black,anchor=west] at (21.5,2.6) {observed forms};
\end{scope}
\node[small,text width=49mm] at (26.5,23.5) {force field $\rightarrow$ dispersion-corrected DFT $\rightarrow$ free energy (finite $T$); multifidelity, delta-learned and active-learning potentials};
\node[ex,text width=49mm] at (1.5,0.8) {threshold clustering tests whether minima are really distinct forms};
\end{scope}

\begin{scope}[shift={(0,-17)}]
\draw[draw=blue!50!black,fill=blue!5,rounded corners=3pt] (0,0) rectangle (162,14);
\node[font=\sffamily\footnotesize\bfseries,anchor=west] at (2,7) {CCDC blind tests};
\node[font=\sffamily\scriptsize,anchor=west,text width=126mm,align=left] at (34,7) {prospective tests over seven cycles; the seventh scores \textbf{generation} (was the observed form produced?) separately from \textbf{ranking} (was it placed near the top?), showing whether a form was missed by sampling or lost in selection};
\end{scope}
\end{tikzpicture}
\caption{Molecular-crystal structure prediction. (a) The prediction task is defined by the molecular identity, its conformational freedom, and the packing of molecules in the cell. (b) Candidate packings are generated by sampling and search methods, by methods that progressively add conformational flexibility, and by recent learned generators. (c) Candidates are ranked on a crystal energy landscape (schematic: relative energy against density) with a hierarchy of energy models; observed forms should fall within a narrow ranking window. The CCDC blind tests score generation and ranking separately.}
\label{fig:molecular-csp}
\end{figure}
 
The supplied molecular description defines the prediction task. A rigid conformer removes an important source of uncertainty but can exclude packings that favor another conformation. Searching internal coordinates increases the accessible space and the cost of each energy evaluation. Local approximate intramolecular models and adaptive refinements address this tradeoff \citep{CSP063,CSP065}. Molecular identity also needs to distinguish charge, protonation, and multicomponent stoichiometry when these are part of the problem; cocrystal prediction explicitly extends the packing task to multiple molecular components \citep{CSP039}.
 
For a given molecular description, search methods differ mainly in how they distribute effort over the packing landscape: quasi-random sampling, basin hopping, and evolutionary niching are three such strategies \citep{CSP066,CSP076,CSP078,CSP104}. Genarris \citep{CSP079,CSP096,CSP171} introduces physically constrained random generation with screening and diversity selection, with later versions refining symmetry handling and packing construction. Population-based high-throughput workflows offer another organization of these stages \citep{CSP164}. The principles resemble those of inorganic search, but rigid-body and conformational variables need their own operators and distance measures.
 
\subsection{Energy ranking and polymorphism}
Once candidate packings have been generated, their relative ranking becomes a separate bottleneck. Distributed multipole electrostatics, tailor-made force fields, and dispersion-aware electronic-structure calculations address different aspects of molecular interactions \citep{CSP003,CSP034,CSP038}. The balance between intermolecular energy and intramolecular deformation can change relative rankings even when every candidate has been locally minimized. A fast model may be suitable for broad screening while a more expensive model is needed to resolve a small final set.
 
Multifidelity learning, learned landscape models, and active-learning potentials make this hierarchy more explicit \citep{CSP081,CSP098,CSP125,CSP150}. Recent work on delta-learned molecular-crystal potentials continues to address the relationship between computational efficiency and reference-level ranking accuracy \citep{CSP186}. The relevant test for these models is the relative energy and ordering of candidate forms, which aggregate errors over unrelated configurations do not capture.
 
Static lattice energies also leave out finite-temperature and environmental effects. Free-energy calculations and studies of crystal-form stability under realistic conditions address part of the discrepancy between optimized minima and observed polymorphs \citep{CSP089,CSP135}. Threshold-based clustering further asks whether multiple optimized structures correspond to meaningfully distinct regions of the landscape \citep{CSP116,CSP136}.
 
\subsection{Learned molecular generators and blind tests}
Recent molecular generative approaches include learned plausibility models, adversarial proposals, all-atom diffusion, conditional unit-cell flows, and rigid-body or coarse-grained flows \citep{CSP132,CSP151,CSP174,CSP188,CSP191,CSP192}. Two 2026 preprints, PackFlow and CG-OMatG, add reinforcement-based alignment. Because these models are conditioned on different inputs (a molecular graph, a fixed conformer, or given crystallographic settings), they leave different amounts of the prediction problem to be solved and are not directly comparable.
 
Shared prospective tests are the main remedy for this incomparability, and the CCDC blind tests provide a long-running framework for them. Seven cycles have been reported \citep{CSP009,CSP014,CSP022,CSP040,CSP051,CSP068,CSP155,CSP156}. The seventh separates structure generation and structure ranking, making it possible to ask whether a form was missed by sampling or lost during selection. Inorganic benchmarks would benefit from the same separation.
 
\section{Benchmarks, datasets, and meaningful comparison}\label{sec:benchmarks-main}
The distinction between generation and ranking in molecular blind tests illustrates a broader evaluation problem. Across atomic and molecular CSP, comparisons must separate candidate recovery, search efficiency, ranking accuracy, and physical assessment. Figure~\ref{fig:benchmark-questions} summarizes these questions and the information each comparison should report.
 
\begin{figure}[!htb]
\centering
\begin{tikzpicture}[x=1mm,y=1mm,font=\sffamily\footnotesize,
ptitle/.style={font=\sffamily\footnotesize\bfseries,anchor=north west},
small/.style={font=\sffamily\scriptsize,text=black!65,align=center,execute at begin node={\hyphenpenalty=10000\exhyphenpenalty=10000}},
note/.style={font=\sffamily\scriptsize,anchor=south west,align=left,execute at begin node={\hyphenpenalty=10000\exhyphenpenalty=10000}},
lim/.style={font=\sffamily\scriptsize,text=orange!55!black,anchor=south west,align=left,execute at begin node={\hyphenpenalty=10000\exhyphenpenalty=10000}},
cand/.style={draw=black!40,fill=white,rounded corners=1pt,minimum width=5.6mm,minimum height=5mm,inner sep=0,font=\sffamily\tiny},
arr/.style={-{Stealth[length=1.5mm]},thick,black!60}]
\newcommand{\panel}[4]{\draw[draw=teal!60!black,rounded corners=3pt,fill=teal!3] (0,0) rectangle (#3,#4); \node[ptitle] at (1.5,#4-0.8) {#1\quad #2};}

\begin{scope}[shift={(0,0)}]
\panel{a}{Recovery: is the reference found?}{79}{50}
\node[small] at (39.5,42) {match if $d(X_i,X^{\mathrm{ref}})<\epsilon$ for a declared matcher};
\foreach \r in {1,...,10} {\node[cand] (c\r) at ({3.5+\r*6.6},31) {\r};}
\node[cand,fill=teal!25,draw=teal!60!black] at (3.5+3*6.6,31) {3};
\node[font=\sffamily\scriptsize,text=black!50] at (75,31) {$\cdots$};
\node[small,anchor=west,text=black!70] at (3,37) {ranked candidates};
\foreach \x in {13.75,40.15} \draw[black!45,densely dashed] (\x,27.5)--(\x,34.6);
\node[small,anchor=north,text=black!60] at (13.75,25.6) {top-1: no};
\node[small,anchor=north,text=teal!50!black] at (40.15,25.6) {top-5: yes};
\node[small,anchor=north,text=teal!50!black] at (69.5,25.6) {top-20: yes};
\node[small,text width=74mm] at (39.5,15.5) {$R_k$ = share of targets with a match in the first $k$; timeouts and failed decodes stay in the denominator};
\node[lim,text width=75mm] at (1.5,1) {higher $k$ raises recovery and cost: report $k$, the relaxation stage and the budget};
\end{scope}

\begin{scope}[shift={(83,0)}]
\panel{b}{Search efficiency: how fast?}{79}{50}
\begin{scope}[shift={(10,10)}]
 \draw[-{Stealth[length=1.2mm]},black!60] (0,0)--(62,0); \node[small,anchor=south east,inner sep=1pt] at (62,0.6) {evaluator calls};
 \draw[-{Stealth[length=1.2mm]},black!60] (0,0)--(0,29) node[small,anchor=south west,inner sep=1pt] {lowest energy found};
 \draw[thick,orange!80!black] (2,26)--(8,26)--(8,18)--(20,18)--(20,14)--(38,14)--(38,6)--(58,6);
 \draw[thick,teal!60!black] (2,26)--(5,26)--(5,13)--(30,13)--(30,10)--(58,10);
 \draw[black!40,dashed] (0,5)--(58,5); \node[small,anchor=west] at (2,6.8) {global minimum};
 \node[small,text=teal!50!black,anchor=west] at (42,11.6) {method B};
 \node[small,text=orange!60!black,anchor=west] at (40.5,7.5) {method A};
\end{scope}
\node[lim,text width=75mm] at (1.5,1) {a common evaluator isolates the search, but may itself be wrong};
\end{scope}

\begin{scope}[shift={(0,-62)}]
\panel{c}{Generalization: what was held out?}{79}{58}
\foreach \y/\t/\d in {46/{random split}/{interpolation},38/{chemistry held out}/{new elements or systems},30/{time held out}/{later discoveries},22/{prototype held out}/{unseen structure families}}{
  \fill[blue!55!black] (4,\y) circle (0.8); \node[note,anchor=west,text width=38mm] at (6,\y) {\textbf{\t}\\{\color{black!65}\d}};}
\draw[draw=orange!75!black,fill=orange!8,rounded corners=2pt] (47,15) rectangle (77,50);
\node[small,text=orange!60!black] at (62,47) {\textbf{leakage routes}};
\node[small,text width=27mm] at (62,32) {generator training data, template catalogue, energy model or reranker, tuning on the test set};
\node[lim,text width=75mm] at (1.5,1) {a component that has seen test structures weakens generalization claims; report unknown exposure as unknown};
\end{scope}

\begin{scope}[shift={(83,-62)}]
\panel{d}{Physical assessment: what does it mean?}{79}{58}
\foreach \y/\t/\d in {45/{matches the reference}/{the reference need not be the evaluator's ground state},35/{lower energy than the reference}/{a better polymorph or an evaluator error; verify it},25/{novel structure}/{needs a dated inventory and stated tolerances},15/{duplicates after relaxation}/{report uniqueness before and after refinement}}{
  \fill[teal!60!black] (4,\y+1.5) circle (0.8);
  \node[note,anchor=north west,text width=70mm] at (6,\y+3.2) {\textbf{\t}\\{\color{black!65}\d}};}
\node[lim,text width=75mm] at (1.5,1) {report these quantities separately before combining them};
\end{scope}

\begin{scope}[shift={(0,-77)}]
\draw[draw=blue!50!black,fill=blue!5,rounded corners=3pt] (0,0) rectangle (162,11);
\node[font=\sffamily\footnotesize\bfseries,anchor=west] at (2,5.5) {Report with every result};
\node[font=\sffamily\scriptsize,anchor=west,text width=118mm,align=left,execute at begin node={\hyphenpenalty=10000\exhyphenpenalty=10000}] at (42,5.5) {target set and version (denominator) $\cdot$ inputs supplied $\cdot$ matcher and tolerances $\cdot$ $k$ and relaxation stage $\cdot$ evaluator and budget $\cdot$ training data and exposure of every component};
\end{scope}
\end{tikzpicture}
\caption{Questions a CSP algorithm comparison must keep apart. (a) Recovery: a target counts as solved at top-$k$ if any of the first $k$ ranked candidates matches the reference under a declared matcher and tolerance; here the first match is at rank 3. Failed or timed-out targets remain in the denominator. (b) Search efficiency: how quickly methods reach low energy under a common evaluator. (c) Generalization: which kind of split was held out, and whether any component of the workflow has seen the test structures. (d) Physical assessment: what a match, a lower energy, a novel structure, or a duplicate does and does not establish. The bottom strip lists information that should accompany every reported result.}
\label{fig:benchmark-questions}
\end{figure}
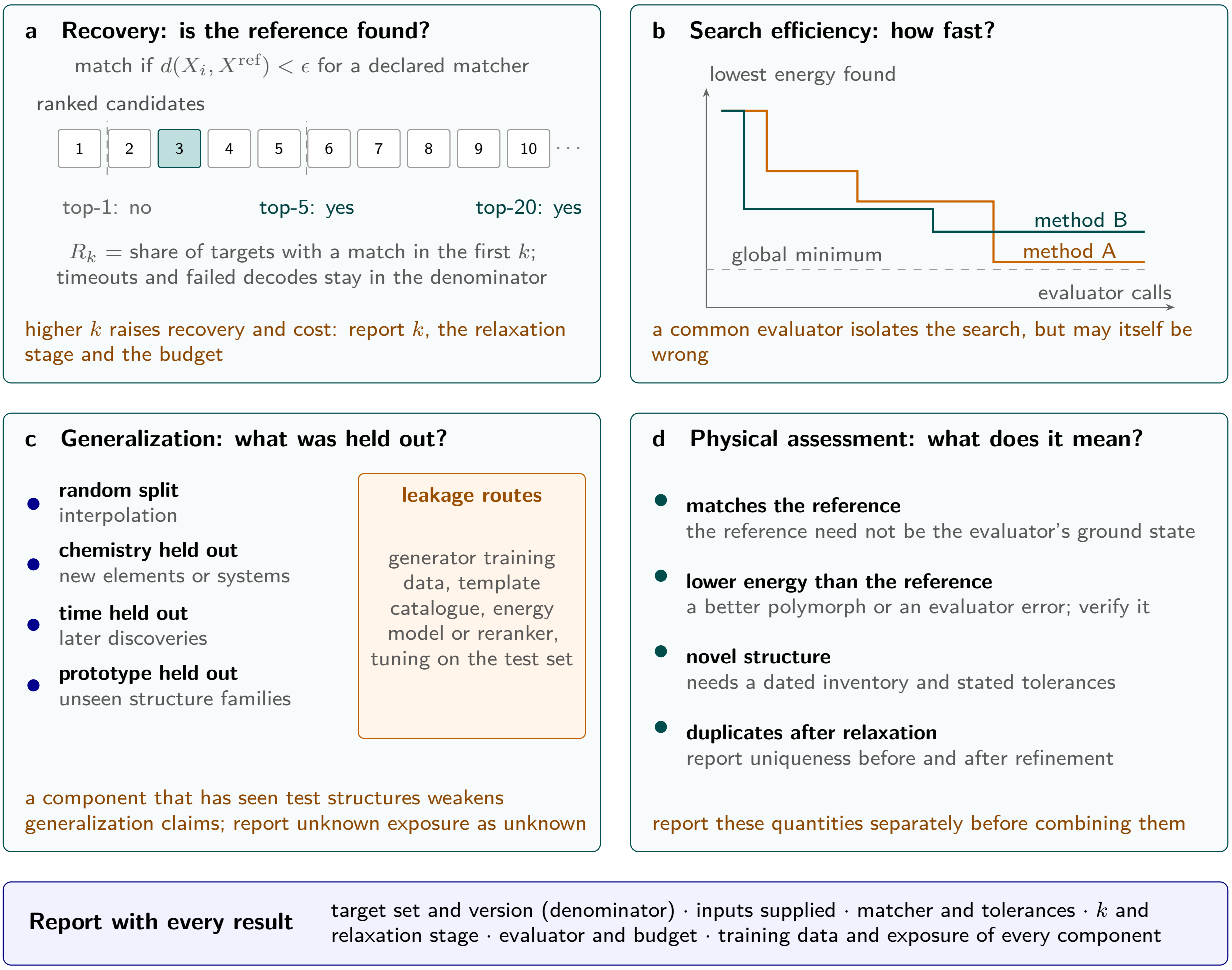
 
\subsection{Structure recovery and search efficiency}
Reference recovery asks whether at least one proposed candidate matches a specified target. For $M$ targets, a top-$k$ recovery rate can be written as
\begin{equation}
 R_k=\frac{1}{M}\sum_{j=1}^{M}\mathbb{1}\!\left[
 \exists i\leq k:\ d(X_{ji},X_j^{\mathrm{ref}})<\epsilon
 \right],
\end{equation}
where $d$ represents a declared structure-matching rule and $\epsilon$ its tolerance; $X_{ji}$ denotes the $i$-th ranked candidate structure proposed for target $j$, and $X_j^{\mathrm{ref}}$ is the reference structure of target $j$. In practice matchers combine several tolerances, lattice transformations, and species constraints, and results depend on the cell convention, normalization, and supercell handling, which should be reported.
 
CSPMetrics \citep{CSPMetrics2024} provides a quantitative framework for comparing predicted structures with references. Its released CSPBenchMetrics implementation combines complementary distances based on atomic coordinates, Wyckoff positions, point sets, graphs, diffraction patterns, fingerprints, and model-predicted energies. Together these distances capture differences that a single binary match or an energy comparison can miss, and reduce reliance on manual inspection, although each depends on the alignment, cell convention, and evaluator used.
 
Recovery depends on the candidate budget and on whether candidates are measured before or after relaxation. A failure to decode, a timeout, or a relaxation error should remain in the task denominator. Increasing $k$ can raise recovery while also increasing cost. Without runtime, evaluator calls, or the cost of selecting $k$ candidates, top-$k$ figures can make methods with very different budgets look comparable.
 
Search efficiency is a different question: how quickly does an algorithm reach low energy under a common evaluator? The early inorganic blind-test appendix used shared interaction models, while comparative work on genetic and minima-hopping methods examined sensitivity to system size and supplied constraints \citep{CSP046,CSP041}. Such designs help isolate search behavior, but a common evaluator may still be systematically wrong. Whole-pipeline comparisons answer the complementary question but make it harder to tell where a difference comes from.
 
\subsection{Crystal-structure sources for model training}
Table~\ref{tab:training-datasets} summarizes major crystal-structure resources relevant to CSP training, including inorganic and molecular collections, derived benchmark splits, and auxiliary potential-training data. For composition-to-structure learning, each example must provide atomic species, lattice vectors, and atomic positions under a defined cell convention. Relaxed computational crystals and experimental structures provide different target distributions: the former depend on the electronic-structure settings, while the latter can contain disorder, partial occupancies, and temperature- or pressure-dependent geometries.
 
A training release should record source identifiers, database version, structural standardization, filtering, and split membership. Database totals are not interchangeable with usable training-set sizes: duplicate representations, polymorphs, trajectories, and structures shared among databases require explicit treatment. Cross-database merging should preserve provenance and prevent equivalent structures or prohibited prototype families from crossing evaluation splits. Trajectory collections primarily support learned potentials for search and relaxation; using every intermediate frame as a target for stable-crystal generation changes the learning objective. Held-out cohorts such as T46 and T180 must be excluded from training and tuning.
 
\begingroup
\small
\setlength{\tabcolsep}{4pt}
\renewcommand{\arraystretch}{1.12}
\begin{longtable}{@{}P{.18\linewidth}P{.25\linewidth}P{.34\linewidth}P{.17\linewidth}@{}}
\caption{Major crystal-structure databases and datasets relevant to CSP training. Entries distinguish direct structural training corpora, derived benchmark splits, repositories, and auxiliary potential-training data. Access information was checked on 17 October 2026; the cited resource documentation identifies the relevant downloads.}\label{tab:training-datasets}\\
\toprule
Resource and citation & Available structures & Training role and main qualification & Access and link \\
\midrule
\endfirsthead
\multicolumn{4}{l}{\textit{Table \thetable\ continued}}\\
\toprule
Resource and citation & Available structures & Training role and main qualification & Access and link \\
\midrule
\endhead
\midrule\multicolumn{4}{r}{\textit{Continued on next page}}\\
\endfoot
\bottomrule
\endlastfoot
\multicolumn{4}{@{}l}{\textbf{Computed inorganic crystal databases}}\\*
\addlinespace[2pt]
Materials Project \citep{MPData2026} & DFT-relaxed inorganic structures with energies and calculated properties. & Broad composition-to-structure training; property conditioning; template libraries. Select a dated release and consistent calculation level. & Free data; API key for the official client. \href{https://docs.materialsproject.org/downloading-data/using-the-api}{Resource} \\
\addlinespace[3pt]
Alexandria \citep{AlexandriaData2026} & DFT-relaxed crystals; separate PBE, PBEsol/SCAN, dimensionality, and trajectory collections. & Large-scale generative training. Select bulk relaxed structures and deduplicate across source databases. & Download / OPTIMADE; CC BY 4.0. \href{https://alexandria.icams.rub.de/datasets.html}{Resource} \\
\addlinespace[3pt]
OQMD \citep{OQMDData2026} & High-throughput DFT structures and energies, including experimental-derived and prototype-decorated candidates. & Broad inorganic training and substitution baselines; preserve prototype provenance. & API and database dumps; CC BY 4.0. \href{https://oqmd.org/download/}{Resource} \\
\addlinespace[3pt]
AFLOW \citep{AFLOWData2026} & Computed crystal structures, properties, and a related crystallographic-prototype library. & Inorganic training and symmetry/prototype representations; distinguish prototypes from calculated entries. & Public data services; AFLUX / REST interfaces. \href{https://aflow.org/documentation/}{Resource} \\
\addlinespace[3pt]
JARVIS-DFT \citep{JARVISData2026} & Relaxed 3D and 2D structures with electronic, elastic, and other property labels. & Property-conditioned training and validation; separate 2D and bulk tasks and calculation settings. & Download tools / OPTIMADE; portal registration. \href{https://jarvis.nist.gov/}{Resource} \\
\addlinespace[3pt]
GNoME release \citep{GNoMEData2026} & Predicted inorganic crystals with DFT energies; public structure archives near the reference convex hull. & Augment hypothetical-crystal coverage; thermodynamic labels do not establish experimental synthesis. & Public archives; data CC BY-NC 4.0. \href{https://github.com/google-deepmind/materials_discovery/blob/main/DATASET.md}{Resource} \\
\addlinespace[3pt]
\multicolumn{4}{@{}l}{\textbf{Experimental and curated crystallographic collections}}\\*
\addlinespace[2pt]
ICSD \citep{ICSDData2026} & Curated inorganic structures, site occupancies, symmetry and structure-type information; also selected theoretical entries. & Experimental priors and prototype retrieval. Filter provenance, disorder, partial occupancy and measurement conditions. & Licensed database; use-specific access terms. \href{https://icsd.products.fiz-karlsruhe.de/en}{Resource} \\
\addlinespace[3pt]
COD \citep{CODData2026} & CIF collection of inorganic, organic, metal-organic and mineral structures. & Open experimental structure source. Validate CIFs and standardize occupancies and duplicate determinations. & Open downloads; CC0 / public domain. \href{https://www.crystallography.net/cod/}{Resource} \\
\addlinespace[3pt]
CSD \citep{CSDData2026} & Curated small-molecule crystal structures and molecular packing information. & Molecular-crystal generators and packing priors; molecular identities and conformers are required conditions. & Licensed collection; individual deposited structures have separate access. \href{https://www.ccdc.cam.ac.uk/solutions/software/csd/}{Resource} \\
\addlinespace[3pt]
\multicolumn{4}{@{}l}{\textbf{Commonly used generative-CSP datasets and splits}}\\*
\addlinespace[2pt]
MP-20 \citep{CSP113,CDVAECode} & Materials Project-derived crystals with at most 20 atoms per cell. & Standard small-cell generator training. Preserve published train/validation/test files and audit prototype overlap. & Released processed splits; retain upstream terms. \href{https://github.com/txie-93/cdvae/tree/main/data/mp_20}{Resource} \\
\addlinespace[3pt]
MPTS-52 \citep{MatbenchGenmetrics2024,DiffCSPData} & Materials Project-derived time-split data with up to 52 atoms per cell. & Chronological generalization and larger-cell CSP; train only on the designated training partition. & Released processed splits; retain upstream terms. \href{https://github.com/jiaor17/DiffCSP/tree/main/data/mpts_52}{Resource} \\
\addlinespace[3pt]
Perov-5 \citep{CSP113} & Five-atom perovskite dataset curated from a computational perovskite collection. & Controlled family-specific generation; narrow prototype coverage limits claims of general CSP. & CDVAE release and upstream CMR data. \href{https://github.com/txie-93/cdvae/tree/main/data/perov_5}{Resource} \\
\addlinespace[3pt]
Carbon-24 \citep{CSP113} & Carbon structures with up to 24 atoms, curated from AIRSS calculations at 10 GPa. & Single-element structural generation; the pressure and chemistry differ from general ambient-pressure training. & CDVAE release and upstream Materials Cloud archive. \href{https://github.com/txie-93/cdvae/tree/main/data/carbon_24}{Resource} \\
\addlinespace[3pt]
\multicolumn{4}{@{}l}{\textbf{Repositories and specialized structural domains}}\\*
\addlinespace[2pt]
NOMAD \citep{NOMADData2026} & Deposited simulation inputs, outputs and normalized atomic structures from heterogeneous workflows. & Extract task-specific training sets with provenance; select periodicity, convergence and final geometries. & Public entries via archive API / OPTIMADE; entry-specific terms. \href{https://nomad-lab.eu/prod/v1/api/v1/extensions/redoc}{Resource} \\
\addlinespace[3pt]
Materials Cloud \citep{MaterialsCloudData2026} & Versioned research datasets, including CIF collections and provenance-rich simulation archives. & Source of specialized structure corpora and reproducible training releases; inspect each deposit. & Public archive; deposit-specific files and licenses. \href{https://archive.materialscloud.org/}{Resource} \\
\addlinespace[3pt]
C2DB \citep{C2DBData2026} & Computed two-dimensional crystals with structural and property data. & 2D crystal generation; vacuum and dimensionality require a task-specific representation. & CMR downloads; CC BY-SA 4.0. \href{https://cmr.fysik.dtu.dk/c2db/c2db.html}{Resource} \\
\addlinespace[3pt]
\multicolumn{4}{@{}l}{\textbf{Structure trajectories for learned relaxation and energy models}}\\*
\addlinespace[2pt]
MPtrj \citep{CSP127} & Materials Project GGA/GGA+U relaxation configurations with energy, force, stress and magnetic information. & Train the potential used in CSP relaxation/search. Intermediate frames are not independent stable crystals. & CHGNet-linked data release; MP terms apply. \href{https://github.com/CederGroupHub/chgnet#mptrj-dataset}{Resource} \\
\addlinespace[3pt]
OMat24 \citep{OMat24Data2026} & Non-equilibrium single-point structures and relaxation trajectories with energies, forces and stresses. & Potential training and geometric augmentation; select relaxed endpoints for equilibrium-structure generators. & Public LMDB downloads; CC BY 4.0. \href{https://huggingface.co/datasets/facebook/OMAT24}{Resource} \\
\addlinespace[3pt]
\end{longtable}
\endgroup

\subsection{Named datasets, test cohorts, and metric packages}
MP-20 \citep{CSP113,CDVAECode} provides a widely used Materials Project-derived training, validation, and test release with at most 20 atoms per cell. MPTS-52 \citep{MatbenchGenmetrics2024,DiffCSPData} originates from the Materials Project time-split framework, now distributed through matbench-genmetrics, and extends evaluation to structures with up to 52 atoms using chronological splits. The number in the name is the size limit. Because release files, preprocessing, and split identifiers determine the actual evaluation set, the dataset name alone does not identify it.
 
CSPBench \citep{CSP180} compares inorganic prediction methods across 180 targets, with a separate 23-target subset for its DFT-based baselines. T180 \citep{CSP185,DeepCSPBenchmarkData2026} and T46 \citep{CSP185,DeepCSPBenchmarkData2026} are separately versioned cohorts in the later deep-generative benchmark release. Their v3.1 manifests specify 180 and 46 targets, respectively, and provide integer elemental counts of the geometric primitive cell as queries. The queries therefore fix the cell multiplicity as well as the composition. Because the T180 targets were revised after the original CSPBench evaluation, results should cite the manifest version. The TCSPv3 algorithm \citep{Feng2026TCSP3} adds a larger 800-target cohort (T800), on which it reports a top-20 match rate of 74.0\%. CSPMetrics (released as CSPBenchMetrics) \citep{CSPMetrics2024} supplies comparison measures and code, not a target set. Table~\ref{tab:named-benchmarks} lists these resources and their roles.
 
\begin{table}[tb]
\centering\small
\caption{Named CSP evaluation resources and their roles.}\label{tab:named-benchmarks}
\begin{tabularx}{\linewidth}{@{}P{.24\linewidth}P{.22\linewidth}Y@{}}
\toprule Resource & Role & Scope and comparison requirement \\
\midrule
MP-20 \citep{CSP113,CDVAECode} & Training and test dataset & Up to 20 atoms per cell; retain the exact split and preprocessing. \\
MPTS-52 \citep{MatbenchGenmetrics2024,DiffCSPData} & Time-split dataset & Up to 52 atoms per cell; distinguish temporal transfer from random-split recovery. \\
CSPBench \citep{CSP180} & Cross-method benchmark & 180-target study; DFT-based comparisons use a separate 23-target subset. \\
T180 \citep{CSP185,DeepCSPBenchmarkData2026} & Versioned test cohort & 180 targets in v3.1; primitive-cell-count queries and MP-20 exposure audit. \\
T46 \citep{CSP185,DeepCSPBenchmarkData2026} & Versioned test cohort & 46 targets in v3.1; use its own references, query manifest, and denominator. \\
CSPMetrics / CSPBenchMetrics \citep{CSPMetrics2024} & Metric package & Complementary structure distances; specify metrics, alignment and any energy model. \\
\bottomrule
\end{tabularx}
\end{table}
 
\subsection{A CSP leaderboard snapshot: recovery and candidate selection}
The CSP leaderboard \citep{CSPLeaderboardSnapshot2026} shows how top-$k$ recovery varies across methods (Figure~\ref{fig:t46-performance}). The leaderboard has benchmark results for 16--20 CSP algorithms depending on the test sets including T46, T180, T240, and T800.
Its T46 cohort contains 46 targets evaluated with 20 algorithms. TCSPv3 \citep{Feng2026TCSP3} recovers 35 targets (76.1\%) at top-1 and 41 (89.1\%) at top-20; Uni-3DAR recovers 32 (69.6\%) and 33 (71.7\%), respectively, while uFlowCSP recovers 27 (58.7\%) and 30 (65.2\%).
 
The comparison reflects each method's recorded implementation and budget. Most entries use ORB-v3 relaxation and energy ranking.
The global searches (USPEX, MAGUS, and XtalOpt) optimize each target directly with their own budgets, so a common limit of 20 candidates does not mean equal computational effort.
 
\begin{figure}[!htb]
\centering
\includegraphics[width=\linewidth]{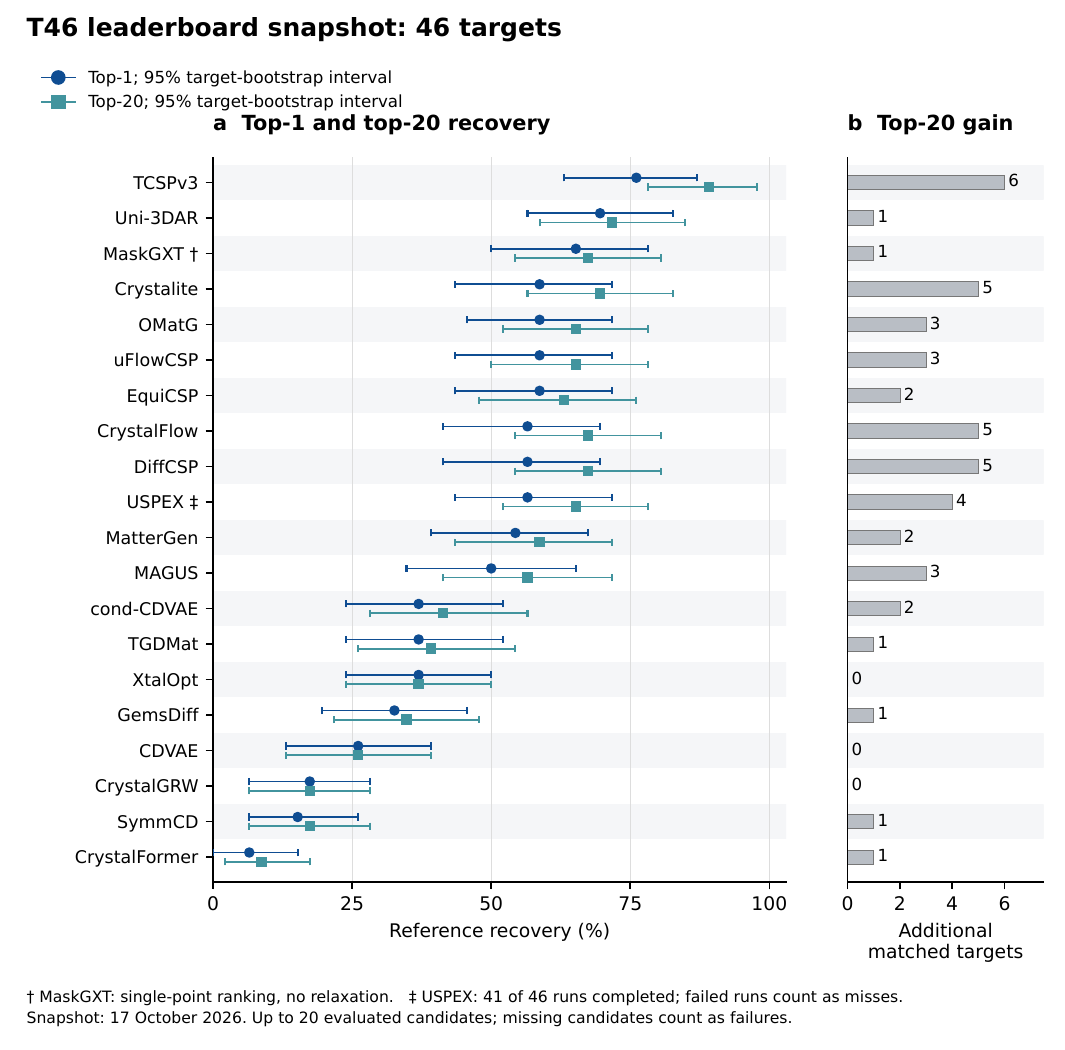}
\caption{T46 performance on the CSP leaderboard as of 17 October 2026 \citep{CSPLeaderboardSnapshot2026}. (a) Top-1 and top-20 reference recovery with the reported 95\% percentile-bootstrap intervals over targets.
(b) Additional matched targets at top-20 relative to top-1. All rates use 46 targets, including failures; one target corresponds to 2.17 percentage points. Up to 20 ranked candidates are evaluated. Matching uses pymatgen StructureMatcher with \texttt{ltol=0.2}, \texttt{stol=0.3}, and a $5^\circ$ angle tolerance, primitive-cell reduction and scaling enabled, and supercell matching disabled.
}
\label{fig:t46-performance}
\end{figure}
 
\subsection{Training distributions and generalization}
Interpreting this CSP performance snapshot also requires examining the training distributions and possible exposure of each workflow component. Frequently used generative datasets such as MP-20 and related crystallographic subsets define specific distributions of composition, cell size, and structural complexity \citep{CSP113,CSP128,CSP146}. Molecular datasets define different distributions, often with restricted molecule sizes or conformational assumptions. The same named benchmark can also be used with different preprocessing, deduplication, splits, training pools, or evaluators.
 
Random train--test splits measure interpolation under the chosen sampling procedure. Chemistry-held-out \cite{CSP185}, time-held-out (e.g. MPTS-52 \cite{DiffCSPData}), and prototype- or family-held-out splits address different extrapolation questions. Structures with different formulas can still share a prototype, and multiple representations or relaxations of one material can cross a naive split. A template baseline is especially informative when a model may be exploiting common structural families. A controlled prototype-removal study addresses this question for selected generative models \citep{CSP185}.
 
Another important issue in CSP benchmarking is to control data leakage, which can enter through more than the generator training set. A pretrained energy model, template catalogue, property predictor, or reranker may have seen the test structures or close derivatives. Such exposure need not invalidate an engineering comparison, but it weakens any generalization claim; for example, a generator trained without a prototype family but ranked by a potential trained on that family is not a family-unseen pipeline. When a component's training data are unknown, exposure should be reported as unknown.
 
\subsection{Stability, novelty, and ranking in CSP benchmarking}
A recovered reference may not be the ground state of the chosen evaluator, and a lower predicted energy may reflect evaluator error instead of a better prediction. Thermodynamic assessment requires compatible competing phases and a defined energy reference. Dynamical assessment requires additional information about local stability. Novelty requires a dated comparison inventory and explicit matching tolerances. These quantities should be reported separately before they are combined into any summary statistic \citep{CSP090,CSP129,CSP180}.
 
For generated candidates, relaxation can merge apparently distinct samples or transform them into known structures. Uniqueness and novelty should therefore be reported both before and after the physically relevant refinement stage. A sampler that concentrates on common low-energy prototypes may improve stable yield while reducing coverage. Conversely, apparently novel outputs may include invalid or poorly relaxed configurations. Only reporting both sides shows the tradeoff.
 
The separate CSPBench and DFT-baseline cohorts introduced above \citep{CSP180} illustrate why each result must retain its own denominator. Molecular blind-test rankings likewise depend on the submitted candidate lists and the rules of each cycle \citep{CSP155,CSP156}. Benchmarks are most informative when their data, information budget, evaluator, and implementation restrictions are treated as part of the result. Table~\ref{tab:evaluation} summarizes the questions that different evaluations answer.
 
\begin{table}[tb]
\centering\small
\caption{Questions answered by different CSP evaluations.}\label{tab:evaluation}
\begin{tabularx}{\linewidth}{@{}P{.22\linewidth}YY@{}}
\toprule Question & Required reporting & Principal limitation \\
\midrule
Reference recovery & Target set, matcher, inputs, top-$k$, failures, relaxation stage & Does not prove global-minimum discovery \\
Optimization efficiency & Repeated searches, common objective, evaluator calls, runtime & Does not establish evaluator accuracy \\
Candidate ranking & Shared candidate pool; relative energy or free-energy reference & Does not test completeness of generation \\
Thermodynamic qualification & Competing phases, corrections, pressure, reference protocol & Does not establish dynamical or kinetic accessibility \\
Diversity and novelty & Frozen inventory, matching rules, pre/post-relax counts & Does not establish physical viability \\
Generalization & Chemistry/family/time splits and all-model exposure & Remains bounded by the tested distribution \\
Prospective prediction & Hidden or future targets and declared prior information & Experimental observation remains condition-dependent \\
\bottomrule
\end{tabularx}
\end{table}
 
\section{From prediction to synthesis: experimentally confirmed discoveries}\label{sec:discoveries}
Benchmarks measure whether a method recovers structures that are already known. The stronger test of CSP is whether its predictions lead to materials that are then made and characterized. Oganov, Pickard, Zhu, and Needs compiled the most widely cited list of such cases up to 2018 \citep{CSP090}, and reviews of high-pressure discovery \citep{Zhang2017HighPressureReview,WangMa2014Perspective} and of superconducting hydrides \citep{FloresLivas2020Hydrides,Pickard2020Hydrides} document many more. Here we extend that list to 2026 and audit each entry. For every case, Table~\ref{tab:discoveries} records the search method, the prediction and confirmation papers, and two labels. The first states whether the prediction preceded the experiment, was made jointly with it, or solved the structure of a material already observed. The second grades the experimental evidence. All identifiers were checked against Crossref; methods that we attributed from a group's established workflow but did not find stated in the paper are marked.
 
\begin{scriptsize}
\setlength{\tabcolsep}{3pt}
\begin{longtable}{P{30mm}P{26mm}P{17mm}P{17mm}P{17mm}cc}
\caption{Experimentally examined CSP predictions. Entries from Table~1 of Oganov et al.~\citep{CSP090} are marked $^\ast$; the others were added for this survey. Relation: P, prediction preceded experiment; J, joint theory--experiment study; S, structure solution of an existing experiment. Evidence: S, structure confirmed by diffraction or microscopy; C, consistent property or partial structural evidence; D, disputed. A dagger ($^\dagger$) marks a search method attributed from the authors' established workflow but not stated in the text we checked; ``structure search'' means the paper reports a first-principles search without naming the code in the abstract.}\label{tab:discoveries}\\
\toprule Material & Search method & Pressure & Prediction & Confirmation & Rel. & Ev.\\ \midrule\endfirsthead
\toprule Material & Search method & Pressure & Prediction & Confirmation & Rel. & Ev.\\ \midrule\endhead
\bottomrule\endfoot
\multicolumn{7}{l}{\textbf{Superconducting hydrides}}\\
CaH$_{6}$ & CALYPSO & 172 GPa & 2012 \citep{DiscWang2012Supercondu} & 2022 \citep{DiscMa2022Hightemper} & P & S\\
H$_{3}$S$^\ast$ & USPEX & $\sim$150-200 GPa & 2014 \citep{DiscDuan2014Pressurein} & 2015 \citep{DiscDrozdov2015Convention} & P & S\\
YH$_{6}$ & CALYPSO & $\sim$166 GPa & 2015 \citep{DiscLi2015Pressurest} & 2021 \citep{DiscTroyan2021Anomalous} & P & S\\
CeH$_{10}$/CeH$_{9}$ superconductivity & structure search & $\sim$95 GPa & 2017 \citep{DiscPeng2017Hydrogen} & 2021 \citep{DiscChen2021Hightemper} & P & C\\
CeH$_{9}$ & structure search & 80-100 GPa & 2017 \citep{DiscPeng2017Hydrogen} & 2019 \citep{DiscSalke2019Synthesis} & P & S\\
LaH$_{10}$$^\ast$ & structure search & $\sim$170-200 GPa & 2017 \citep{DiscLiu2017Potential} & 2018 \citep{DiscGeballe2018Synthesis} & P & S\\
YH$_{9}$ & structure search & $\sim$201 GPa & 2017 \citep{DiscPeng2017Hydrogen} & 2021 \citep{DiscKong2021Supercondu} & P & S\\
ThH$_{10}$ & USPEX & 170-175 GPa & 2018 \citep{DiscKvashnin2018Hightemper} & 2020 \citep{DiscSemenok2020Supercondu} & P & S\\
UH$_{7}$$^\ast$ & USPEX & $\sim$22-63 GPa & 2018 \citep{DiscKruglov2018Uranium} & same & J & S\\
(La,Y)H$_{10}$ & structure search & $\sim$183 GPa & 2021 \citep{DiscSemenok2021Supercondu} & same & J & S\\
(La,Ce)H$_{9}$ & DFT on substitutional alloy, template from CSP & $\sim$100 GPa & 2022 \citep{DiscBi2022Giant} & same & J & S\\
LaBeH$_{8}$ & structure search & 80 GPa & 2022 \citep{DiscZhang2022Design} & 2023 \citep{DiscSong2023Stoichiome} & P & S\\
(La,Ce)H$_{9}$-x & DFT on substitutional alloy & $\sim$100 GPa & 2023 \citep{DiscChen2023Enhancemen} & same & J & S\\
\multicolumn{7}{l}{\textbf{Other high-pressure compounds}}\\
cg-N & ab initio candidate enumeration, pre-global-search & >110 GPa, $\sim$2000 K & 1992 \citep{DiscMailhiot1992Polymeric} & 2004 \citep{DiscEremets2004Singlebond} & P & S\\
Transparent dense Na & USPEX & $\sim$200 GPa & 2009 \citep{DiscMa2009Transparen} & same & J & S\\
Li oC40 & CALYPSO & 60-80 GPa & 2011 \citep{DiscLv2011Predicted} & 2011 \citep{DiscGuillaume2011Cold} & J & S\\
HLP-N & CALYPSO$^\dagger$ & $\sim$244 GPa, $\sim$3300 K & 2012 \citep{DiscWang2012Cagelike} & 2019 \citep{DiscLaniel2019Hexagonal} & P & S\\
NaCl$_{3}$, Na$_{3}$Cl and other Na-Cl & USPEX & 20-80 GPa & 2013 \citep{DiscZhang2013Unexpected} & same & J & S\\
XeFe$_{3}$ / XeNi$_{3}$ & CALYPSO + AIRSS & $\sim$200 GPa & 2014 \citep{DiscZhu2014Reactions} & 2018 \citep{DiscStavrou2018Synthesis} & P & S\\
tI24-CN$_{2}$ & USPEX & >100 GPa, recovered & 2015 \citep{DiscDong2015The} & 2023 \citep{DiscLaniel2024Synthesis} & P & S\\
FeO$_{2}$ & USPEX & $\sim$76 GPa & 2016 \citep{DiscHu2016Feo} & same & J & S\\
Na$_{2}$He & USPEX & >113 GPa & 2017 \citep{DiscDong2017A} & same & J & S\\
WN$_{6}$ & ML-accelerated evolutionary search & 126-165 GPa & 2018 \citep{DiscXia2018A} & 2021 \citep{DiscSalke2021Tungsten} & P & S\\
LaB$_{3}$C$_{3}$ carbon-boron clathrate & computational modelling (analogue of SrB3C3) & high pressure & 2020 \citep{DiscStrobel2021A} & same & J & S\\
SrB$_{3}$C$_{3}$ carbon-boron clathrate & CALYPSO$^\dagger$ & high pressure & 2020 \citep{DiscZhu2020Carbonboro} & same & J & S\\
bp-N & DFT on candidate$^\dagger$ & $\sim$146 GPa & 2020 \citep{DiscLaniel2020Highpressu} & same & J & S\\
CaS$_{2}$ and CaS$_{3}$ & structure search & high pressure & 2021 \citep{DiscWang2021Synthesis} & same & J & S\\
\multicolumn{7}{l}{\textbf{Superhard and hard materials}}\\
M-carbon$^\ast$ & USPEX & $\sim$15-20 GPa (cold compression) & 2006 \citep{CSP027} & 2012 \citep{DiscWang2012Crystal} & S & D\\
gamma-B$_{28}$$^\ast$ & USPEX & $\sim$9-20 GPa synthesis (recoverable) & 2009 \citep{DiscOganov2009Ionic} & same & J & S\\
FeB$_{4}$$^\ast$ & other: ab initio evolutionary search (Kolmogorov; precursor of MAISE) & 8 GPa synthesis (recoverable) & 2010 \citep{DiscKolmogorov2010New} & 2013 \citep{DiscGou2013Discovery} & P & D\\
MnB$_{4}$ / MnB$_{3}$$^\ast$ & USPEX & ambient & 2014 \citep{DiscNiu2014Variableco} & same & J & S\\
TiN$_{2}$$^\ast$ & USPEX & $\sim$73 GPa synthesis (recoverable) & 2015 \citep{DiscYu2015Phase} & 2016 \citep{DiscBhadram2016Highpressu} & P & S\\
tI14-C$_{3}$N$_{4}$$^\ast$ & AIRSS & >100 GPa, recovered & 2016 \citep{DiscPickard2016Carbon} & 2023 \citep{DiscLaniel2024Synthesis} & J & S\\
Na$_{2}$B$_{30}$$^\ast$ & USPEX & ambient & 2018 \citep{DiscHe2018Predicting} & -- & S & C\\
W$_{2}$CrB$_{2}$ / W$_{4}$CrB$_{3}$$^\ast$ & unknown & ambient & 2018 \citep{DiscHu2018Atomicscal} & same & J & S\\
\multicolumn{7}{l}{\textbf{Energy-storage materials}}\\
delta-Mg(BH$_{4}$)$_{2}$$^\ast$ & USPEX & 2.1 GPa & 2012 \citep{DiscZhou2012Firstprinc} & 2011 \citep{DiscFilinchuk2011Porous} & S & S\\
Li$_{7}$Ge$_{3}$$^\ast$ & AIRSS & ambient & 2014 \citep{DiscMorris2014Thermodyna} & 2015 \citep{DiscJung2015Elucidatio} & P & S\\
beta-Li$_{15}$Si$_{4}$$^\ast$ & USPEX & $\sim$7 GPa synthesis (recoverable) & 2015 \citep{DiscZeng2015A} & same & J & S\\
\multicolumn{7}{l}{\textbf{Other inorganic crystals}}\\
Li/Na carbonophosphates & substitution/data mining & ambient & 2012 \citep{DiscChen2012Carbonopho} & same & J & C\\
Polar orthorhombic hafnia & structure search & ambient (ALD films) & 2014 \citep{DiscHuan2014Pathways} & 2026 \citep{DiscLee2026Emergence} & P & S\\
Zr$_{2}$Co$_{11}$$^\ast$ & other: adaptive genetic algorithm (AGA) & ambient & 2014 \citep{DiscZhao2014Exploring} & -- & S & C\\
LaWN$_{3}$ & high-throughput + global structure prediction & ambient (sputtering + annealing) & 2015 \citep{DiscSarmientoPerez2015Prediction} & 2021 \citep{DiscTalley2021Synthesis} & P & S\\
Ni$_{3}$InAs$^\ast$ & AIRSS & ambient & 2015 \citep{DiscSchusteritsch2015Firstprinc} & -- & S & C\\
ZrO$^\ast$ & AIRSS & ambient & 2015 \citep{DiscNicholls2015Crystal} & same & J & C\\
Ba$_{5}$Y$_{13}$[SiO$_{4}$]$_{8}$O$_{8.5}$ / two new structure types & probe-structure assembly + DFT & ambient & 2017 \citep{CSP070} & same & J & S\\
Sr$_{5}$P$_{3}$ electride$^\ast$ & USPEX & ambient & 2017 \citep{DiscWang2017Exploratio} & same & J & S\\
TaFeSb half-Heusler & structure search (inverse design)$^\dagger$ & ambient & 2019 \citep{DiscZhu2019Discovery} & same & J & S\\
i-MAB borides, Mo-(Sc,Y)-Al-B & chemical-ordering search (DFT) & ambient (1400 C, Ar) & 2020 \citep{DiscDahlqvist2020Theoretica} & same & J & S\\
La$_{2}$O$_{2}$S$_{1.5}$ and oA-La$_{2}$O$_{2}$S & evolutionary search & ambient (topochemical) & 2021 \citep{DiscSasaki2021Design} & same & J & S\\
Li$_{6}$SiO$_{4}$Cl$_{2}$ & layer-stacking models + energy minimization & ambient & 2021 \citep{DiscMorscher2021Lisubsubsi} & same & J & S\\
New phases from ML-chosen element combinations & ML (unsupervised element selection) & ambient & 2021 \citep{DiscVasylenko2021Element} & same & J & S\\
\multicolumn{7}{l}{\textbf{Molecular and organic crystals}}\\
Derivative-DNTT$^\ast$ & unknown & ambient & 2011 \citep{DiscSokolov2011From} & same & J & S\\
All-organic polymer dielectrics$^\ast$ & minima hopping & ambient & 2014 \citep{DiscSharma2014Rational} & same & J & C\\
Pharmaceutical polymorphs via CSP + high-pressure crystallization & GRACE dispersion-corrected DFT CSP & ambient (high-pressure crystallization) & 2015 \citep{DiscNeumann2015Combined} & same & P & S\\
epsilon-resorcinol$^\ast$ & USPEX & ambient & 2016 \citep{DiscZhu2016Resorcinol} & same & J & S\\
Coumarin II-V$^\ast$ & USPEX & ambient & 2017 \citep{DiscShtukenberg2017Powder} & same & J & S\\
Glycine dihydrate$^\ast$ & USPEX & ambient & 2017 \citep{DiscXu2017The} & same & J & S\\
T$_{2}$-gamma porous organic molecular crystal & CSP: lattice-energy global search, energy-structure-function maps & ambient & 2017 \citep{CSP071} & same & P & S\\
2D hybrid perovskite with cis-1,3-cyclohexanediammonium & random structure search + MLIP & ambient & 2024 \citep{DiscKarimitari2024Accurate} & same & J & S\\
Seventh CSP blind test targets & multiple CSP methods & ambient & 2024 \citep{CSP155} & same & P & S\\
\multicolumn{7}{l}{\textbf{Clusters, surfaces and 2D materials}}\\
B$_{36}$$^-$$^\ast$ & other: global-minimum cluster search (basin hopping / coalescence kick; Wang-Boldyrev groups) & gas phase & 2014 \citep{DiscPiazza2014Planar} & same & J & C\\
B$_{40}$$^-$$^\ast$ & other: global-minimum cluster search (basin hopping / TGMin; Wang-Li groups) & gas phase & 2014 \citep{DiscZhai2014Observatio} & same & J & C\\
Rutile TiO$_{2}$(110) reconstructions$^\ast$ & USPEX & ambient (UHV) & 2014 \citep{DiscWang2014New} & -- & S & C\\
Rutile TiO$_{2}$(011)-(2x$_{1}$)$^\ast$ & USPEX & ambient (UHV) & 2016 \citep{DiscWang2016The} & -- & S & C\\
2D tellurium$^\ast$ & unknown & ambient & 2017 \citep{DiscZhu2017Multivalen} & same & J & S\\
\multicolumn{7}{l}{\textbf{Machine-learning-era claims}}\\
A-Lab synthesized targets & ML-guided synthesis on Materials Project/GNoME targets & ambient & 2023 \citep{DiscMerchant2023Scaling} & 2023 \citep{DiscSzymanski2023An} & P & D\\
GNoME stable crystals & ML generative/GNN screening & ambient & 2023 \citep{DiscMerchant2023Scaling} & same & S & D\\
TaCr$_{2}$O$_{6}$ & ML generative (MatterGen) & ambient & 2025 \citep{CSP159} & same & J & C\\
\end{longtable}
\end{scriptsize}

\paragraph{What the record shows.} The 68 entries, one per material, span superconducting hydrides, unusual high-pressure chemistry, superhard and energy materials, molecular polymorphs, clusters, surfaces, and recent machine-learning claims (Figure~\ref{fig:discovery-timeline}). Fifty-two have structures confirmed by diffraction or microscopy. Only 23 are strictly prediction-first; most entries are joint theory--experiment studies (37) or structure solutions of materials that had already been synthesized (8). This distinction matters: a joint study shows that CSP can identify a structure consistent with experiment, whereas a prediction-first case shows that CSP can direct an experiment to a material that would otherwise not have been sought. Among the prediction-first cases, the median time from prediction to confirmation is three years, and the longest are a decade or more, as for CaH$_6$ \citep{DiscWang2012Supercondu,DiscMa2022Hightemper} and cubic-gauche nitrogen \citep{DiscMailhiot1992Polymeric,DiscEremets2004Singlebond}.
 
\paragraph{High-pressure hydrides as the clearest successes.} The most convincing prediction-first record is in compressed hydrides. H$_3$S was predicted with USPEX \citep{DiscDuan2014Pressurein} in 2014 and found to superconduct at 203~K \citep{DiscDrozdov2015Convention}; the clathrate LaH$_{10}$ was predicted by structure search \citep{DiscLiu2017Potential} in 2017 before its synthesis \citep{DiscGeballe2018Synthesis}; and CaH$_6$, first predicted with CALYPSO \citep{DiscWang2012Supercondu} in 2012, was confirmed with a $T_c$ of about 215~K at 172~GPa \citep{DiscMa2022Hightemper} and independently by a second group \citep{DiscLi2022Supercondu}. YH$_6$, YH$_9$, CeH$_9$, ThH$_{10}$, and the ternary LaBeH$_8$ followed the same path from prediction to synthesis \citep{DiscLi2015Pressurest,DiscTroyan2021Anomalous,DiscPeng2017Hydrogen,DiscKong2021Supercondu,DiscSalke2019Synthesis,DiscKvashnin2018Hightemper,DiscSemenok2020Supercondu,DiscZhang2022Design,DiscSong2023Stoichiome}. High pressure favours CSP. The enthalpy term $pV$ dominates differences between candidates, kinetic barriers matter less in diamond-anvil syntheses, and the target compositions are small. Sixteen of the 23 prediction-first entries are high-pressure phases, as are other striking results such as transparent dense sodium, ``forbidden'' sodium chlorides, and Na$_2$He \citep{DiscMa2009Transparen,DiscZhang2013Unexpected,DiscDong2017A}. CALYPSO-based searches contributed several of these high-pressure cases: Xe--Fe/Ni intermetallics predicted with CALYPSO and AIRSS \citep{DiscZhu2014Reactions} were later synthesized as Xe(Fe,Fe/Ni)$_3$ and XeNi$_3$ \citep{DiscStavrou2018Synthesis}, which supports the predicted compound formation rather than every structural detail; CaS$_2$ and CaS$_3$ were synthesized in a study motivated by their prediction \citep{DiscWang2021Synthesis}; and the lithium oC40 (Aba2-40) phase was predicted \citep{DiscLv2011Predicted} and observed \citep{DiscGuillaume2011Cold} independently, in papers published two days apart, so it is a case of agreement rather than a prediction-led synthesis. Recent single-crystal and Raman studies have also confirmed earlier predictions of dense nitrogen-rich phases: the layered polymeric nitrogen phase HLP-N \citep{DiscLaniel2019Hexagonal}, tungsten hexanitride WN$_6$ with armchair N$_6$ rings, predicted by a machine-learning-accelerated evolutionary search \citep{DiscXia2018A,DiscSalke2021Tungsten}, and the recoverable carbon nitrides tI14-C$_3$N$_4$ and tI24-CN$_2$, whose structures were solved by single-crystal diffraction seven to eight years after their prediction \citep{DiscPickard2016Carbon,DiscDong2015The,DiscLaniel2024Synthesis}.
 
\paragraph{Ambient-pressure discovery is harder.} At ambient pressure, metastable phases are accessible and synthesis routes rather than ground-state energies often decide what forms. CSP has nonetheless contributed to superhard borides, battery-electrode phases, electrides, and molecular polymorphs \citep{DiscNiu2014Variableco,DiscMorris2014Thermodyna,DiscWang2017Exploratio,DiscNeumann2015Combined}. Several of the best ambient successes combine prediction with a chemist's choice of where to search: probe-structure assembly led to new Y--Ba--Sr--Ca oxides \citep{CSP070}, machine-learned element selection guided the synthesis of new lithium conductors \citep{DiscVasylenko2021Element}, and energy--structure--function maps identified a porous molecular crystal before it was made \citep{CSP071}. Structure prediction has also guided functional ambient-pressure materials: the TaFeSb half-Heusler thermoelectric, an unreported composition within a known prototype, was identified by inverse design and synthesized with a record figure of merit \citep{DiscZhu2019Discovery}, and a random structure search driven by a machine-learning potential predicted the structure of a new two-dimensional hybrid perovskite that synthesis then confirmed \citep{DiscKarimitari2024Accurate}. Earlier predictions have also been realized without high pressure: LaWN$_3$, predicted in 2015 by high-throughput screening combined with global structure prediction to be a polar distorted perovskite, was synthesized as a piezoelectric nitride perovskite by reactive sputtering \citep{DiscSarmientoPerez2015Prediction,DiscTalley2021Synthesis}, and the polar Pmn2$_1$ phase identified for HfO$_2$ by a structure search in 2014 was observed in ultrathin Hf$_{0.5}$Zr$_{0.5}$O$_2$ films in 2026 \citep{DiscHuan2014Pathways,DiscLee2026Emergence}. Joint studies extend the list to chemically ordered i-MAB borides, metastable La$_2$O$_2$S phases obtained topochemically, and a hexagonal argyrodite, Li$_6$SiO$_4$Cl$_2$ \citep{DiscDahlqvist2020Theoretica,DiscSasaki2021Design,DiscMorscher2021Lisubsubsi}, although their searches were restricted to ordering, evolutionary or layer-stacking models rather than unconstrained global search. A ferrielectric PbZrO$_3$ structure, proposed and evaluated by DFT rather than found by a CSP search, was later observed by electron microscopy; because it was not produced by a CSP algorithm, it is not included in Table~\ref{tab:discoveries} or in the counts above \citep{DiscAramberri2021On,DiscLiu2023Translatio}. The seventh blind test shows that molecular CSP now routinely places experimental forms among its lowest-energy predictions \citep{CSP155}.
 
\paragraph{Claims of the machine-learning era.} Large-scale ML screening has produced claims of a different scale. GNoME reported hundreds of thousands of new stable crystals \citep{DiscMerchant2023Scaling}, and the A-Lab reported the autonomous synthesis of 41 targets \citep{DiscSzymanski2023An}. Subsequent analyses questioned whether these were new, ordered compounds or known and disordered phases, and whether the phase identification was reliable \citep{Leeman2024Challenges,Cheetham2024Artificial}. MatterGen reported the synthesis of a generated composition, TaCr$_2$O$_6$ \citep{CSP159}. We grade these entries as disputed or consistent rather than confirmed, and note that no generative model surveyed in Section~\ref{sec:generative-models} yet has an independently confirmed prediction-first discovery.
 
\paragraph{Verification standards.} The field's own failures show why the evidence grade is needed. Room-temperature superconductivity claims in a carbonaceous sulfur hydride and in nitrogen-doped lutetium hydride, and a related yttrium superhydride report, were retracted \citep{Snider2020CSH,Dasenbrock2023LuNH,Snider2021YH9}. These were experimental claims rather than CSP predictions, but they arose in the area where CSP has been most successful, and they show that agreement between a predicted property and a measurement is not the same as structural confirmation. Within Table~\ref{tab:discoveries}, FeB$_4$ and M-carbon illustrate the same point: the FeB$_4$ structure was confirmed but its superhardness was contested, and the M-carbon diffraction data fit several candidate structures similarly well \citep{DiscKolmogorov2010New,DiscGou2013Discovery,CSP027,DiscWang2012Crystal}.
 
\paragraph{Implications for method evaluation.} Most confirmed discoveries came from DFT-based global search with USPEX, AIRSS, or CALYPSO, often in close collaboration with experimental groups. Template and generative methods, which dominate recent benchmark tables, are almost absent from the confirmed record. This does not show that they cannot discover materials; it reflects that experimental validation takes years and that few of their predictions have yet been tested. It does mean that benchmark recovery, the metric used in Section~\ref{sec:benchmarks-main}, measures a necessary but not sufficient condition for discovery. A useful next step would be prospective benchmarks in which predictions are registered before synthesis is attempted.
 
\begin{figure}[!htb]
\centering
\includegraphics[width=\linewidth]{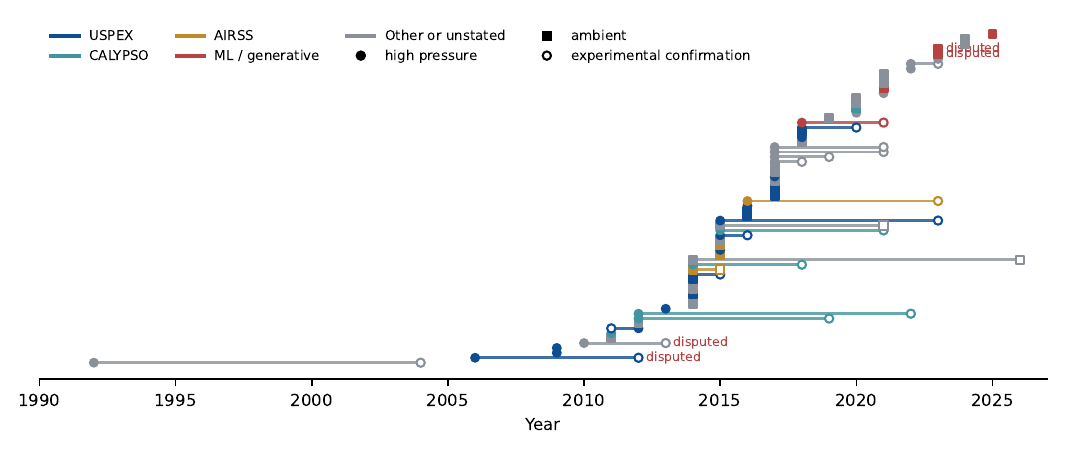}
\caption{Prediction-to-confirmation timeline for the entries in Table~\ref{tab:discoveries}. Each row is one entry; filled markers give the prediction year and open markers the first experimental confirmation (joint studies show a single marker). Circles denote high-pressure phases and squares ambient-pressure ones; colour gives the search method family. Entries graded as disputed are labelled.}
\label{fig:discovery-timeline}
\end{figure}
 
\section{Software, model releases, and reproducibility}
Reproducing these comparisons requires the full workflow to be specified, including the search code, relaxation engine, model parameters, training data, cell bounds, and stopping criteria. XtalOpt's release series \citep{CSP053,CSP158,CSP196} exemplifies why version identification matters; modular environments such as CrySPY \citep{CSP102} and AGOX \citep{CSP112} show how several algorithms may coexist within one package.
 
The companion repository's software directory lists 129 implementations and software components (snapshot of 17 October 2026) with their source code, license, checkpoints, datasets, and benchmarks. The accompanying CSV preserves a separate 122-entry export. Public code is not always open-source licensed, a downloadable checkpoint does not show that the released inference path works, and a checkpoint for an optional evaluator is not the generator's own weights; these differences matter most for hybrids assembled from several packages.
 
Classical searches usually have no training data or checkpoint to release. For learned methods, a useful reproduction record identifies the exact data release and split, configuration, model weights, and any additional teacher or reward models. For all methods, reporting hardware and wall time should be accompanied by algorithmically meaningful costs such as relaxation counts or first-principles evaluations.
 
These implementation requirements also inform practical method selection within the task restrictions. If a suitable structural family is known, template construction or bounded ordering can provide informative baselines. If broad fixed-composition exploration is needed, random, population, and landscape methods offer distinct ways to spend the evaluator budget. If a trained generative model covers the chemistry and cell regime, it can supply proposals that are then ranked and refined. Molecular packing requires corresponding conformational and intermolecular models. Newer methods add options for these tasks without replacing the older ones.
 
\section{Discussion and conclusions}
The methodological and reproducibility considerations reviewed here point to a common development: CSP increasingly combines reusable components across algorithmic layers. Symmetry and chemical construction support random, evolutionary, and generative proposals. Learned potentials can accelerate several search families. Adaptive policies can allocate expensive evaluations, and trained generators can supply starting structures for subsequent optimization. As a result, the line between a search algorithm and a complete prediction workflow matters more when methods are compared.
 
The main conceptual divide is between exploring an objective landscape for a particular target and learning a proposal distribution from previous structures. Classical searches spend substantial effort on the target-specific evaluator, whereas learned generators transfer information from their training corpus. Hybrids combine these sources of knowledge. Their value depends on the regime: prior structural regularities may make proposals efficient, while unfamiliar chemistries, larger cells, or new structural families can expose limitations of the learned distribution.
 
Exact conditioning is a second organizing principle. Formula ratios, cell counts, space groups, templates, conformers, and experimental observations are different forms of information. They should be treated as part of the task definition and reported at the point where they enter the algorithm. Conditional deep generative models belong within CSP when they predict the remaining structural degrees of freedom under that definition. Unconditional generation and reconstruction are related tasks, but success on them does not demonstrate prescribed-formula prediction.
 
\subsection{Weaknesses of current CSP algorithms}\label{sec:weaknesses}
The CSP methods reviewed here share several limitations that recur across classical search, template methods, and learned generators. The most immediate limitation concerns how candidates are judged and ranked. Many methods generate a correct structure but do not rank it first: on the T46 leaderboard, the leading method recovers 35 targets at top-1 but 41 within its top 20 candidates \citep{CSPLeaderboardSnapshot2026}. Most generators show smaller gaps (Figure~\ref{fig:t46-performance}), partly because their candidate lists rarely contain a correct structure that the first-ranked candidate misses, so improving ranking and improving candidate diversity are separate problems. Final selection usually rests on a single static-lattice energy from DFT or a machine-learning potential, although competing polymorphs often differ by only a few meV per atom. Machine-learning interatomic potentials make large searches affordable \citep{CSP085}, but their errors are largest for unusual bonding, high pressures, and structures far from their training data, which is precisely where new phases are sought, and few workflows estimate when a learned energy should not be trusted.
 
A second group of weaknesses concerns what is predicted and how far it scales. Most methods target ordered, defect-free, zero-temperature ground states of a fixed composition. Vibrational and configurational entropy, partial occupancy, defects, interfaces, metastability, and the synthesis route are seldom part of the objective, yet experiments often produce disordered or kinetically selected phases (Section~\ref{sec:discoveries}). The idealized targets are also small. Global-search cost grows rapidly with the number of atoms and species, and many deep generative models are trained on MP-20, whose cells contain at most 20 atoms, so they cannot represent larger cells at all; reported accuracy therefore concentrates on small, low-arity structures. Learned generators are further tied to their training distribution: controlled tests of prototype dependence indicate that recovery is linked to structure types represented in the training data \citep{CSP185}, so success on benchmarks that overlap the training data does not demonstrate the discovery of new structure types.
 
A final group concerns whether results can be trusted, compared, and validated. Implementations can fail silently: an audit of TCSP~2.0 found that 20\% of its selected structures had the wrong composition because of element-mapping and text-substitution errors \citep{Feng2026TCSP3}, and similar errors in formula handling, cell multiplicity, or file conversion could affect any pipeline but are rarely tested. Comparisons are hard to reproduce because benchmarks differ in targets, matcher tolerances, candidate budgets, relaxation protocols, and the treatment of failed runs \citep{CSP180,CSPMetrics2024}, and computational cost is rarely reported in a comparable form (Section~\ref{sec:benchmarks-main}). Most importantly, prospective validation remains scarce. Most experimentally confirmed predictions came from DFT-based global search at high pressure, and no deep generative model yet has an independently confirmed prediction-first discovery (Section~\ref{sec:discoveries}).
 
\subsection{Future CSP research directions}\label{sec:todo}
The weaknesses of current CSP algorithms discussed above suggest a concrete agenda, and its first step is honest measurement. A CSP result should state what information the method receives, count every evaluator call and candidate, report wall time and hardware, and keep failed or unsupported targets in the denominator; only then can methods with very different internal costs be compared. Within that accounting, generation and ranking should be reported separately, through top-1 and top-$k$ recovery side by side, because they fail for different reasons and call for different remedies. Ranking in particular needs evaluators that combine free-energy corrections with calibrated uncertainty, so that expensive calculations are spent on the candidates a cheap model cannot resolve.
 
Better measurement also requires better tests. Molecular CSP already has blind tests that separate generation from ranking \citep{CSP155}, and inorganic CSP needs the equivalent: targets withheld from all training data, audited for leakage, and registered before their experimental structures are released. Such tests should report performance by cell size, number of elements, and structure type, including prototypes absent from training, so that extrapolation is measured separately from interpolation. The targets themselves should move closer to what experiments produce. Finite-temperature free energies, metastability windows, disorder, defects, and synthesis constraints belong in the prediction task rather than in post hoc analysis.
 
Meeting these targets will likely require hybrid methods. Learned proposals embedded in population or landscape searches \citep{Zheng2026LatentGA,Shibayama2026DiversityGA}, reinforcement post-training of generators \citep{CSP182}, and active-learning potentials \citep{CSP085} each combine data-driven efficiency with physical search, and they are most useful when every component is validated and its cost counted. Because such pipelines have many parts, they should also be tested as software, with automatic checks that every candidate preserves the requested composition and cell content, and released with versioned code, checkpoints, data, and matcher settings. Above all, the field needs to close the loop with experiment by registering predictions before synthesis is attempted, publishing negative as well as positive outcomes, and maintaining a shared record of examined predictions, for which Table~\ref{tab:discoveries} is a starting point.
 
\subsection{Concluding remarks}
Finally, progress in CSP depends on how it is evaluated. The separation of generation and ranking in molecular blind tests, controlled comparisons under shared evaluators, and recent attention to prototype dependence each clarify a different source of success or failure. Future comparisons will be more informative when they retain all failed targets, distinguish model exposure from true extrapolation, and connect computational cost to independently assessed structural quality. Comparisons of this kind would let the complementary strengths of the algorithms reviewed here be measured directly.
 
\section*{Data availability}
The curated catalogue of the surveyed papers (identifier, year, title, method family, domain, publication status, DOI and source URL), the directory of software implementations with their code, license, checkpoint and dataset links, the benchmark and dataset records, and the structural-prior and formula-conditioning classifications of the methods are maintained at \url{https://github.com/usccolumbia/awesome-crystal-structure-prediction}.
 
\section*{Author contributions}
Conceptualization, J.H.; methodology, J.H.; investigation, J.H., Y.F.; data curation, J.H., Y.F.; visualization, J.H.; writing--original draft preparation, J.H.,Y.F.; writing--review and editing, J.H., Y.F.; supervision, J.H.; funding acquisition, J.H.
 
\section*{Acknowledgement}
The research reported in this work was partially supported by the National Science Foundation under grant no. 2311202. The views, perspectives and content do not necessarily represent the official views of the NSF. We would like to thank Anthropic's Claude Team plan for scientists, a specialized program for academic and nonprofit research labs across all scientific fields: math, physics, chemistry, biology, and computer science. We also greatly appreciate the support from OpenAI for the ChatGPT for Academic Researchers program. We thank Remy Rogers for careful proofreading and improvements to the manuscript.
 
\section*{Competing interests}
The authors declare no competing interests.
\bibliographystyle{unsrtnat}
\bibliography{references}
\end{document}